\documentclass[usegraphicx,usedcolumn,useAMS,usenatbib]{mn2e}

\title[Metal Enrichment of the ICM]{Metal Enrichment of the ICM:
a 3-D Picture of Chemical and Dynamical Properties}

\author[S. A. Cora]{Sof\'{\i}a A. Cora$^{1,2}$\thanks{E-mail:
sacora@fcaglp.unlp.edu.ar}\\
$^{1}$Facultad de Ciencias Astron\'omicas y Geof\'{\i}sicas de la Universidad 
Nacional de La Plata and Instituto de Astrof\'{\i}sica de La Plata,\\ 
Observatorio Astron\'omico, 
Paseo del Bosque S/N, 
1900 La Plata, Argentina\\
$^{2}$Consejo Nacional de Investigaciones Cient\'{\i}ficas y T\'ecnicas,
Rivadavia 1917, Buenos Aires, Argentina} 
\begin{document}

\date{}

\pagerange{\pageref{firstpage}--\pageref{lastpage}} \pubyear{2005}

\maketitle

\label{firstpage}

\begin{abstract}
We develop a model for the metal enrichment of the intracluster
medium (ICM) that combines a cosmological non-radiative hydrodynamical
{\em N}-Body/SPH simulation of a cluster of galaxies, and a
semi-analytic model of galaxy formation.  
The novel feature of our hybrid model is that the chemical properties 
of the diffuse gas in the underlying simulation
are dynamically and consistently generated from stars in the galaxies.
We follow the production of several chemical elements,
provided by low- and
intermediate-mass stars, core collapse and type
Ia supernovae.
We analyse the spatial distribution of metals in the ICM,
investigate the way in which the chemical enrichment proceeds,
and use iron emissivity as a tracer of gas motions.  
The similar abundance patterns developed by O
and Fe indicate that both types of SNe pollute the
ICM in a similar fashion.
Their radial abundance profiles
are enhanced in the inner $100\, h^{-1}$~kpc in the last Gyr
because of the convergence of enriched gas clumps to the cluster centre;
this increment cannot be explained by the metal ejection of cluster
galaxies which is quite low at the present epoch.
Our results support a scenario in which
part of the central intracluster gas
comes from gas clumps that, in the redshift range of $z \sim 0.2$ to $\sim 0.5$,
have been enriched to solar values and
are at large distances from the cluster centre
(from $\sim 1$ to $\sim 6 \,h^{-1}$~Mpc) moving at very
high velocities (from $\sim 1300$ to $\sim 2500 \,{\rm km} \,{\rm s}^{-1}$).
The turbulent gas motions within the cluster,
originated in the inhomogeneous gas infall 
during the cluster assembly, are manifested in 
emission-weighted velocity maps as gradients 
that can be as large as $\sim 1000 \,{\rm km} \, {\rm s}^{-1}$ over
distances of a few hundred kpc. 
Gradients of this magnitude are also
seen in velocity distributions along sightlines through the cluster
centre. Doppler shifting and broadening suffered by the Fe ${\rm
K}_{\alpha}$ 6.7 keV emission line along such sightlines could be used to
probe these gas large-scale motions when they are produced 
within an area
characterised by high iron line emissivity.
\end{abstract}

\begin{keywords}
galaxies: formation - galaxies: evolution - galaxies: cluster: general
\end{keywords}

\section{Introduction}\label{sec_intro}

Clusters of galaxies are the largest virialised objects in the
Universe.  Within the hierarchical structure formation scenario of the
cold dark matter (CDM) model, they arise from the rare highest peaks
of the initial density perturbation field.  During cluster assembly,
the dissipative gas component is shock heated by the potential well of
the cluster, reaching very high temperatures ($T\sim 10^{7-8}$ K).
This hot and diffuse gas radiates energy through thermal
bremsstrahlung in the X-ray band of the electromagnetic spectrum. The
gravitationally dominating collisionless dark matter component on the
other hand shows considerable substructure, as revealed by high
resolution simulations that follow cluster assembly
(e.g. \citealt{springel01} and references therein). These
substructures are the surviving remnants of dark matter haloes that
have fallen in at earlier times, and plausibly mark the location of
the cluster galaxies.

X-ray observations have first revealed the presence of hot gas trapped
in the potential well of galaxy clusters, and provide a wealth of
information about the dynamical, thermal and chemical properties of
clusters.  A strong feature in their spectra is the presence of line
emission produced by highly ionised iron, mainly, the H- and He-like
iron lines at 6.9 and 6.7 keV.  The content of iron in the
intracluster medium (ICM) is approximately one third of the solar
value (\citealt{fukazawa98}; Ettori, Allen \& Fabian 2001), 
suggesting that some
of its hot gas must have originated in the galaxies that reside in the
cluster.  The processes considered for the supply of this enriched gas
include galactic winds driven by supernovae explosions
(\citealt{white91,renzini97}), ram pressure stripping
\citep{moriburkert00}, early enrichment by hypernovae associated
with population type III stars \citep{loewenstein01},
and intracluster stars \citep{zaritsky04}.

Spatially resolved maps of the temperature and metal abundance
distributions in the intracluster medium have been obtained by X-ray
surveys carried out with {\em ROSAT}, {\em ASCA}, {\em BeppoSAX}
(Finoguenov, David \& Ponman 2000; Finoguenov, Arnaud \& David 2001;
\citealt{degrandi01}), and more recently with {\em XMM-Newton} and
{\em Chandra} (\citealt{gastaldello02,finoguenov02,
sandersfabian02,matsushita03};
\citealt{tamura04}). As a result, abundance profiles of several
chemical species with different origins have been obtained, namely for
O and Si generated by core collapse supernovae (SNe CC) and Fe,
mainly produced by type Ia supernovae (SNe Ia).  These profiles
reflect physical mixing processes in the baryonic component due to gas
cooling, star formation, metal production, and energetic and chemical
feedback, all coupled to the hierarchical build up of the galaxies and
the clusters.  These observations can therefore be very valuable for
constraining models of galaxy formation and 
chemical enrichment of the ICM, allowing to 
evaluate the relative importance of different 
types of supernovae for metal pollution.
The spatial distribution of metals obtained from models
of the ICM chemical enrichment
can be analysed by constructing X-ray weighted metal maps
which trace the
interactions between the ICM and cluster galaxies; thus,
when   
they are compared with observed metal maps, they
provide information about the mechanisms involved in the enrichment 
process and may reveal the dynamical state of a galaxy cluster
(\citealt{kapferer05,domainko05}).

X-ray observations also provide important information about the
dynamics of the intracluster gas through radial velocity measurements
in clusters \citep{dupke01}, allowing in principle further tests of
structure formation models.  Gaseous bulk flows affect the
characteristics of X-ray spectra of the ICM by Doppler broadening and
shifting of emission lines. The prominent Fe ${\rm K}_{\alpha}$ 6.7
keV iron emission line present in X-ray spectra of galaxy clusters
could be used as a tracer of these motions. This possibility was
evaluated by \citet{Sunyaev03}, who showed that future X-ray missions
like {\em CONSTELLATION-X} and {\em XEUS}, with an energy resolution
of a few eV, should be able to detect several Doppler-shifted
components of this emission line in the core of the cluster.

All this wealth of data calls for a theoretical interpretation of the
ICM chemical enrichment in the framework of a detailed cluster
formation model.  The approaches used so far to this end include
semi-analytic models of galaxy formation coupled to {\em N}-Body
simulations of galaxy clusters (De Lucia, Kauffmann \& White 2004;
\citealt{nagashima05}), hydrodynamical simulations of cluster
formation which include, self-consistently, star formation, SNe
feedback and metal enrichment from SNe CC and Ia
(e.g.~\citealt{valdarnini03,tornatore04}), and  
an intermediate approach that combines {\em N}-Body and 
hydrodynamic simulations together
with a phenomenological galaxy formation model and a prescription of the
effect under study, such as galactic winds and
merger-driven starbursts \citep{kapferer05} and
ram-pressure stripping \citep{schindler05, domainko05}.
In this work,
we present a different intermediate approach which combines a cosmological
non-radiative hydrodynamical {\em N}-Body/SPH simulation of a cluster
of galaxies, with a semi-analytic model of galaxy formation.  The
simplicity of the semi-analytic model has the advantage of reaching a
larger dynamic range than fully self-consistent hydro-simulations, at
a far smaller computational cost. In particular, it allows us to
explore more easily the range of parameters that characterise
appropriate chemical enrichment models.

The long cooling time of the bulk of the gas in rich clusters
justifies the assumption of a non-radiative gas (\citealt{evrard90,
frenk99}).  However, this time-scale becomes smaller than the
Hubble time in the core of the cluster, where the densities are
higher.  Here is where the semi-analytic model plays an important
part, taking into account radiative cooling, and models for star
formation, chemical enrichment and energetic feedback from galaxies.
The semi-analytic model used here is based on previous works
(\citealt{springel01}; \citealt{lucia04}, DL04 hereafter), but was
extended with a new chemical implementation that tracks the abundance
of different species resulting from mass loss through stellar winds of
low-, intermediate- and high-mass stars, and from the explosions of
SNe CC and Ia supernovae. The principal new feature of our model is
however the link between semi-analytic model results and the chemical
enrichment of gas particles in the underlying {\em N}-Body/SPH
simulation. We pollute gas particles with metals ejected from the
galaxies, consistent with the modelling of the semi-analytic
model. These metals are then carried around and mixed by the
hydrodynamic processes during cluster formation.  Thus, we are not
limited to an analysis of mean chemical properties of the intracluster
gas, like in the original model of DL04.  Instead, the enrichment of
gas particles allows a study of the spatial distribution of metals in
the ICM, and to use iron emissivity as a tracer of gas motions.

The aim of our model is, on one hand, to understand the way in which
the chemical enrichment patterns characterising the intracluster gas
develop, and, on the other, to detect the imprints of gas bulk motions
in the shape of X-ray emission lines, thus providing guidance for the
interpretation of future spectroscopic X-ray data.  The first part of
the present study is therefore devoted to an analysis of chemical
enrichment of the ICM, focusing on the spatial distribution of
different chemical elements.  In a second part we analyse the link
between the occurrence of multiple components in the Fe ${\rm
K}_{\alpha}$ 6.7 keV emission line and the range of velocities of the
gas that generates this emission.  This is directly relevant to the
question whether metal emission-line-weighted velocity maps of the
cluster could be used to infer the spatial coherence of these bulk
motions.  Gathering all the information available in the form of
projected maps and detailed spectra along lines of sight through the
ICM, we are able to extend the information in 2-D provided by observed
projected distributions of gas temperature, abundances and surface
brightness, into a 3-D picture of the ICM properties.

This paper is organised as follows.  Section~\ref{sec_hybrid}
describes our hybrid model used to study the ICM chemical enrichment,
summarises the properties of the hydrodynamical simulation used and
presents the main features of the semi-analytic model.
Section~\ref{sec_ChemImpl} contains the details of the chemical model
implemented in the semi-analytic model. It also describes the way in
which metals released by stellar mass losses and supernovae explosions
are spread among gas particles of the {\em N}-Body/SPH simulation.
In Section~\ref{sec_CharSAM}, we fix the parameters that characterise
our version of the semi-analytic model and compare its results with
local observations.  The following three sections present the main
results of this work.  Section~\ref{sec_radialprof} gives the radial
abundance profiles of different chemical elements, comparing them with
observations. 
In Section~\ref{sec_Maps}, we
visualise the thermodynamical and chemical properties of the ICM by
projecting the corresponding mass-weighted or emission-weighted
related quantities.  These two techniques (radial profiles and
projected maps) are used to analyse the history of the ICM chemical
enrichment as well as the dynamical evolution of the intracluster gas, 
as discussed in Section~\ref{sec_EvolMaps}.
Section~\ref{sec_Spectra} describes the use of Fe ${\rm K}_{\alpha}$
6.7 keV emission line as a probe of the ICM dynamics.  Finally, in
Section~\ref{sec_Conclu}, we give a summary and discussion of our
results.

Throughout this paper we
express our model chemical abundances in terms of the solar photospheric ones
recently recalibrated by
Asplund, Grevesse \& Sauval (2005).
When necessary, we refer the chemical
abundances obtained in different observational works to these standard solar
values.

\section[]{Hybrid  model of the ICM chemical enrichment}\label{sec_hybrid}

Our hybrid model for studying the chemical enrichment of the
intracluster medium consists of a combination of a non-radiative {\em
N}-Body/SPH simulation of a galaxy cluster and a semi-analytic model
of galaxy formation.  Dark matter haloes and substructures that emerge
in the simulation are tracked by the semi-analytic code and used to
generate the galaxy population \citep{springel01}.  The key additional
component of our model is that we also use the diffuse gas component
of the hydrodynamical simulation to follow the enrichment process. By
enriching the gas particles locally around galaxies, we can account
for the spreading and mixing of metals by hydrodynamical processes,
thereby obtaining a model for the evolution of the spatial
distribution of metals in the ICM.  This in turn allows a more
detailed analysis of the ICM metal enrichment than possible with
previous models based on semi-analytic techniques (\citealt{kauff98};
DL04). In this section, we provide a detailed description of both
parts of the modelling.

\subsection[]{Non-radiative {\em N}-Body/SPH cluster simulation}\label{sec_NBody}

In the present study, we analyse a non-radiative hydrodynamical
simulation of a cluster of galaxies. The
cluster was selected from the GIF-$\Lambda$CDM model \citep{kauff99},
characterised by cosmological parameters $\Omega_{\rm o}$=0.3,
$\Omega_{\Lambda}$=0.7 and ${\rm H}_{\rm o}= 100 \, h \, {\rm km} \,
{\rm s}^{-1} \, {\rm Mpc}^{-1}$, with $h=0.7$; it has spectral shape
$\Gamma=0.21$, and was cluster-normalised to $\sigma_8=0.9$.  The
second most massive cluster in this parent cosmological simulation (of
mass $M_{\rm vir}=8.4 \times 10^{14}\, h^{-1} \,{\rm M}_{\odot}$) was
resimulated with higher resolution in the Lagrangian region of the
object and its immediate surroundings \citep{springel01},
with a number $N_{\rm hr}=1999978$ of high resolution particles.  
Within this
region, each particle was split into dark matter 
($m_{\rm dm}=1.18 \times 10^{9} \, h^{-1} {\rm M}_{\odot}$)
and gas 
($m_{\rm gas}=1.82 \times 10^{8} \, h^{-1} {\rm M}_{\odot}$)
according to $\Omega_{\rm b}=0.04$ and $\Omega_{\rm DM}=0.26$,
consistent with the baryon density $\Omega_{\rm b} h^{2}=0.02$
required by Big-Bang nucleosynthesis constraints.  However, the
identification of dark matter haloes for the semi-analytic model was
based only on the dark matter particles, with their mass increased to
its original value.  The boundary region, where mass resolution
degrades with increasing distance from the cluster, extends to a total
diameter of $141.3 \, h^{-1} \, {\rm Mpc}$.  The simulation was
carried out with the parallel tree {\em N}-Body/SPH code {\small
GADGET} (\citealt{springel01, springel05}). 
The starting redshift and the gravitational softening length are
$z_{\rm start}=50$ and $\varepsilon=3.0 \, h^{-1} \, {\rm kpc}$, respectively.
We note that this simulation did not include gas cooling.  This is a
reasonable approximation for cluster simulations since the bulk of the
halo gas has very long cooling times. However, the cooling time
becomes shorter than the age of the Universe in a central region with
radius $\sim 100 - 200$ kpc, from where the bulk of X-ray radiation
is emitted \citep{sarazin86}.

\subsection[]{Semi-analytic model of galaxy formation}\label{sec_SAM}

The properties of galaxies in the semi-analytic model are determined
by the included physical processes.  We consider the effects of
cooling of hot gas due to radiative losses, star formation, feedback
from supernovae explosions, metal production, and merging of galaxies.
Except for the metal production, the parametrization of these physical
processes corresponds to the model described by
\citet{springel01}. However, we modified the model to track the
enrichment cycle of metals between the different baryonic components
of the haloes, i.e., cold gas, hot diffuse gas, and stars, in the most
consistent way possible within our framework.  Our specific chemical
implementation is similar to the semi-analytic model discussed by
DL04.
In the present work, we refine the chemical enrichment prescription of
DL04 by tracking the mass evolution of different chemical elements as
provided by three kinds of sources: low- and intermediate-mass stars,
core collapse supernovae, and type Ia supernovae.  The first group of
sources yields metals through mass losses and stellar winds.  In the
following, we briefly summarise the details of the semi-analytic
code. In Section~\ref{sec_ChemImpl}, we then describe the new chemical
implementation we introduce in this work.

Based on the merging trees, the mass of hot gas is calculated at the
beginning of the evolution between consecutive outputs of the
simulation. We assume that the hot gas always has a distribution that
parallels that of the dark matter halo, whose virial mass changes from
one output to another due to the hierarchical growth of structure.
Once a fraction of hot gas has cooled and star formation and feedback
processes are triggered, the mass of hot gas is given by
\begin{equation}
M_{\rm hot}=f_{\rm b}\,M_{\rm vir} - \sum_i{[ M_{\rm stellar}^{(i)} + M_{\rm 
cold}^{(i)}]}, 
\label{eqSAM1}
\end{equation} 
where $M_{\rm vir}$ is the virial mass of the dark matter halo,
$M_{\rm hot}$ is the mass of the hot gas associated with it and
$M_{\rm stellar}$ and $M_{\rm cold}$ are the masses of stars and cold
gas of each galaxy contained in the halo.  The virial mass is given by
$M_{\rm vir}= 100 H^2 R_{\rm vir}^3/G$.  This is the mass enclosed by
the virial radius $R_{\rm vir}$, which is defined as the radius of a
sphere with mean density $200 \rho_{\rm crit}$ centred on the most
bound particle of the group. Here $\rho_{\rm crit}$ is the critical
density. 
The virial velocity is given by $V_{\rm vir}^2=G M_{\rm
vir}/R_{\rm vir}$.

The mass of hot gas that cools at each snapshot is given by the
cooling rate
\begin{equation}
\frac{{\rm d}M_{\rm cool}}{{\rm d}t}= 4 \pi \rho_{\rm g} r_{\rm cool}^2
\frac{{\rm d}r_{\rm
cool}}{{\rm d}t}, \label{eqSAM2}
\end{equation}
where $\rho_{\rm g}$ is the density profile of an isothermal sphere
that has been assumed for the distribution of hot gas within a dark
matter halo, and $r_{\rm cool}$ is the cooling radius.
The local cooling time is defined as the ratio of the specific
thermal energy content of the gas, and the cooling rate per unit
volume $\Lambda (T,Z)$.  The latter depends quite strongly on the
metallicity $Z$ of the hot gas and the temperature
$T=35.9 (V_{\rm vir}/{\rm km} \, {\rm s^{-1}})$ of the halo, and is
represented by the cooling functions computed by \citet{suth93}.

The star formation rate is given by 
\begin{equation}
\frac {{\rm d}M_{\star}}{{\rm d}t} = \frac {\alpha M_{\rm cold}} {t_{\rm dyn}^{\rm gx}}, 
\label{eqSAM3}
\end{equation}
where $t_{\rm dyn}^{\rm gx}= 0.1 \, R_{\rm vir}/V_{\rm vir}$ is the
dynamical time of the galaxy, and $\alpha$ is a dimensionless
parameter that regulates the efficiency of star formation.  We adopt
the variable star formation efficiency introduced by DL04 that depends
on the properties of the dark halo, $\alpha(V_{\rm vir})=\alpha_0
\,(V_{\rm vir}/220\,{\rm km}\,{\rm s^{-1}})^{n}$, with $\alpha_0$ and
$n$ as free parameters.

Each star formation event generates a stellar mass $\Delta M_{\star} =
\dot{M_{\star}} \Delta T / N_{\rm steps}$, where $N_{\rm step}=50$ are
timesteps of equal size used to subdivide the intervals between
simulation outputs, $\Delta T$, and to integrate the differential
equations that describe the changes in mass and metals of each
baryonic component over these timespans.  Each solar mass of stars formed
leads to a number $\eta_{\rm CC}$ of core collapse supernovae.  This
class of supernovae includes those of type Ib/c and II, the former
being generated when the progenitor looses its hydrogen-rich envelope
before the explosion. Ordinary type II explosions are the most
abundant ones however.  The energy $E_{\rm SNCC}$ released by each
core collapse supernova is assumed to reheat some of the cold gas of a
galaxy, with a mass of
\begin{equation}
 \Delta M_{\rm reheat}=\frac{4}{3} \epsilon \frac {\eta_{\rm CC}\, E_{\rm
SNCC}}{V_{\rm vir}^2}\, \Delta M_{\star},
\label{eqSAM4}
\end{equation}
where $\epsilon$ is a dimensionless parameter that regulates the
efficiency of the feedback process.

The fate of the reheated gas is quite uncertain. We adopt a model in
which the transfer from the cold to the hot phase does induce galactic
outflows, i.e.~the reheated mass is kept within its host halo. This
model is referred to as retention model.
The resulting cycle of metal enrichment is consistent with the way in
which metals provided by the semi-analytic model are deposited in the
gas particles around each galaxy in the underlying {\em N}-Body/SPH
simulation, described in Section~\ref{sec_injection}.
The effect of other feedback schemes (ejection and wind models)
on the chemical enrichment of the ICM is discussed by DL04.
They found that these three prescriptions (including the retention model)
predict a similar time evolution of the ICM metal pollution, which mainly
occurs at redshifts larger than 1. As we show in the present study, 
gas dynamical processes driven during the cluster assembly occurring at 
present epochs contribute significantly to determine the
spatial distribution of metals in the ICM, very likely erasing any signature 
of the feedback scheme involved.

In a hierarchical scenario of structure formation, mergers of galaxies
are a natural consequence of the accretion and merger processes of
dark matter haloes in which they reside. We directly use the simulation
outputs to construct the merger histories of dark haloes.  To this end
we first identify dark haloes as virialised particle groups by a
friend-of-friend (FOF) algorithm. The {\rm SUBFIND} algorithm
\citep{springel01} is then applied to these groups in order to find
self-bound dark matter substructures, and the resulting set of
gravitationally bound structures is tracked over time to yield merging
history trees.

In this subhalo scheme, we distinguish three types of galaxies when
tracking galaxy formation along the merging trees.  The largest
subhalo in a FOF group hosts the `central galaxy' of the group; its
position is given by the most bound particle in that subhalo. Central
galaxies of other smaller subhaloes that are contained in a FOF group
are referred to as `halo galaxies'.  The (sub)haloes of these galaxies
are still intact after falling into larger structures.  The third
group of galaxies comprises `satellite' galaxies.  This type of
galaxies is generated when two subhaloes merge and the galaxy of the
smaller one becomes a satellite of the remnant subhalo.  These
galaxies are assumed to merge on a dynamical time-scale with the
halo-galaxy of the new subhalo they reside in (\citealt{springel01};
DL04). In these previous versions of the semi-analytic model,
satellite galaxy positions were given by the most-bound particle
identified at the last time they were still a halo galaxy.  In the
present work, we instead assume a circular orbit for these satellite
galaxies with a velocity given by the virial velocity of the parent
halo and decaying radial distance to the corresponding central
galaxy. This provides a more robust estimate of the position of
satellite galaxies, consistent with the dynamical friction formula
used \citep{binney87}.

With respect to the spectro-photometric properties of galaxies, we
apply the calculations made by DL04, considering evolutionary
synthesis models that depend on the metallicity of the cold gas from
which the stars formed \citep{BruzualCharlot93}.  Our morphological
classification of galaxies is based on the criterion adopted by
\citet{springel01}, who used `shifted' values of the Hubble-type $T$
of galaxies to obtain a good morphology density relation.  Thus, we
classify as S0 galaxies those with $0 < T < 5$, and as elliptical and
spirals galaxies those with lower and higher values, respectively.

\section[]{Chemical implementation}\label{sec_ChemImpl}

One of the main aims of this work is to reach a better understanding
of how the chemical enrichment of the ICM proceeds by establishing a
connection between the spatial distribution of several properties of
the ICM, such as surface mass density, mass-weighted temperature
distribution, X-Ray surface brightness and chemical abundances.

Recent observations of radial abundance profiles of different elements
(\citealt{finoguenov00,degrandi01,gastaldello02,degrandi03,tamura04}) 
provide valuable constraints on the physical
processes involved, especially those related to the feedback
mechanisms that inject metals into the diffuse phase.  The available
observational data calls for a model that can explain the spatial
distribution of different chemical elements. Thus, it is essential to
implement a chemical model that takes into account several different
elements, and discriminates between different sources, such as SNe CC
and SNe Ia.

In our model, information about the spatial distribution of the ICM
properties can be obtained from the gas particles in the
hydro-simulation. However, the non-radiative simulation used only
give the thermodynamic properties of the diffuse hot gas.  In order to
obtain the chemical properties of gas particles, we have to establish
a link between the chemical implementation in the semi-analytic model
and the metal enrichment of the gas particles. In the following, we
will first describe in detail how we model the enrichment cycle of
metals among the different baryonic components in the semi-analytic
model, and then explain how these quantities are tied to the gas
dynamics in the simulation.

\subsection[]{Enrichment cycle of chemical elements in the semi-analytic model}\label{sec_Circ}

In our model, stars can contaminate the cold and hot gas because of
mass loss during their stellar evolution and metal ejection at the end
of their lives.  The hot gas has primordial abundances initially
($76$ per cent of hydrogen and $24$ per cent of helium), but becomes chemically
enriched as a result of the transfer of contaminated cold gas to the
hot phase due to reheating by supernovae explosions.  This chemical
enrichment has a strong influence on the amount of hot gas that can
cool, since we are using metal dependent cooling rates. This process
in turn influences the star formation activity which is ultimately
responsible for the chemical pollution.

Our semi-analytic model considers mass losses by stars in different
mass ranges, from low and intermediate stars to quasi-massive and
massive stars, taking into account stellar lifetimes. Massive stars
give raise to SNe CC.  Ejecta from SNe Ia are also included. 
The cold gas component of each galaxy
becomes gradually more chemically contaminated as star formation
proceeds and, consequently, new stars that are formed are
progressively more metal rich. This fact calls for the use of metallicity 
dependent stellar yields. However, variations of yields with stellar
metallicities are very small, so we only consider the mass dependence of
stellar yields. 
We adopt yields of different chemical elements estimated 
for stars with solar heavy element abundance. 

Thus,
for each stellar mass $\Delta M_{\star}$ formed,
we determine the fraction of mass of stars contained in a given mass
range by assuming an Initial Mass Function (IMF) $\Phi(M)$.
The mass of chemical element $j$ ejected by stars with masses in the range
centred in $m_k$, with lower and upper limits $m_k^l$ and $m_k^u$, is given by
\begin{equation}
\Delta M_{{\rm ej}_k}^j=
[R_k \, X^j + Y_k^j] \, \Delta M_{\star},
   \label{eqCE1} 
\end{equation}
where $X^j$ is the solar abundance of element $j$,
and $R_k$ and $Y_k^j$ are  
the recycled fraction 
and the yield of element $j$ for the {\em k}-th mass range,
respectively.
The recycled fraction is given by
\[
R_{k}= \int \limits_{m_k^{\rm l}}^{m_k^{\rm u}} \Phi(M) \, r_{k} \, dM,
\]
with $r_k$ defined as the difference between the mass
$m_k$ of the star at birth and its remnant mass after mass loss due to
stellar winds and/or supernova explosions. The yield of chemical element $j$
newly formed is
\[
Y_k^j= \int \limits_{m_k^{\rm l}}^{m_k^{\rm u}} \Phi(M) \, p_k^j \, dM,
\]
where $p_k^j$ is the stellar yield
of a chemical element {\em j} produced
by a star with mass in the range centred around $m_k$.
The variable $j$ may refer to H, He or chemical species with atomic number
larger than $2$.

The mass of species $j$
ejected by each SN Ia is labelled $m^j_{\rm Ia}$.
In this case, we are dealing with the total mass ejected because
this type of supernovae leaves no remnants, thus 
\begin{equation}
\Delta M_{{\rm ej (Ia)}_k}^j= \eta_{{\rm Ia}_k} \, m^j_{\rm Ia} \, 
\Delta M_{\star}, \label{eqCE2} 
\end{equation}
where $\eta_{{\rm Ia}_k}$ is the number of SNe Ia
per stellar mass formed in the mass interval $k$, which depend on the
model adopted for this type of SNe.
Note that SNe Ia do not contribute to H or He, hence
$M_{\rm ej (Ia)_k}^j$ is zero
when referring to these two chemical elements.

The transport of different chemical species between hot gas, cold gas,
and stars, can be expressed as variation of the mass of the chemical
element $j$ present in each baryonic component, namely
\begin{eqnarray}
\Delta M_{\rm hot}^j &=& - \Delta M_{\rm cool} \, A_{\rm hot}^j + \Delta M_{\rm reheat} \, A_{\rm cold}^j, 
\label{eqC1}
\end{eqnarray}
\begin{eqnarray}
\Delta M_{\rm cold}^j &=& + \Delta M_{\rm cool} \, A_{\rm hot}^j - \Delta M_{\star} \, A_{\rm cold(SF)}^j \nonumber \\
                && + \Delta M_{\rm ej}^j + \Delta M_{\rm ej (Ia)}^j - \Delta M_{\rm reheat} \, A_{\rm cold}^j, 
\label{eqC2}
\end{eqnarray}
\begin{eqnarray}
\Delta M_{\rm stellar}^j &=& + \Delta M_{\star} \, A_{\rm cold(SF)}^j -  R \,  A_{\rm star}^j \, \Delta M_{\star},
\label{eqC3}
\end{eqnarray}
where 
$A_{\rm B}^j=M_{\rm B}^j/M_{\rm B}$,
is the abundance of different chemical
elements in each baryonic component,
which involves the total mass of the baryonic component $M_{\rm B}$ 
and the mass of species $j$ contained in it, 
$M_{\rm B}^j$.
The suffix B may refer to hot gas, cold gas or stars; 
$A_{\rm cold(SF)}^j$
denotes the abundance of the element $j$ in the cold gas at the time
of birth of $\Delta M_{\star}$. 
The above set of equations takes into account the accumulated contribution 
of different mass ranges that affect the baryonic components at a given time,
as a result of the combination of the star formation rate of each galaxy and
the return time-scale of the ejecta from all sources considered.

\subsection[]{Injection of metals in the diffuse gas}\label{sec_injection}

We describe now the procedure used to distribute the chemical elements
generated by the galaxies in the semi-analytic model among the gas
particles of the {\em N}-Body/SPH simulation. For each snapshot, we
identify the gas particles contained within spheres of radius $100 \,
h^{-1}$ kpc centred on each galaxy.
Once the set of $N_{\rm gas}$ gas particles around a given galaxy is
found, we apply our chemical enrichment model to them, based on
the semi-analytic computation of galaxy formation.  The mass of
chemical element $j$ transfered to the hot phase is evenly distributed
among the set of $N_{\rm gas}$ gas particles as
\begin{eqnarray}
\Delta  m_{\rm gas}^j = + \frac{1}{N_{\rm gas}}  \Delta M_{\rm reheat} \, A_{\rm cold}^j . 
\label{eqD1}
\end{eqnarray}
This prescription arises from equation (\ref{eqC1}), which involves
the hot gas component of the semi-analytic model.  Note that the
effect of gas cooling on gas particles is suppressed in our model
because of the use of a non-radiative simulation.  This process should
however not change the mass fraction of different species, since
cooling would reduce the mass in them proportionally to their
abundances.

The increment of the masses of chemical elements in a gas particle in
principle modifies its total mass.  However, the mass of gas particles
does not change in the {\em N}-Body/SPH simulation. Since we are
interested in the resulting chemical abundances of gas particles after
this enrichment process, we redefine the masses of species $j$
contained in gas particles such that their sum gives the original mass
of the gas particle $m_{\rm gas}$, as given in the simulation, and the
mass fraction of each chemical element keeps the value reached after
the contribution given by Equation~(\ref{eqD1}) has been added.

The chemical elements injected into gas particles are redistributed
among neighbouring particles simulating a small-scale mixing process.
We apply a smoothing technique to the chemical properties of the gas
particles within the virial radius of the cluster, 
considering 32 neighbours around each of them.
The radius of the sphere containing these neighbouring particles
is less than $\sim 100 \, h^{-1} \, {\rm kpc}$, therefore the large-scale
distribution of metals in the ICM is not affected by this procedure.
After smoothing, the mass of chemical species $j$ of particle $i$ is given
by
\begin{equation}
m_{{\rm gas},i}^j =\sum_{l=1}^{32} {m_{{\rm gas},l}^j}/{\rho_{{\rm gas},l}} \, 
W(r_{li},h_i),
\label{eqD2}
\end{equation}
where $m_{{\rm gas},l}^j$ and $\rho_{{\rm gas},l}$ are the mass of
species {\em j} of the {\em l}-th neighbour and its density,
respectively.  $W(r_{li},h_i)$ is the kernel used, $r_{li}$ is the
relative distance between particles, and $h_i$ is the smoothing length
of particle {\em i} for which the smoothed quantity is being
estimated. The smoothing kernel is
the one used for SPH calculations in the {\em N}-Body/SPH simulation
\citep{springel01a}, that has been obtained from a {\em B}-spline.
This pocedure is applied at every output of the simulation.

\section[]{Characteristics of the semi-analytic model}\label{sec_CharSAM}

Our semi-analytic model is characterised by several parameters that
regulate the way in which gas cooling, star formation, supernovae
feedback and galaxy mergers proceed.  A small number of them are
considered free parameters and their appropriate values are determined
by observational constraints. These are the star formation efficiency
$\alpha$ and the feedback efficiency $\epsilon$, which regulate the
star formation rate and the core collapse supernovae rate,
respectively, and a parameter characterising the contribution of
supernovae type Ia.
This section describes the fixed parameters adopted and the conditions
we impose to determine the free ones. We then compare
some properties of our model with observational results.

\subsection[]{Fixed parameters of the model}

Our chemical implementation involves an IMF and stellar yields of
different species. We do not analyse the impact of changing these
parameters on the chemical enrichment of the ICM; they are
considered fixed as well as the baryon fraction $f_{\rm b}$.
For the later, we adopt a value of $0.13$, which was chosen to
match that used in the {\em N}-Body/SPH simulation.  

We consider a Salpeter IMF, normalised by fixing the fraction 
$\xi$ of stars with masses larger than $1\,{\rm M}_{\odot}$, which are the
major contributors to the chemical enrichment
(Portinari, Chiosi \& Bressan 1998).  
Thus, we obtain the
condition
\begin{equation}
 \int \limits_{1 M_{\odot}}^{M_{\rm u}} \Phi(M) M {\rm d}M = \xi, \label{eqMP5}
\end{equation}
with the upper limit $M_{\rm u}=100 {\rm M}_{\odot}$. The lower limit
$M_{\rm l}$ is determined by the value of $\xi$ adopted, such that the
integration between $M_{\rm l}$ and $M_{\rm u}$ equals unity. 
We adopt $\xi=0.5$, a high enough value to 
reach the right metallicity in all baryonic components.

For low- and
intermediate-mass stars $(0.8 {\rm M}_{\odot} \la M \la 5 - 8\, {\rm
M}_{\odot})$, we adopt the stellar yields of \citet{marigo01}, 
and for quasi-massive $(5\, {\rm M}_{\odot} \la M \la 8\,
{\rm M}_{\odot})$ and massive stars $(8\, {\rm M}_{\odot} \la M \la
120\, {\rm M}_{\odot})$, we use yields from models of
\citet{portinari98}.  The first two ranges of masses
eject mainly $^4{\rm He}$, $^{12}{\rm C}$, $^{14}{\rm N}$, and
possibly $^{16}{\rm O}$, through stellar winds.  The last two stellar
mass ranges contribute to the chemical enrichment of the interstellar
medium by stellar winds, and supernovae explosions triggered by core
collapse.
SNe CC are the main contributors of
$\alpha-$elements. 
In our calculations,
we follow the production of 8 chemical elements 
(H, $^4{\rm He}$, $^{12}{\rm C}$,
$^{14}{\rm N}$, $^{16}{\rm O}$, $^{24}{\rm Mg}$,
$^{28}{\rm Si}$, $^{56}{\rm Fe}$)
generated by stars with masses distributed in 27 mass ranges, from $0.8
\, {\rm M}_{\odot}$ to $100 \, {\rm M}_{\odot}$. The mass of
the rest of the elements produced
is stored in a separate variable.

Since we consider yields from stars with solar metallicity, and  
taking into account that the solar heavy element abundance has been 
recalibrated to $Z_{\odot}=0.0122$ \citep{asplund05}, 
we estimate average yields from those given for metallicities 0.02 and 0.008
by \citet{marigo01} and \citet{portinari98}.
Taking into account that
the combination of standard IMF and yields do not satisfy
observational constraints, such as
abundances in solar neighbourhood stars, metallicity-mass
relation and ICM metallicity (Moretti, Portinari \& Chiosi 2003), and
considering the uncertainties that affect the stellar yields \citep{Gibson97}, 
Mg and O are increased by a factor of 4 and 1.5, respectively.
When all mass ranges are considered, we get a
net yield of 0.043  
($\sim 3.5 \, Z_{\odot}$), and a corresponding recycled 
fraction of 0.39.

Integrating the normalised IMF between $8\, {\rm M}_\odot$ and $100
\,{\rm M}_{\odot}$, which is the mass range of stars leading to SNe CC,
gives $\eta_{\rm CC}=0.009$.  The energy ejected by each SNe CC is set
to $1.2 \times 10^{51}$, following the results by \citet{ww95}, on
which the nucleosynthesis calculations of total ejecta of massive
stars made by \citet{portinari98} are based.

We include the contribution of SNe Ia which 
are characterised by high iron production ($\sim 0.6 \,{\rm
M}_{\odot}$) and long delay time from the formation of the progenitor
to the supernova explosion. 
The currently accepted scenarios for the occurrence of SNe
Ia are, on one hand, carbon deflagration in C-O white dwarfs in binary
systems (single degenerate systems), and, on the other, mergers of two
white dwarfs (double degenerate systems). The former would account for
$\sim 80$ per cent of the events, with mean time delays of $0.5 - 3$~Gyr,
while the later would constitute the remaining $\sim 20$ per cent, with
lower time delays of $\sim 0.3$~Gyr (\citealt{tutukov94,
yoshii96,ruizlapuente98,hachisu99,
greggio05}).
We adopt the single degenerate model to estimate the SNe Ia rate, 
following the scheme of \citet{greggio83}, where type Ia supernovae originate 
in binary systems whose components have masses between 
$0.8$ and $8 \,{\rm M}_{\odot}$. Calculations are based on
the formalism described in \citet{lia02}, but assuming that
SNe CC originate from single stars with masses  
larger than $8\,{\rm M}_{\odot}$.  
For the chemical contribution of SNe Ia,
we adopt the nucleosynthesis
prescriptions from the updated model W7 by \citet{Iwamoto99}.

We use the stellar lifetime given by \citet{padovani93} to model
the return time-scale of the ejecta from all sources considered,
being specially relevant for the single stars in the 
low- and intermediate-mass range and for the progenitors of SNe Ia.
In the last case, the mass range of secondary stars in binary systems
gives explosion times for SNe Ia comprised between
$\sim 2.9 \times 10^{7}$ and  $\sim 1.4 \times 10^{10}$~yrs, with the SNe Ia rate
reaching a maximun within $\sim 0.1 - 0.7$~Gyr for a single stellar population. 

This chemical model shares similar characteristics with 
that implemented in the semi-analytic
code developed by \citet{nagashima05}, being an improvement with
respect to other semi-analytic models (\citealt{kauff98,
cole00,Somerville01}, DL04) which typically
consider both the fraction of metals ejected by the stars formed and
the recycled mass that has been locked in stars at their birth and
re-ejected as free parameters.

\subsection[]{Parameters governing the SF and SNe CC}

The parameters involved in the star formation and feedback processes
have a strong influence on the star formation rate.  The SNe CC rate
closely resembles the star formation rate density (SFR) 
because it is dominated by high mass
stars with short lifetimes ($\la 3 \times 10^7$~Gyrs).
The value of the index $n$ involved in the
star formation law adopted (Equation~\ref{eqSAM3}) is set equal to the
value suggested by DL04, $n=2.2$. However, the values of $\alpha_0$
and $\epsilon$ used by DL04 for the retention model do not constitute
the best option for our semi-analytic model. The different chemical
model implemented in our study demands a new determination of the
appropriate values.  We use as observational constraints
the Milky Way properties, the luminosity
function, the Tully-Fisher, color-magnitude and mass-metallicity relations
(\citealt{kauff99,springel01}; DL04), obtaining the best fit for
\begin{enumerate}
\item $\alpha_0=0.1$,
\item $\epsilon=0.2$.
\end{enumerate}
The agreement of our model results with the observational data 
used to constrain these free parameters 
is quite similar
to that showed by the models of DL04, therefore we refer the reader to 
their paper for further discussion on this topic.

The evolution of SFR in our
model is broadly consistent with the observational results recently
compiled by \citet{Somerville01},
with the peak of the SFR however shifted
to higher redshifts.
This behaviour is similar to results for the cosmic
star formation history in hydrodynamic simulations by
\citet{Springel03} (see their figure 12), although the normalisation
is less certain in our model here since our estimate of the comoving
volume is made uncertain by the irregular shape of our high-resolution
zone within the simulated volume, which is used to estimate 
SF and SNe density rates.

We note that our simulation method is really quite different from that
applied by \citet{Springel03}.  The present work is based on a
non-radiative simulation combined with a semi-analytic model, while
\citet{Springel03} use a fully dissipative simulation with a
subresolution model for the ISM and feedback effects.  However,
despite the huge differences in methodology, the principal qualitative
results are pretty similar, which is an encouraging consistency.  
Even though our simulation is similar in volume 
to simulation `G5' of \citet{Springel03},
the SFR density obtained from the semi-analytic model is higher than in 
`G5', specially at higher redshifts, being in
better agreement with their simulations with higher mass resolution
(see their figure 10).
One possibility to explain this is that the semi-analytic model is
less affected by resolution effects all the way down to the mass
resolution limit of the haloes, while this is not the case for the full
hydro-simulations.  While a halo with $\sim 30$ particles will be
able to yield a good representation of the SFR when the semi-analytic
model is applied, one has to go substantially above the halo detection
threshold in the full hydrodynamical simulations in order to recover a
numerically converged result for the hydrodynamics. Effectively, this
means that the dynamic range of the semi-analytic method is higher.

A convergence test of cluster simulations similar to the one used in the
present study has been done  by Ciardi, Stoehr \& White (2003). 
They analysed the
redshift evolution of the total star formation rate, normalised to the
total baryonic mass, obtained by applying a semi-analytic model to
simulations `S1'-`S4' performed by \citet{springel01}.
They show that simulation `S2', with the same mass resolution than
our hydrodynamical simulation,
converged at $z \sim 9$ to the higher resolution simulation `S3'
($m_{\rm dm}=2.4 \times 10^{8} \,h^{-1} {\rm M}_{\odot}$)
which already accounts for all significant star formation.
Therefore, the simulation used in our hybrid model is suitable for 
studying the chemical enrichment of the ICM. 

\subsection[]{Parameters governing the SNe Ia rate}\label{sec_SNeIa}

The contribution of SNe Ia to the metal production of
the model is characterised by the fraction $A$ of binaries originating 
this type of supernovae.
Apart from the iron content, this
parameter do not
significantly affect the properties of the baryonic components in the
model, hence the constraints imposed for determining the star formation and
feedback efficiencies are almost insensitive to the value of $A$.
This value has to satisfy other kind of requirements.

During the last five years, great efforts have been undertaken to
detect SNe over cosmological distances, with the aim of studying the
evolution of SNe CC and SNe Ia rates with redshift
(\citealt{Pain96}; Cappellaro, Evans \& Turatto 1999, \citealt{Hardin00,
Pain02,Madgwick03,Tonry03,
Dahlen04,Strolger04}).
We adopt $A=0.1$ in order to 
reproduce this set of observations
and to achieve the observed iron distribution in the ICM,
that will be discussed in detail in Section~\ref{sec_radialprof}.

The combination of the SF history of the model and the 
scenario chosen for the origin of SNe Ia 
naturally leads to SNe CC and SNe Ia rates whose ratio evolves with
redshift.  
The ratio  between the number of SNe Ia and SNe CC 
has a fairly
constant value of $\sim 0.4$ to $z \sim 0.7$,
similar to the behaviour suggested by observations of \citet{Dahlen04}.
The time delay in SNe Ia explosions makes this ratio
smaller at higher redshifts. We find that
it decreases to $\sim 0.25$ at $z\sim
2.5$, where SNe Ia rate reaches its maximum,
and continues decreasing afterwards, manifesting the minor contribution
of SNe Ia at large lookback times.

\subsection[]{Comparison of model results with observations}\label{sec_CompObs}

\begin{figure*}
  \centering
  \begin{minipage}[c]{.33\textwidth}
   \centering \includegraphics[width=60mm]{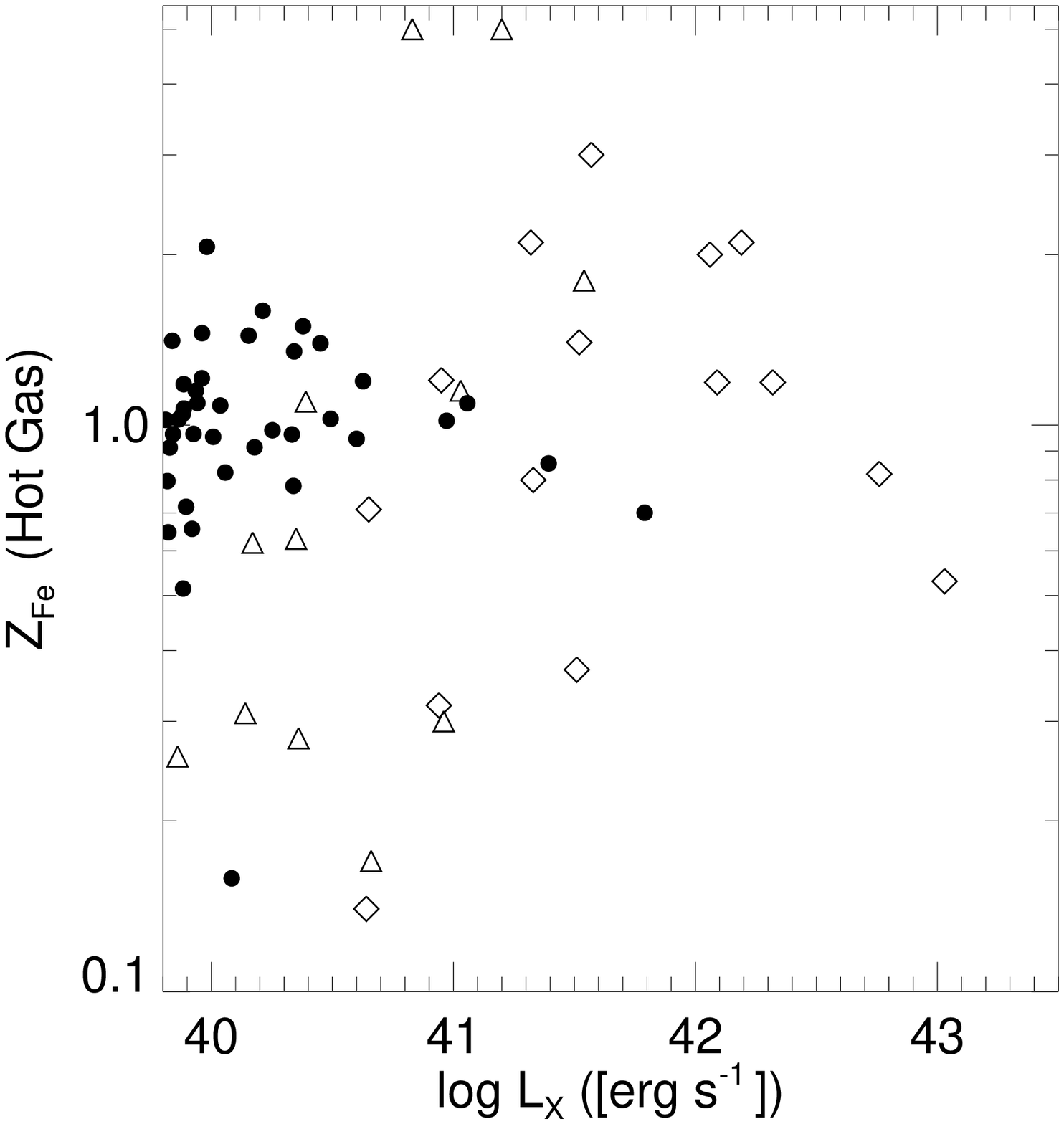}
  \end{minipage}%
  \begin{minipage}[c]{.33\textwidth}
   \centering \includegraphics[width=60mm]{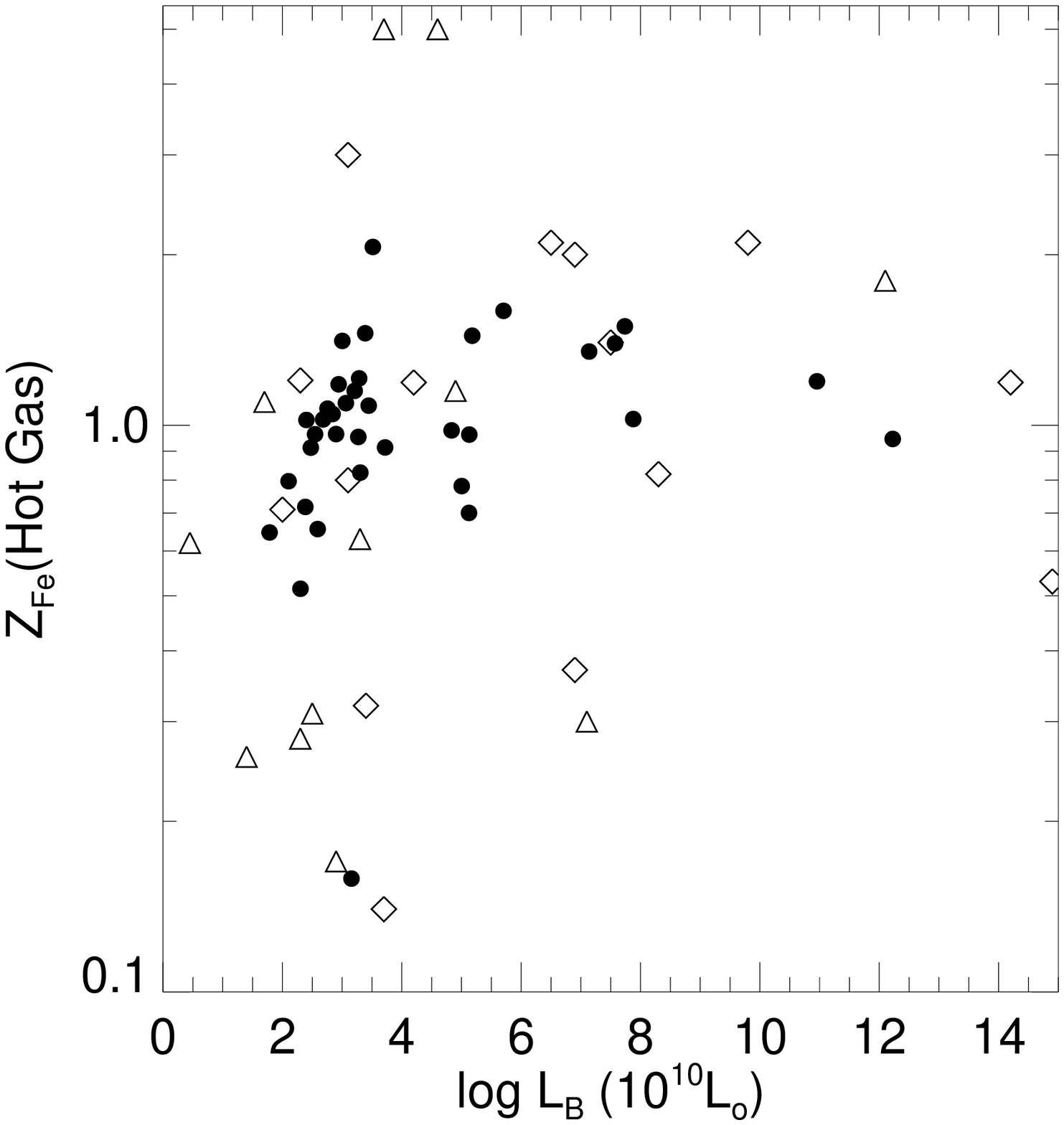}
  \end{minipage}%
  \begin{minipage}[c]{.33\textwidth}
   \centering \includegraphics[width=60mm]{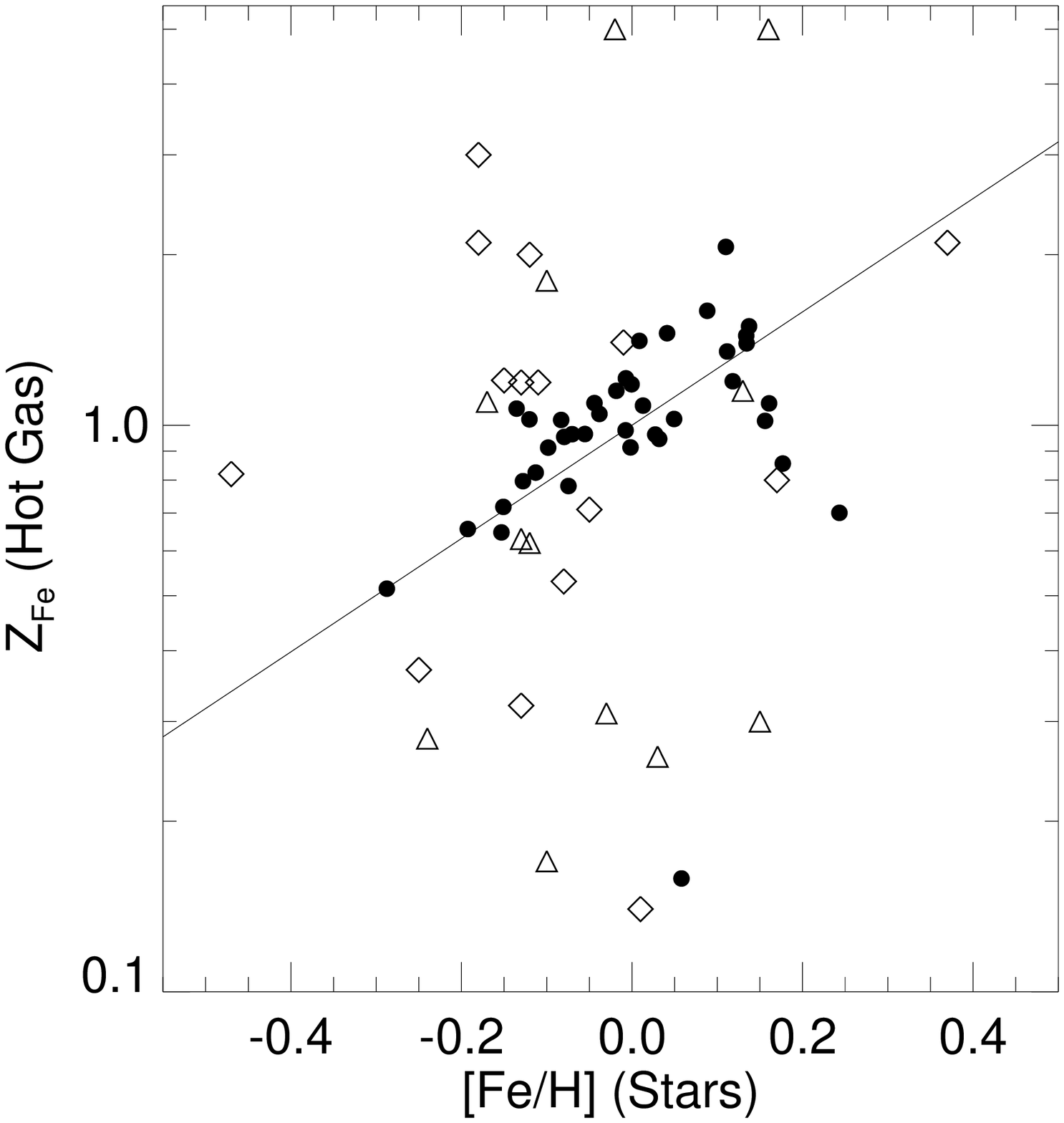}
  \end{minipage}%
\caption{Hot gas iron abundance, $Z_{\rm Fe}$, as a function of
the X-ray luminosity, $L_{\rm X}$ (left panel), {\em B}-band luminosity of the
galaxy, $L_{\rm B}$ (middle panel), and iron abundance of their stars
$[{\rm Fe}/{\rm H}]$ (right panel). Filled circles
represent central galaxies in the simulation selected according to the
X-ray luminosities of their associated hot gas
($39.8 < {\rm log}(L_{\rm X}) < 43.5$),
as the range span by the sample of galaxies analysed by
\citet{humphrey05}. Their observational results are represented by empty
squares and triangles, where the later indicate lower limit determinations
of iron abundances. The orbservational points are consistent with
the $Z_{\rm Fe}(\rm {gas})=[{\rm Fe}/{\rm H}](\rm {stars})$ line,
plotted in the right panel, within their error bars, which are not included
in these plots for clarity (we refer the reader to the figures 5 and 6 of
\citealt{humphrey05}).
}
\label{fig1_N1}
\end{figure*}

Having specified the parameters that define our model,
we make further comparisons with observations in order
to check that 
it produces a proper circulation of metals among the 
different baryonic components. 

An important constraint on cluster chemical models is the
fraction of heavy elements ejected by both types of SNe 
which end up in stars and in the intergalactic medium.
A recent study of metal enrichment of early type galaxies 
based on the X-ray properties obtained with {\em Chandra} has been done by  
\citet{humphrey05}.
The sample of galaxies span $\sim 3$ orders of magnitude in 
X-ray luminosity ($L_{\rm X}$), from group-dominant to 
low-$L_{\rm X}$ galaxies. We apply this selection criterium
to the central galaxies in our simulation and estimate the
iron abundance $Z_{\rm Fe}$ of the hot gas contained in the
dark haloes in which they reside. The three panels in Figure~\ref{fig1_N1} 
present the dependence of $Z_{\rm Fe}$ with luminosities $L_{\rm X}$
(which calculation is based on the formula presented by
\citealt{valdarnini02}) and 
$L_{\rm B}$, and the iron metallicity 
$[{\rm Fe}/{\rm H}]$ of their stars; they are compared with the
values reported by \citet{humphrey05}.
The general trend of
the properties of our model galaxies is consistent with that showed by their 
observational results. On one hand, there is no evidence of a correlation
between $Z_{\rm Fe}$ and $L_{\rm X}$ or $L_{\rm B}$.
On the other, 
the iron abundance for the gas and 
stellar components of each of our model galaxies 
closely follow the 
$Z_{\rm Fe}(\rm {gas})=[{\rm Fe}/{\rm H}](\rm {stars})$ line, 
around which many of the observed galaxies are clustered.
This fact strongly supports the circulation of metals among the different 
baryonic components achieved by our model.

The fraction of iron in hot gas and stars that is synthesized by different 
type of SNe is indicative of the different time-scales of star formation and 
of pollution of the hot gas with SNe Ia products.
Adopting standard SNe Ia and CC stellar yields and performing a fitting 
analogous to that of \citet{gastaldello02}, 
\citet{humphrey05} infer an SNIa iron enrichment fraction 
of $\sim 0.66$  
in the hot gas
of their observed galaxies, and 
$\sim 0.35$ in their stars.
The model galaxies considered in Figure~\ref{fig1_N1} are 
characterised by similar
fractions; from the total amount of iron originated in SNe Ia,
$\sim 65$ per cent is contained in the hot gas of these galaxies while the remaining 
$\sim 35$ per cent is locked in the stars.
The corresponding fractions for iron produced by SNe CC are $\sim 62$  
and $\sim 38$ per cent for the hot gas and stars, respectively.
We see that the iron mass fraction in stars is a bit higher
if we consider iron from SNe CC instead of that produced by SNe Ia.
This fact reflects the increment of  
the ratio  between the number of SNe Ia and SNe CC
to lower redshifts, thus the hot gas continues being polluted
by SNe Ia ejecta after most of the stars have been formed.
However, the bulk of the star formation have occurred when the hot
gas have already been considerably polluted by SNe Ia products, as
it is evident from the small difference between these fractions.

A similar conclusion can be obtained from the fraction of iron 
originated in different sources that end up in 
the baryonic components of our simulated
cluster. We have that $\sim 78$ per cent of the iron 
produced by SNe Ia is contained in the hot gas and $\sim 21$ per cent
is locked in the stars, 
while for SNe CC we get
$\sim 72$ and $\sim 28$ per cent for hot gas and stars, respectively.
The interpretation about the influence of the evolution of SNe rates on 
these percentages becomes clearer from the inspection of
Figure~\ref{figEjEvol2} and the associated discussion, which focalise on
the ejected iron mass rate for 
both types of supernovae produced by the cluster galaxies.

Our simulated cluster has
cold baryon
fraction   
$f_{\rm c,cls} = {M}_{\rm stellar}/({M}_{\rm stellar}+{M}_{\rm ICM}) \sim 0.081$, 
iron yield 
$Y_{\rm cls,Fe}={M}_{\rm Fe}/{M}_{\rm stellar}= (M_{\rm ICM}^{\rm Fe}+M_{\rm stellar}^{\rm Fe})/M_{\rm stellar} \sim 3.2$ times solar and iron mass fraction 
$f_{\rm Fe}=M_{\rm Fe}/M_{\rm vir} \sim 5.3 \times 10^{-5}$,
where ${M}_{\rm stellar}$ is the mass of stars of the cluster galaxies,
$M_{\rm ICM}$ is the mass of the intracluster hot gas, and
$M_{\rm stellar}^{\rm Fe}$ and $M_{\rm ICM}^{\rm Fe}$ are the masses
of iron contained in these two baryonic components.
These values are compared with estimates from a combination of
near-infrared properties of galaxies within clusters for
which X-ray imaging data is available (Lin, Mohr \& Standford 2003). 
The properties of
galaxies are obtained from the Two Micron All Sky Survey
(2MASS). \citet{lms03} present several relations for low redshift
clusters ($0.016 \la z \la 0.09$), spanning a range of X-ray
emission-weighted temperatures of
$2.1\,{\rm keV} \la k_{\rm B} T_{\rm X} \la 9.1\,{\rm keV}$
(with $k_{\rm B}$ the Boltzmann constant)
that corresponds to about an order of magnitude in
cluster virial mass ($0.8 -9 \times 10^{14} {\rm M}_{\odot}$).  The
simulated cluster has a virial temperature of $k_{\rm B}\, T \sim 7.2$~keV.
The approximate values of the quantities mentioned previously, 
obtained by \citet{lms03} for a cluster with this temperature
are 
$f_{\rm c,cls} \sim 0.09$,
$Y_{\rm cls,Fe}\sim 3.5$ times solar and
$f_{\rm Fe} \sim 9 \times 10^{-5}$. 
Model values are lower than these ones but are within the error
bars of the observations (see figure 9 of \citealt{lms03}).

The number abundance of iron relative to hydrogen for the ICM respect
to the solar value
is $[\rm Fe/H]_{\rm ICM} \sim 0.28$, close to the lower
limit of observed ICM abundances which range from $\sim 0.3$ to $\sim 0.5$
(e.g., \citealt{ettori01}).
The iron mass-to-light ratio is $\sim 0.014\,\Gamma_{\odot}$, 
well within the range obtained by
\citet{renzini93} for $h=0.7$, that is, $0.0085 - 0.017\,
\Gamma_{\odot}$.  

The total {\em B}-band luminosity of the cluster is $L_{\rm B}=3.5 \times
10^{12} L_{\odot}$.  The {\em B}-band and {\em V}-band mass-to-light
ratios are $\Gamma_{\rm B} \sim 437 \,\Gamma_{\odot}$ and $\Gamma_{\rm V} \sim 345\,
\Gamma_{\odot}$, respectively.  Observational determinations of these
quantities give $\gamma_{\rm V} \sim 175-252$
(\citealt{Carlberg96, Girardi00}), and 
$\gamma_{\rm B} \sim 200-400$ (\citealt{Kent83, Girardi00}).
The intra-cluster mass to light ratio of our cluster
is $M_{\rm ICM}/L_{\rm B} \sim 40 \Gamma_{\odot}$, quite similar
to the typical observed value of $\sim 36 \Gamma_{\odot}$ indicated by 
\citet{moretti03}.

The analysis of the physical properties of gas particles included in
the hybrid model provides more detailed features of the ICM, regarding
the spatial distribution of different chemical elements, and the
connection between the X-ray spectra generated by the diffuse gas with
its large-scale motions, as we will show in the next sections.

\section[]{Metal distribution in the ICM}
\label{sec_radialprof}

The increase of spectral and spatial resolution of the last generation
of X-ray satellites, like {\em ASCA} and {\em BeppoSAX}, and more
recently, {\em Chandra} and {\em XMM-Newton}, has provided a great
deal of information about cluster centres. In particular, it has
become possible to obtain detailed abundance measurements of several
chemical elements within radial bins, allowing a determination of
their spatial distribution (\citealt{finoguenov00,
degrandi01,gastaldello02,degrandi03,
tamura04}).
Radial distributions of these abundances
support the presence of metallicity gradients in
the ICM, a feature that has been shown to be common in groups and
clusters of galaxies.
These radial profiles are of great importance to infer the origin of
the different species found in the cluster.

In our approach, the spatial distribution of chemical and
thermodynamical properties of the ICM is provided by the gas particles
in the {\em N}-Body/SPH simulation.  The hybrid model combines the
dynamics of gas particles with the chemical information supplied by
the semi-analytic model. Thus, at each snapshot, gas particles carry
the information about masses of H, He and heavy elements, with the
possibility of isolating the contribution due to SNe CC and SNe Ia.
These data are analysed by estimating radial profiles of the chemical
elements abundances.

Radial profiles are constructed
by taking into account the gas particles lying inside
the virial radius of the cluster. We then divide the volume limited
by this radius into concentric spherical shells centred on the
dominant cluster galaxy, each of them containing the same number
$N_{\rm gas}^{\rm shell}$ of particles.  For each shell, we estimate
the mass-weighted and X-ray emission-weighted
mean abundance relative to hydrogen for several species.
The later is estimated using the Fe ${\rm K}_{\alpha}$ 6.7 keV emission line,
which calculation for each gas particle is described in section \ref{sec_Maps}.
These
quantities are then expressed in terms of the solar value.
These mean abundances are
assigned to the mean radius of each shell.
                                                                                
\begin{figure}
 \centering
    \centering \includegraphics[width=85mm]{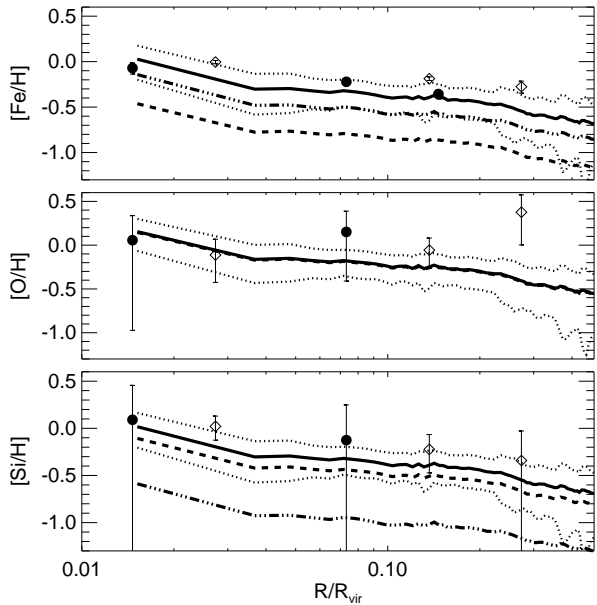}
\caption{Fe ${\rm K}_{\alpha}$ 6.7 keV emission-line-weighted
abundance profiles of Fe, O and Si relative to H, referred
to the solar value \citep{asplund05}.  Total abundances, with the
contribution of both types of SNe, are given by solid lines; root mean
square standard deviations are indicated by dotted lines. Separate
contributions of SNe CC and SNe Ia are shown by dashed and
dashed-dotted lines, respectively. Symbols with error bars are
observational data from \citet{tamura04} for hot clusters 
($k_{\rm B} T \ga 6\, {\rm keV}$, filled circles)
and medium-temperature clusters ($3 \la k_{\rm B} T \la 6\, {\rm keV}$, 
open diamonds).}
\label{figRP1}
\end{figure}

Figure~\ref{figRP1} shows the radial distributions of
Fe ${\rm K}_{\alpha}$ 6.7 keV emission-line-weighted iron, oxygen and
silicon abundances by number for our model, which are preferred to mass-weighted
ones in order to compare with observations.
The radius is normalised with $R_{\rm vir}$.
These results are characterised by the presence of negative gradients
in the radial distribution of all the chemical elements considered.
The emission-weighted values are $\sim 0.1 - 0.15$ dex higher than
mass-weighted abundances, specially in the inner region ($\la 0.1 \, R/R_{\rm vir}$).

There are ambiguous
results regarding the dependence of iron abundance on cluster
temperature. \citet{baumgartner05} find decreasing abundances from
$\sim 0.7$ to $\sim 0.3\, {\rm (Fe/H)_{\odot}}$ in the
temperature range $k_{\rm B} T \sim 3 - 10$ keV, while a
roughly constant value of $\sim 0.3 \, {\rm (Fe/H)_{\odot}}$ is claimed by
\citet{renzini03}. An intermediate behaviour was detected previously
by \citet{fukazawa98} and \citet{finoguenov01}, with iron abundances
of $\sim 0.3 \, {\rm (Fe/H)_{\odot}}$ for cool clusters and 
$\sim 0.2 \, {\rm (Fe/H)_{\odot}}$ for hotter ones. 
Thus, it is better to compare observed and
simulated results on cluster abundances within the same temperature
range.

\citet{tamura04} analysed {\em XMM-Newton} observations of 19 galaxy
clusters.  They obtained accurate abundance distributions from
deprojected spectra for clusters in different temperature ranges, and
for three spherical regions with radii between 0-50, 50-200 and
200-500 $h^{-1} \, {\rm kpc}$.  
Our simulated cluster has temperatures of the order of 
$k_{\rm B} T \sim 7.5$ keV (see mass-weighted and emission-weighted maps in
Figures~\ref{figMaps1} and \ref{figMaps2}).
Then, results from our model are
compared with data for hot clusters ($k_{\rm B T} \ga 6\, {\rm keV}$)
and intermediate-temperature clusters ($3 \la k_{\rm B} T \la 6\, {\rm keV}$),
represented in Figure~\ref{figRP1} by filled circles and 
open diamonds, respectively.
The later set of clusters is used since they
are better determined, providing better constraints to the behaviour 
of O and Si.
Following \citet{degrandi01}, we estimate $R_{\rm vir}$ for observed data 
from an ICM temperature-dependent formula.

From Figure~\ref{figRP1}, we can see that 
the contribution of SNe Ia (dashed-dotted
lines) is necessary to recover the observed Fe abundance distribution.
The contribution of Si from SNe Ia is not as large as for the Fe
production, but helps to reach the mean observed values. However, the
error bars in the Si determinations are much larger than those for Fe
and do not establish a reliable constraint for the model parameters.
This is also the case for O; the large observational
uncertainties in the determination of these chemical abundances 
for hot clusters allow
any type of behaviour of the radial profiles, while
data points for medium temperature clusters 
are consistent with a flat profile with a slight
increasing trend with radius.  Note that the model abundances show
radial profiles with negative gradients for all the elements
considered. However, the uncertainties both in observed and simulated
abundances
do not allow us to infer strong constraints from the comparison at
this point. 
Only the slopes of iron abundance profiles are better established by 
observational data. The good agreement between the iron abundance distribution
in our model and that observed in hot clusters makes this model
suitable for further analysis. We now carry out a deeper inspection 
of our results 
in order to gain insight in the enrichment history and dynamical
evolution of the ICM.

\section[]{Maps of ICM properties} \label{sec_Maps}

Our hybrid model for investigating the ICM metal enrichment provides
information about the chemical and thermodynamical properties of the
diffuse intracluster gas in terms of the gas particle properties
(positions, velocities, density $\rho$, and internal energy per unit
mass $u$) of the underlying hydrodynamic simulation.  From these
quantities we can obtain their temperature as $T=\mu\, m_{\rm p}/k_{\rm B}
\,(\gamma-1)\, u $, where $\mu=0.6$ is the mean molecular weight,
$m_{\rm p}$ the proton mass and $\gamma=5/3$
the adiabatic exponent.  We can learn about the density, temperature and
abundance distributions, and velocity structure of the ICM by
constructing maps of these properties using planar projections of the
gas particles within the cluster virial radius.  This kind of analysis
is matched to the recent progress in X-ray observations with {\em
Chandra} and {\em XMM-Newton}, which provides temperature and
abundance maps of cluster cores (\citealt{schmidt02,
sandersfabian02, sanders04, fukazawa04}).
In order to compare with observations we will analyse not only
mass-averaged properties, but also X-ray emission-weighted ones.  As a
result of the high temperature and low density of the ICM, the most
prominent feature in its X-ray spectrum is a blend of emission lines
from iron (mainly FeXXV and FeXXVI) with photon energies between 6.7
and 7 keV.

\begin{figure*}
 \centering
  \begin{minipage}[c]{0.378\textwidth}
    \centering \includegraphics[width=70.5mm]{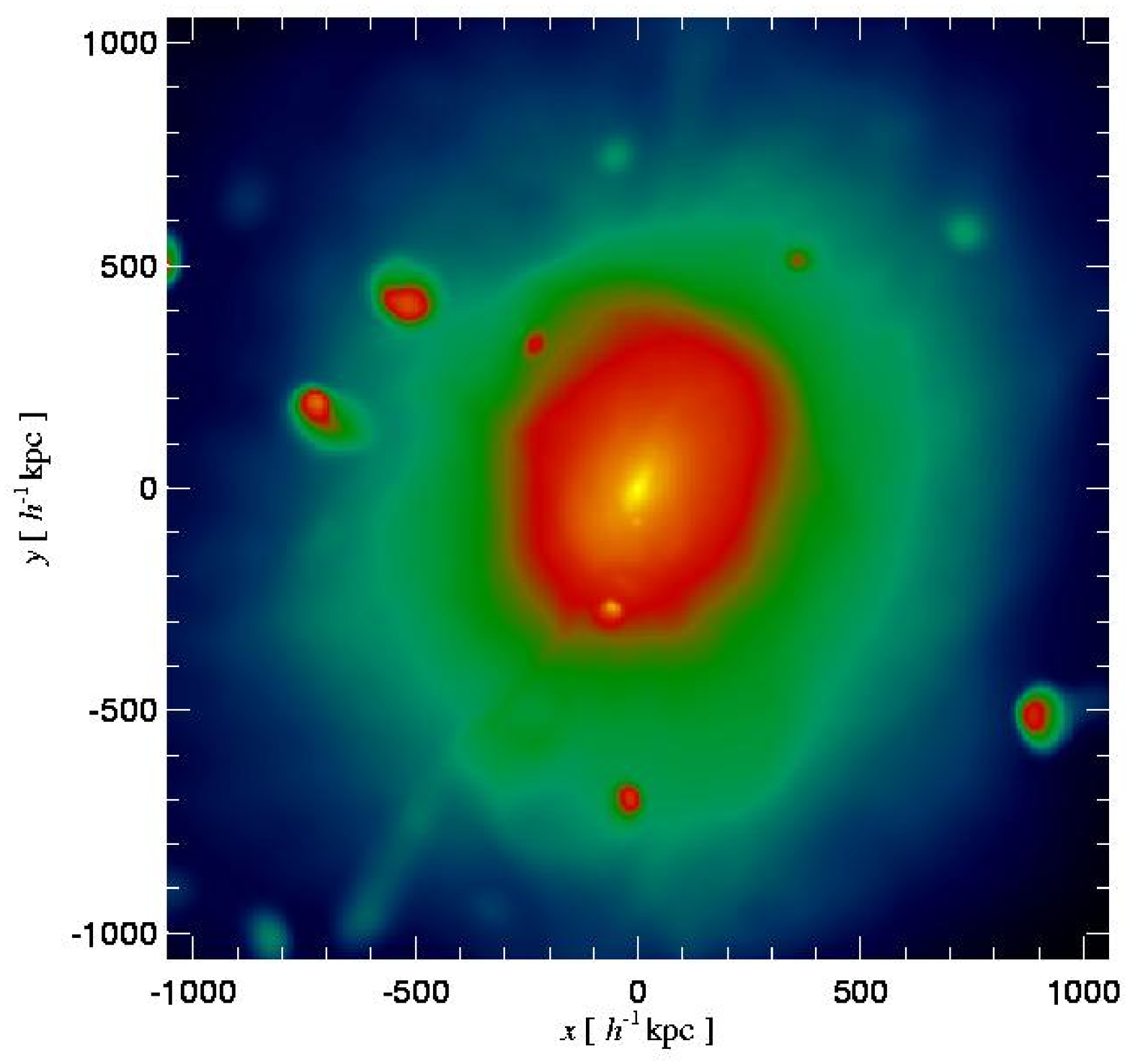}
  \end{minipage}%
  \begin{minipage}[c]{0.1\textwidth}

    \centering \includegraphics[width=13mm]{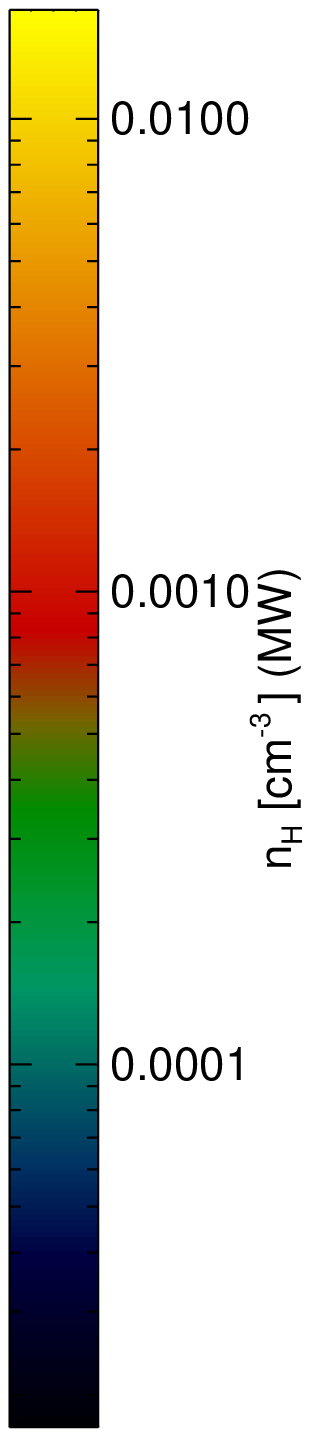}
  \end{minipage}%
  \hspace{2mm}
  \begin{minipage}[c]{0.378\textwidth}
    \centering \includegraphics[width=70mm]{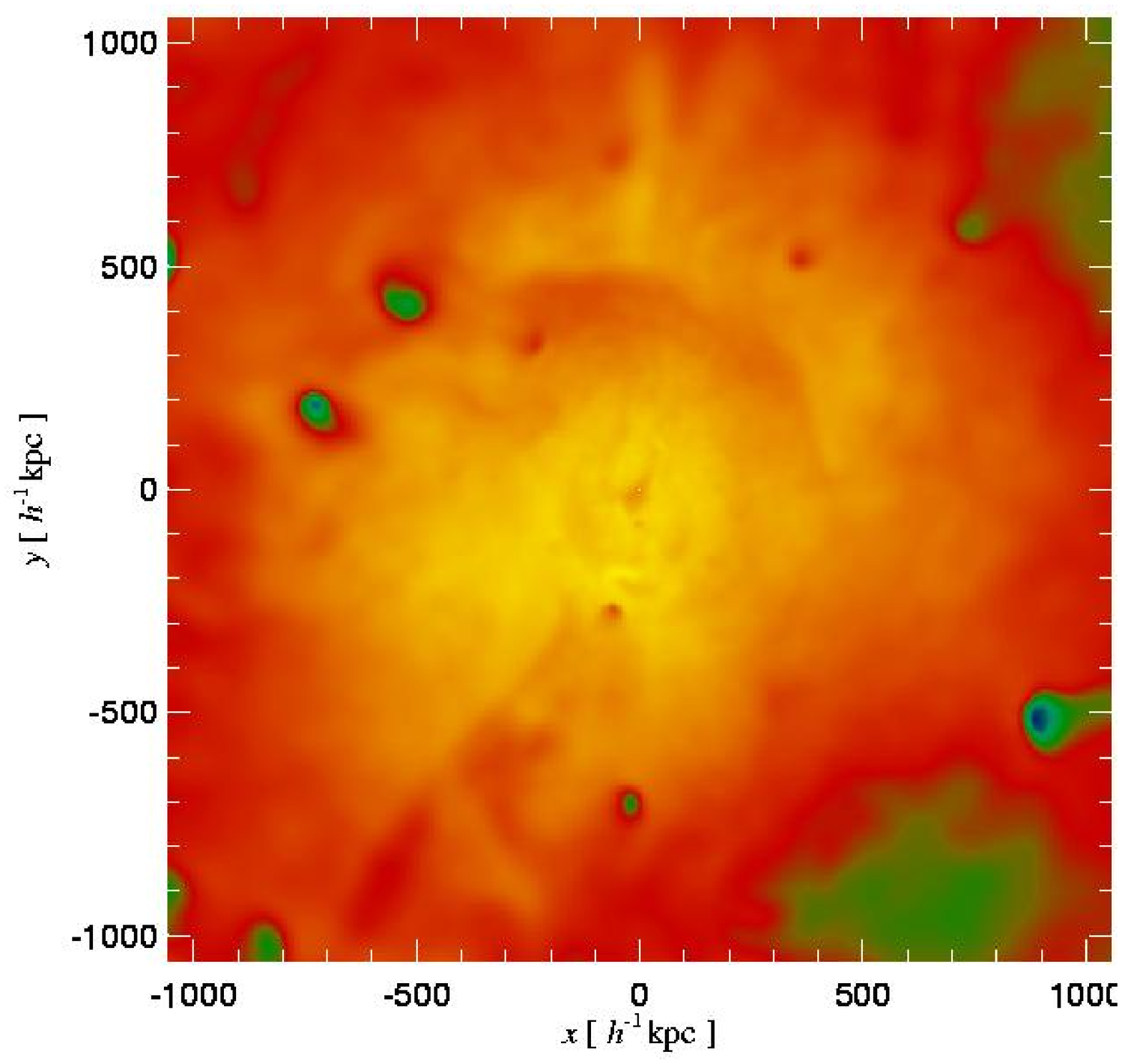}
  \end{minipage}%
  \begin{minipage}[c]{0.1\textwidth}
   \centering \includegraphics[width=13mm]{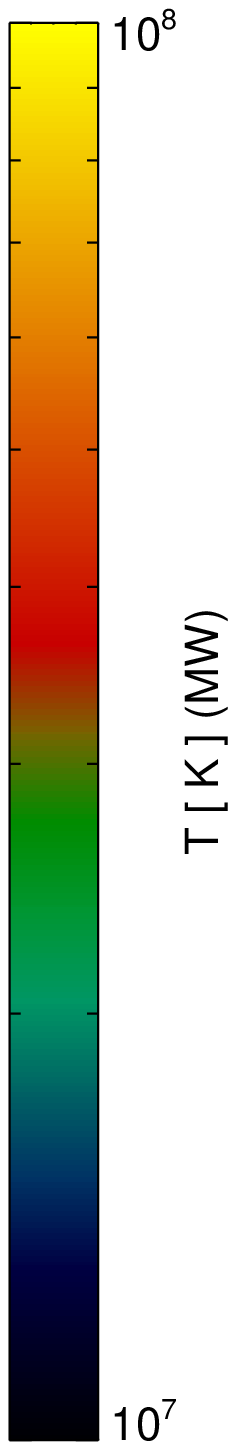}
  \end{minipage} \\%
  \begin{minipage}[c]{0.378\textwidth}
    \centering \includegraphics[width=71mm]{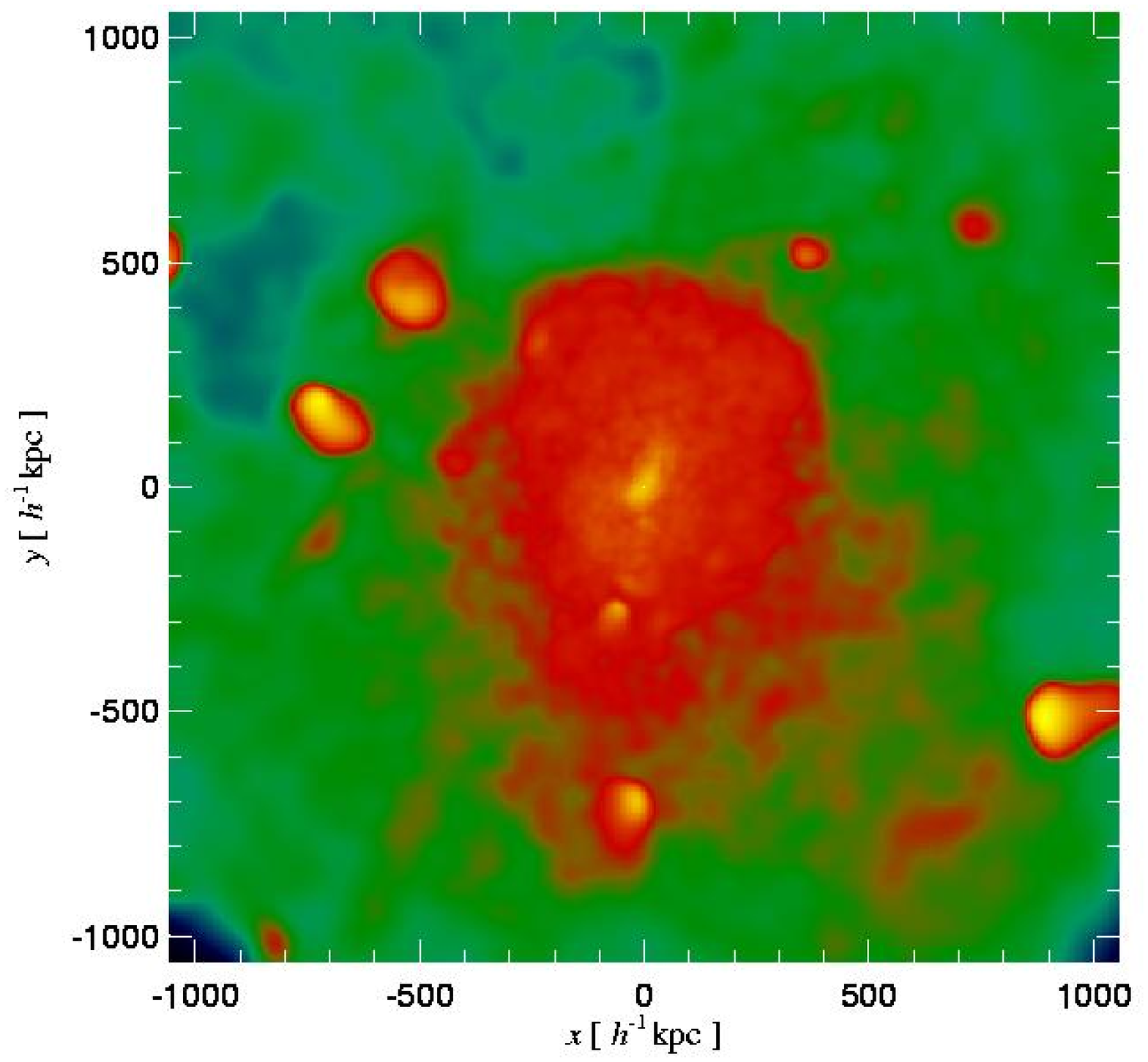}
  \end{minipage}%
  \begin{minipage}[c]{0.1\textwidth}
   \centering \includegraphics[width=13mm]{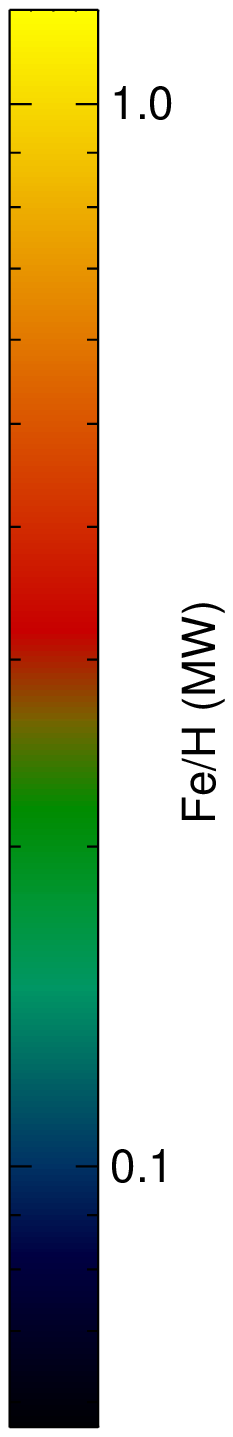}
  \end{minipage}%
  \hspace{2mm}
  \begin{minipage}[c]{0.378\textwidth}
    \centering \includegraphics[width=71mm]{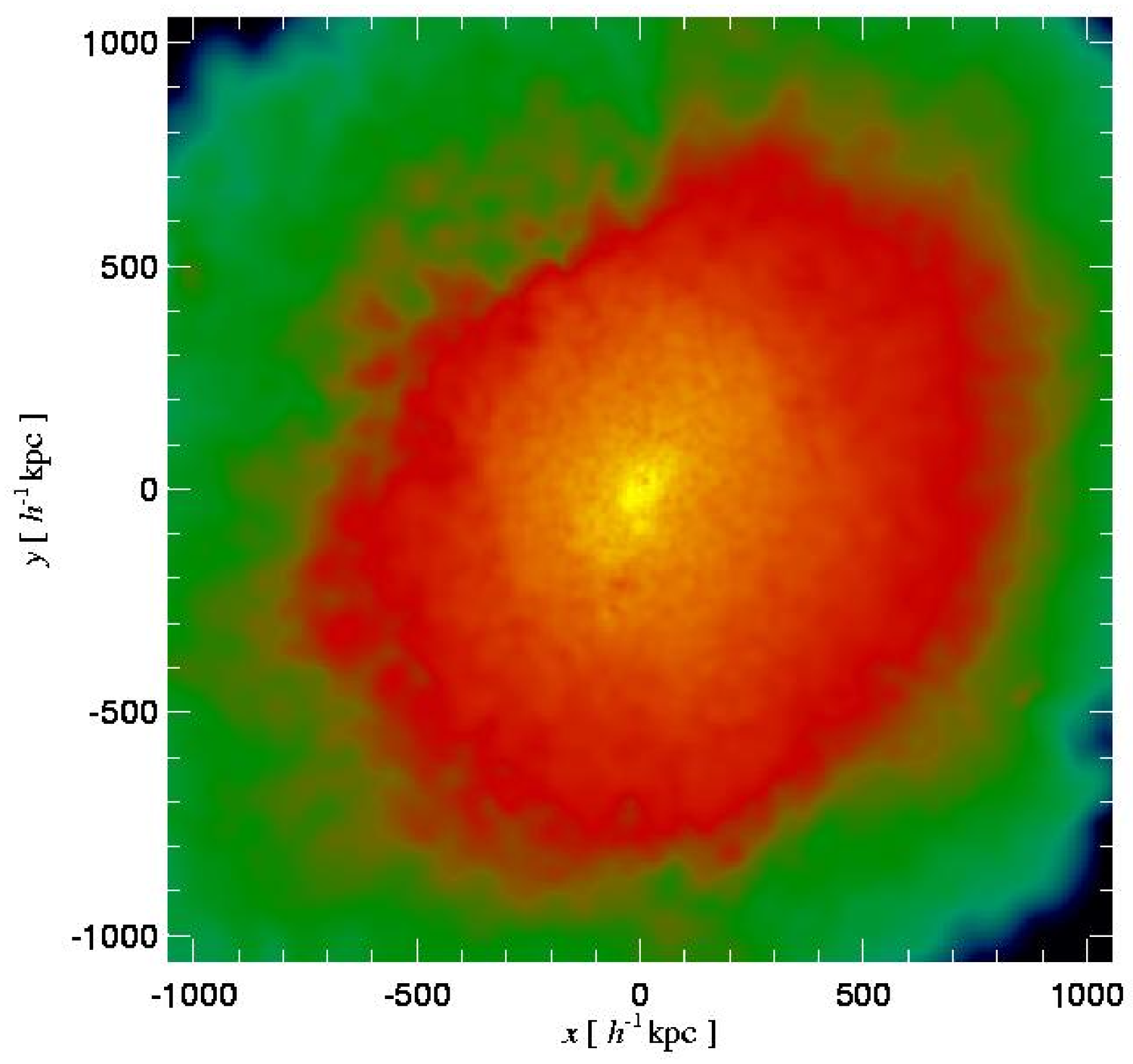}
  \end{minipage}%
  \begin{minipage}[c]{0.1\textwidth}
    \centering \includegraphics[width=13mm]{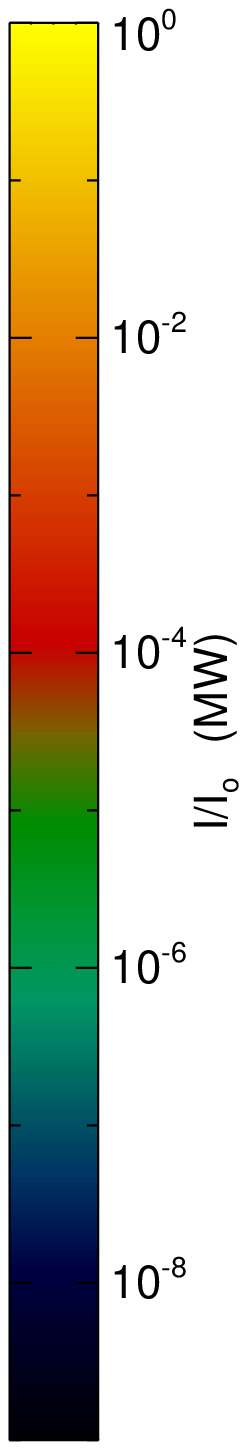}
  \end{minipage}
\caption{Projection of mass-weighted (MW) properties of gas particles
contained within a sphere of radius $1\, h^{-1}\,{\rm Mpc}$ centred on
the dominant cluster galaxy: hydrogen number density (top-left panel),
temperature (top-right panel), iron abundance by number Fe/H referred to the
solar value (bottom-left panel) and emissivity of the Fe ${\rm
K}_{\alpha}$ 6.7 keV emission line normalised with the maximum
mass-weighted emissivity $I_{\rm o}$
(bottom-right panel).}
\label{figMaps1}
\end{figure*}

\begin{figure*}
%
  \begin{minipage}[c]{0.33\textwidth}
   \includegraphics[width=58mm]{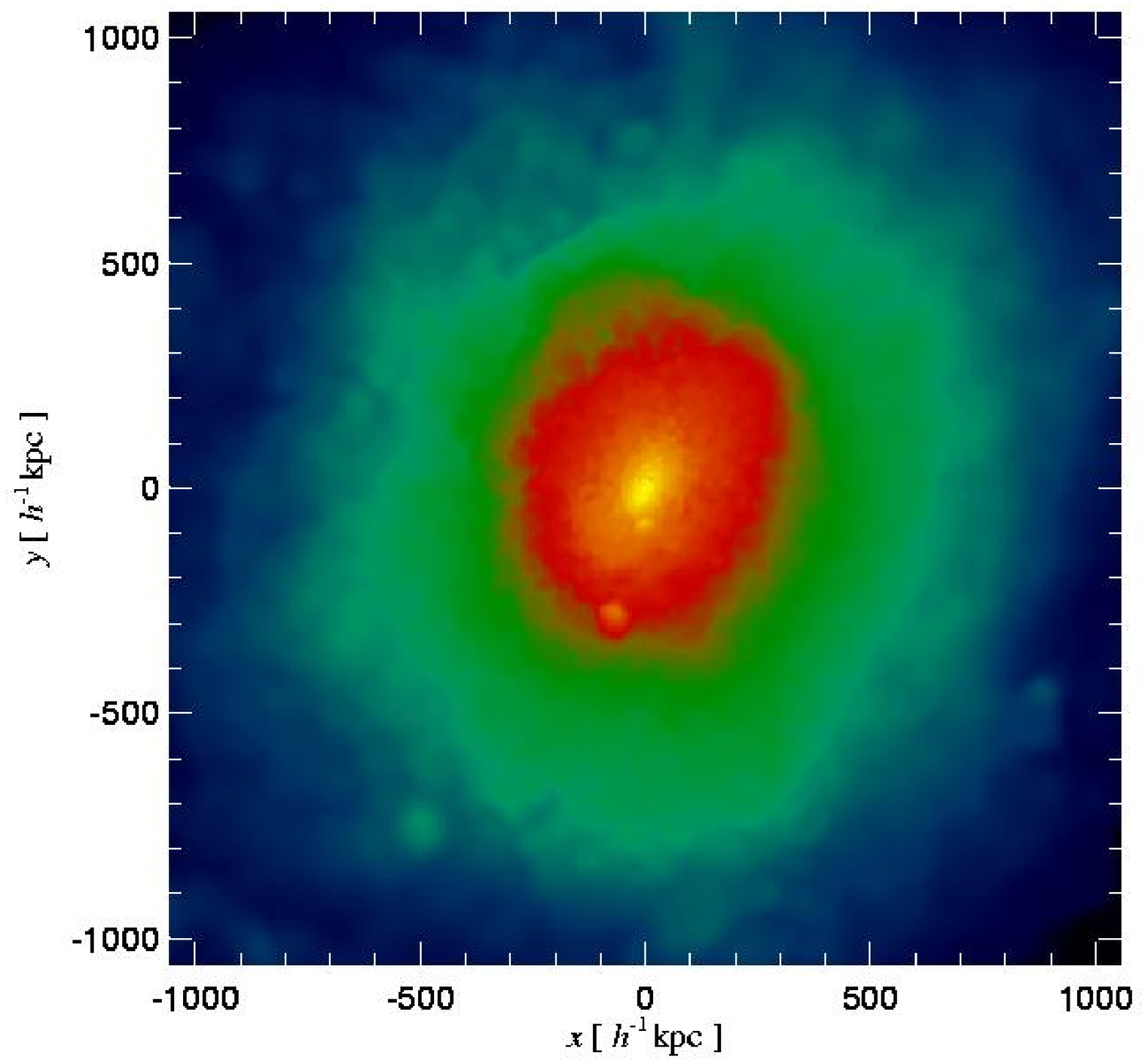}
  \end{minipage}%
  \begin{minipage}[c]{0.33\textwidth}
   \includegraphics[width=58mm]{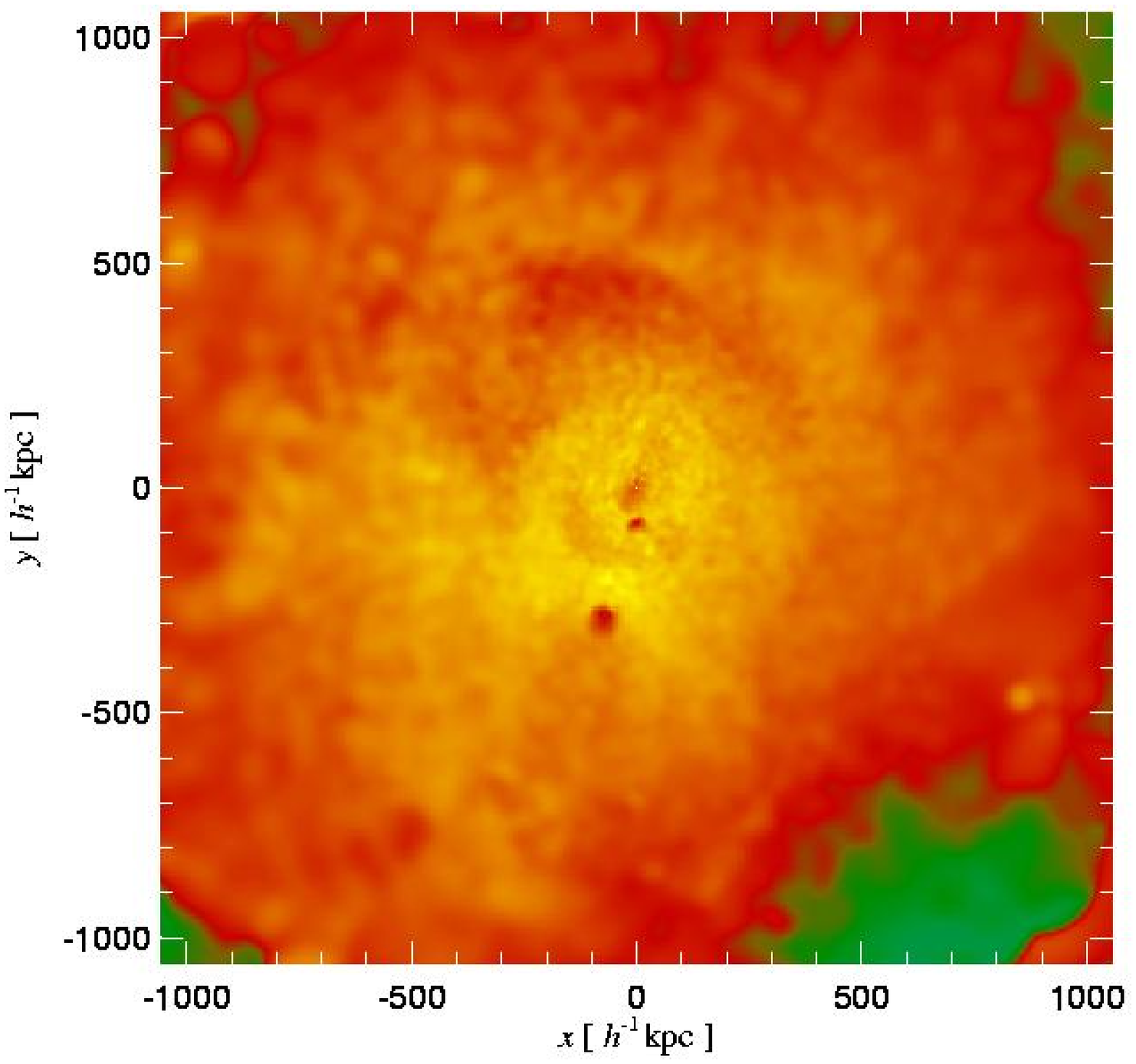}
  \end{minipage}%
  \begin{minipage}[c]{0.33\textwidth}
     \includegraphics[width=58mm]{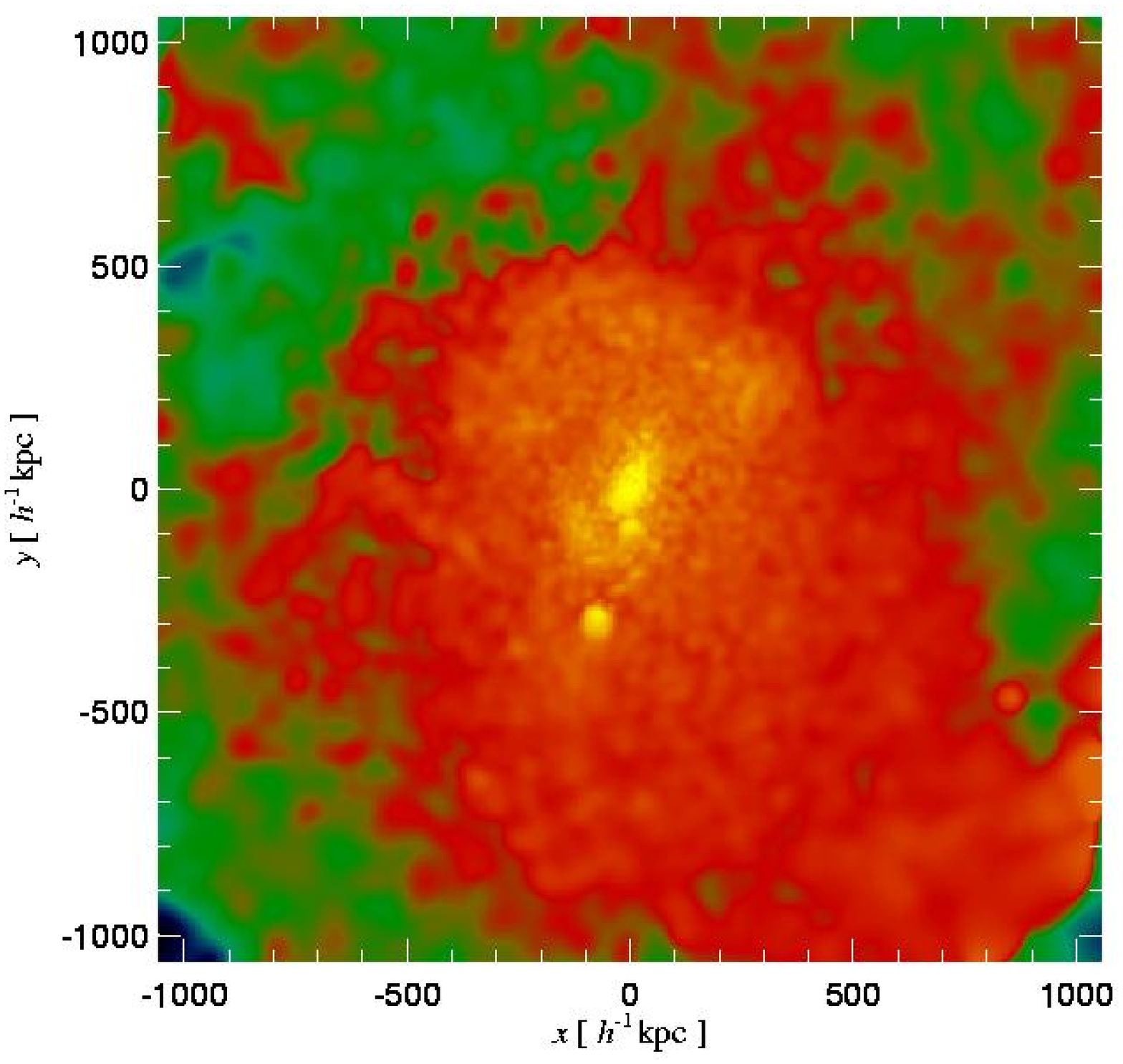}
  \end{minipage} \\%
  \hspace{5mm}
  \begin{minipage}[c]{0.30\textwidth}
    \centering \includegraphics[width=10mm,angle=270]{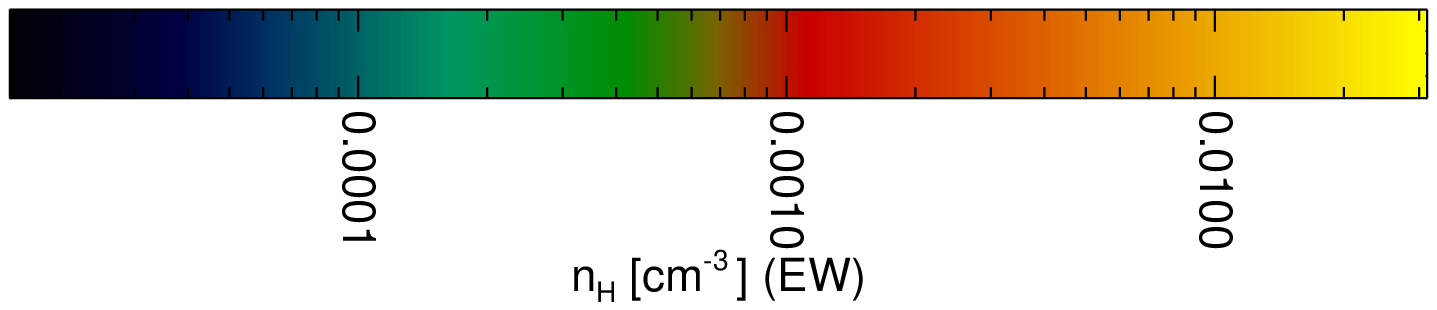}
  \end{minipage}%
  \hspace{5mm}
  \begin{minipage}[c]{0.30\textwidth}
   \centering \includegraphics[width=10mm,angle=270]{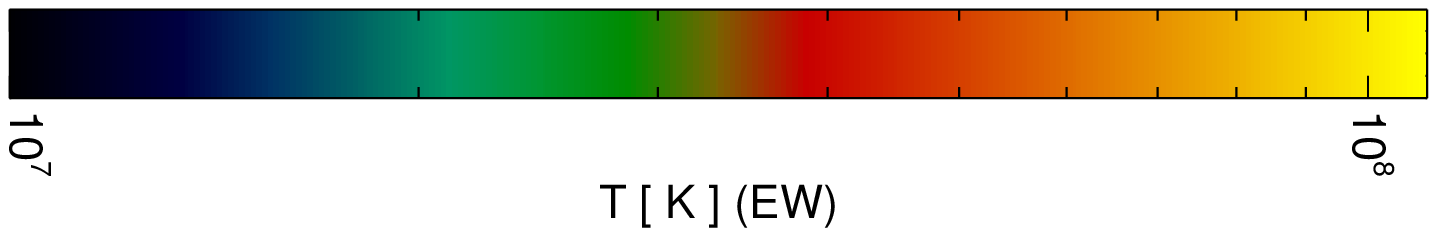}
  \end{minipage}%
  \hspace{5mm}
  \begin{minipage}[c]{0.30\textwidth}
   \centering \includegraphics[width=10mm,angle=270]{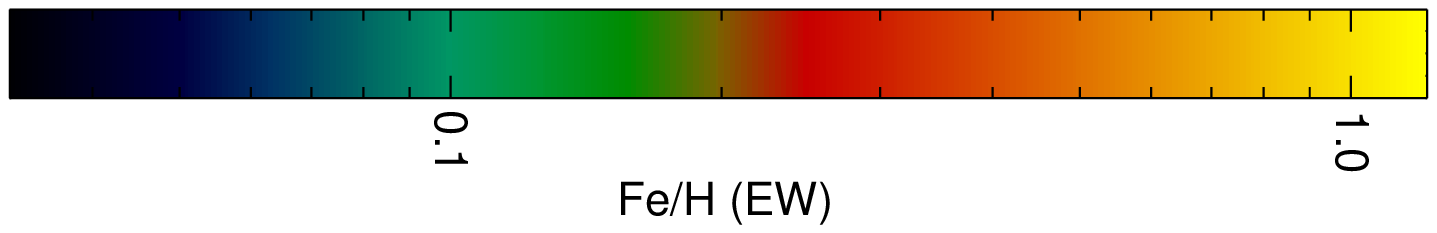}
  \end{minipage} %
\caption{Projection of Fe ${\rm K}_{\alpha}$ 6.7 keV emission
line-weighted (EW) properties of gas particles contained within a
sphere of radius $1\, h^{-1}\,{\rm Mpc}$ centred on the dominant
cluster galaxy: hydrogen number density (left panel), temperature
(middle panel) and iron abundance by number Fe/H relative to the solar value
(right panel).}
\label{figMaps2}
\end{figure*}

The first step in computing line-emission weighted properties of the
ICM from our simulations is to estimate the intensity of the Fe ${\rm
K}_{\alpha}$ 6.7 keV emission line for each gas particle.  The atomic
processes involved in the generation of the emission are quite
sensitive to the hydrogen volume density $n_H$, temperature $T$, and
chemical abundances of the gas.  Following the procedure performed by
\citet{furlanetto04}, we build a grid where these quantities are
varied, and we compute the intensity of the emission line for each
point in the grid.  For this purpose we use the radiative-collisional
equilibrium code {\small CLOUDY} \citep{ferland00,ferland01}. Given
the thermodynamical conditions of the ICM, we can well approximate it
as a low density hot plasma in collisional equilibrium, without being
affected by an ionizing backgroud.  Therefore, photoionization is
discarded in the models we generate with {\small CLOUDY}.

The iron line emissivity of the gas particles that make up the ICM is
obtained by interpolating the emissivities assigned to our grid of gas
states, where the parameters defining the three-dimensional grid in
density, temperature and iron abundances cover the range of values
exhibited by the gas inside the virial radius of the cluster at $z=0$.
An important detail of our chemical implementation is that we follow
the evolution of the abundances of several species.  Hence, instead of
assuming solar metallicity for gas particles, or varying this quantity
as a global parameter like in other works (e.g., \citealt{suth93,
furlanetto04}), our grid of models depends on sets of
abundances of the chemical elements considered in the semi-analytic
model.  As gas particles become more metal rich, the abundances of all
elements increase roughly proportionally to each other. We take
advantage of this feature to construct the sets of abundances
characterised by the mean value of iron abundance within a certain
range of values.
Since we only consider the most abundant elements in our semi-analytic
model, the code {\small CLOUDY} automatically gives solar values
(meteroritic abundances of \citet{grevesse89} with extensions by
\citet{grevesse93}) for the rest of the elements.
                                                                                
The range of values covered by the grid parameters are $ 1\times 10^6
< T < 1\times 10^9 \,$ K for gas temperature, and $3.16 \times 10^{-6}
< n_{\rm H} < 0.1 \, {\rm cm}^{-3}$ for the hydrogen volume density.
This last range of values is estimated from the density $\rho$ of gas
particles in the {\em N}-Body/SPH simulation and from their hydrogen
abundance $A_{\rm gas}^{\rm H}$ in our model.  In general, this
quantity ranges from the initial value of 0.76 to $\sim 0.70$,
which is reached by the most metal rich particles.  Iron abundances by
number range from Fe/H$\sim 5 \times 10^{-8}$ to $\sim 5 \times 10^{-5}$,
which are $\sim 1.8 \times 10^{-3}$ and $\sim 1.8$ times the solar value,
respectively.  

We choose density spacings $\Delta \log(n_{\rm H}/{\rm cm}^{-3})=0.25$. 
Since emissivities are quite sensitive to 
temperature variations, we
adopt temperature spacings of $\Delta \log(T/{\rm
K})=0.1 $, for $10^7 {\rm K} < T < 10^{8.5} {\rm K}$, and $\Delta
\log(T/{\rm K})=0.25$ otherwise.  
We consider seven bins for the
iron abundances with spacings $\Delta \log({\rm Fe}/{\rm H})=0.5$ from
$\log({\rm Fe}/{\rm H})=-7.75$ to $-4.25$.  In order to produce the
whole grid of models in a systematic way, we use a user-friendly
package of IDL routines that facilitates the use of the code {\small
CLOUDY} (in version C96B4), called {\small MICE} (MPE IDL Cloudy
Environment), developed by Henrik Spoon.

Figure~\ref{figMaps1} shows the projected mass-weighted maps of
hydrogen density (top-left panel), temperature (top-right panel), iron
abundance (bottom-left panel) and emissivity of the Fe ${\rm
K}_{\alpha}$ 6.7 keV line  normalised with the maximum 
mass-weighted emissivity $I_{\rm o}$ (bottom-right panel) for gas particles
within a radius of $1\, h^{-1}\,{\rm Mpc}$ from the cluster centre at
$z=0$.  From the first three maps we can connect the iron abundances
with the distributions of gas density and temperature in the ICM.  We
can see that those regions with high levels of chemical enrichment are
associated with cool and high density areas; this fact is particularly
evident in the small substructures in the outskirts of the cluster
that have been recently accreted.  These substructures do not produce
strong X-ray emission due to their low temperature (fourth map), and
the line emission map results in rather smooth maps, following mainly
the coarse features of the density distribution.
                                                                                
Figure~\ref{figMaps2} is equivalent to Figure~\ref{figMaps1}, but the
Fe ${\rm K}_{\alpha}$ 6.7 keV emission-line-weighted quantities are
shown instead.  We can see that the denser, cooler and more metal rich
substructures found in the outskirts of the mass-weighted maps of
Figure~\ref{figMaps1} do not appear in the corresponding plots of
Figure~\ref{figMaps2}, since the intensity of the lines emitted from
these regions is very low (see bottom-right panel of
Figure~\ref{figMaps1}).

\section[]{Enrichment history and dynamical evolution of the ICM} \label{sec_EvolMaps}

In order to 
understand the way in which the chemical enrichment patterns characterising
the intracluster gas develop,
we analyse the
temporal evolution of maps of the projected chemical properties of the
gas and the corresponding radial abundance profiles. 
This allows us 
to evaluate the relative contribution of SNe CC and Ia to 
the metal enrichment, as well as to investigate the dynamical evolution 
of the intracluster gas which plays a fundamental role in the
spatial distribution of metals.

\begin{figure*}
  \centering
  \begin{minipage}[c]{.3\textwidth}
   \centering \includegraphics[width=57mm]{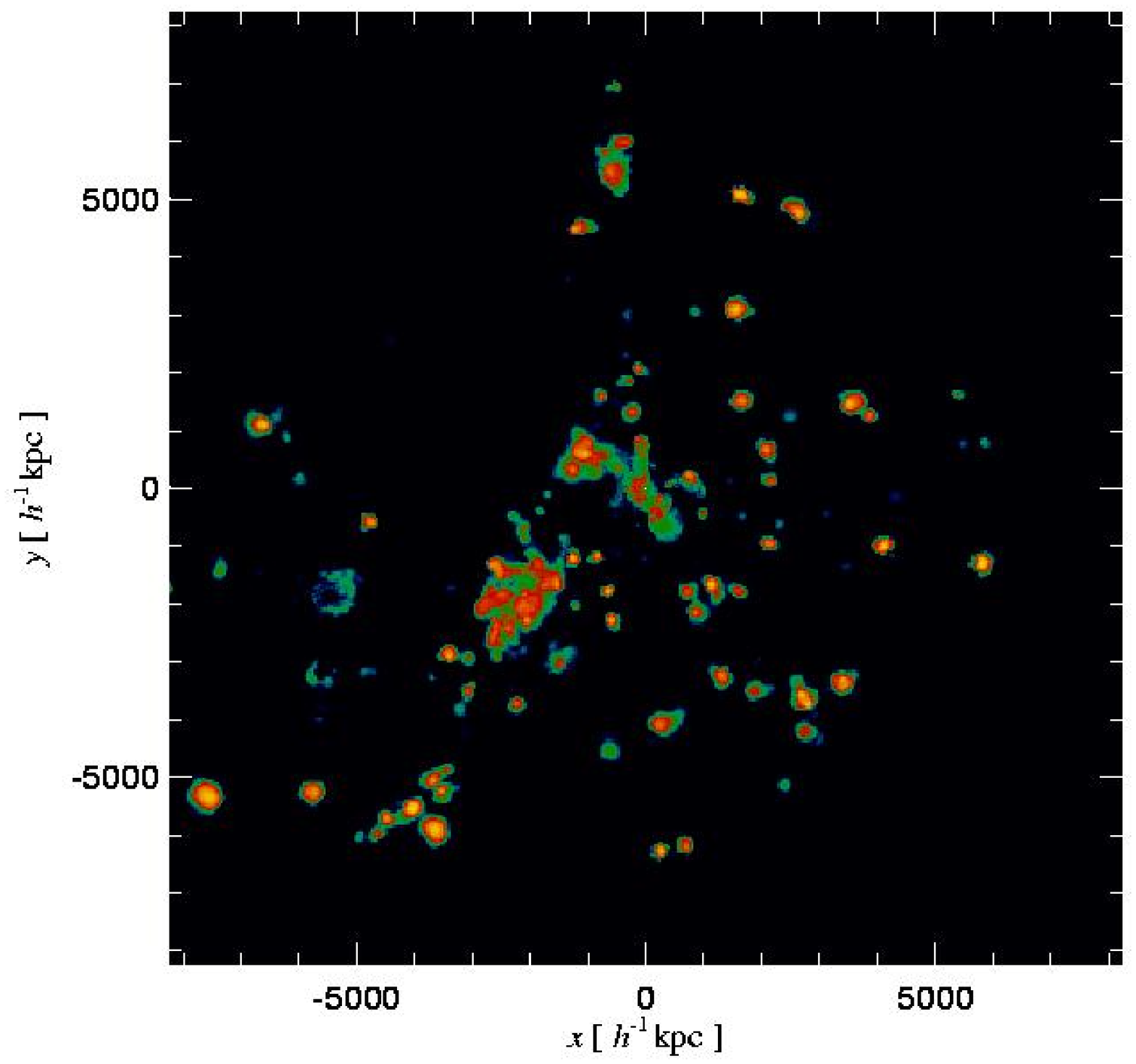}
  \end{minipage}%
  \hspace{4mm}
  \begin{minipage}[c]{.3\textwidth}
   \centering \includegraphics[width=57mm]{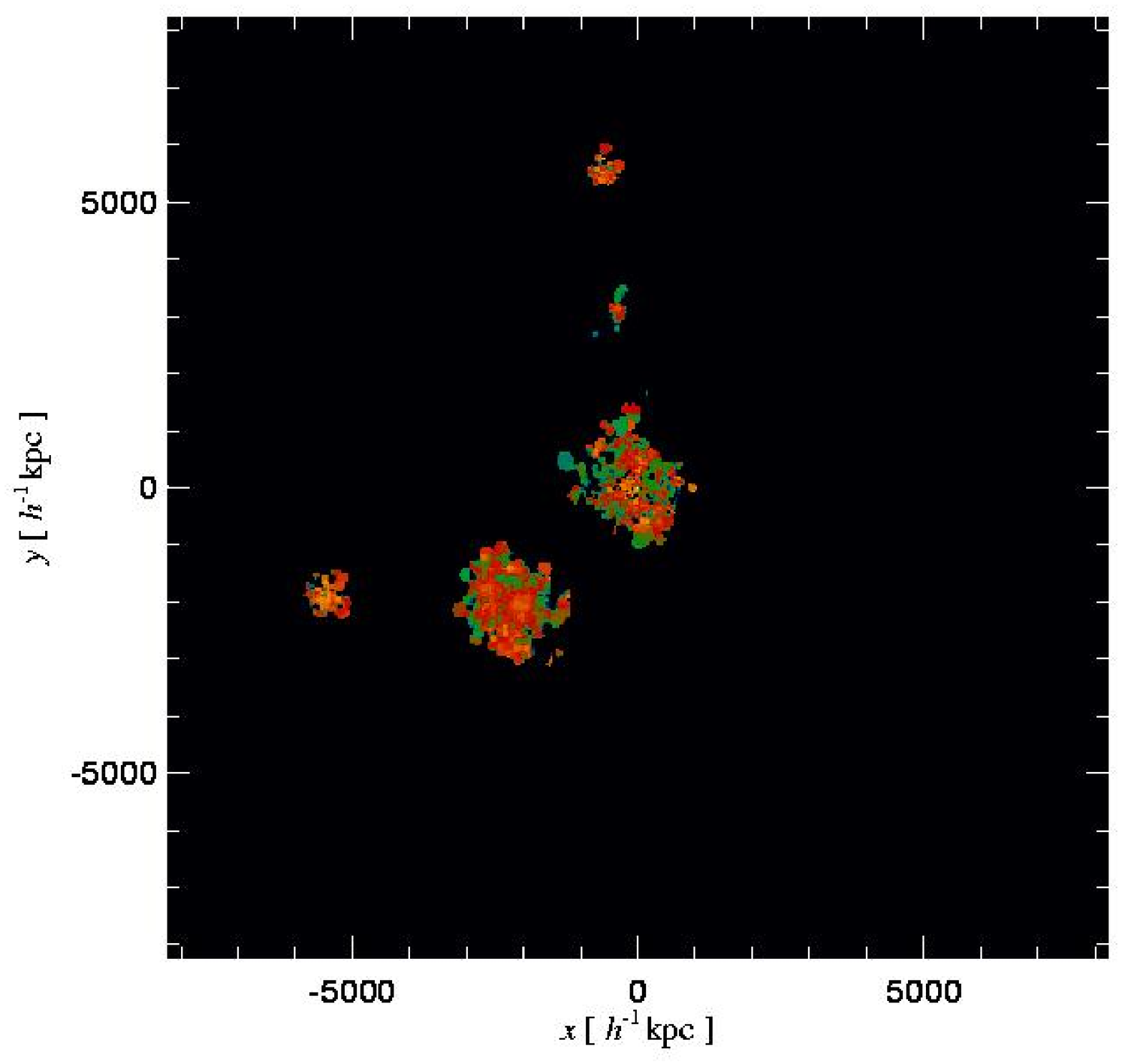}
  \end{minipage}%
  \hspace{4mm}
  \begin{minipage}[c]{.3\textwidth}
   \centering \includegraphics[width=57mm]{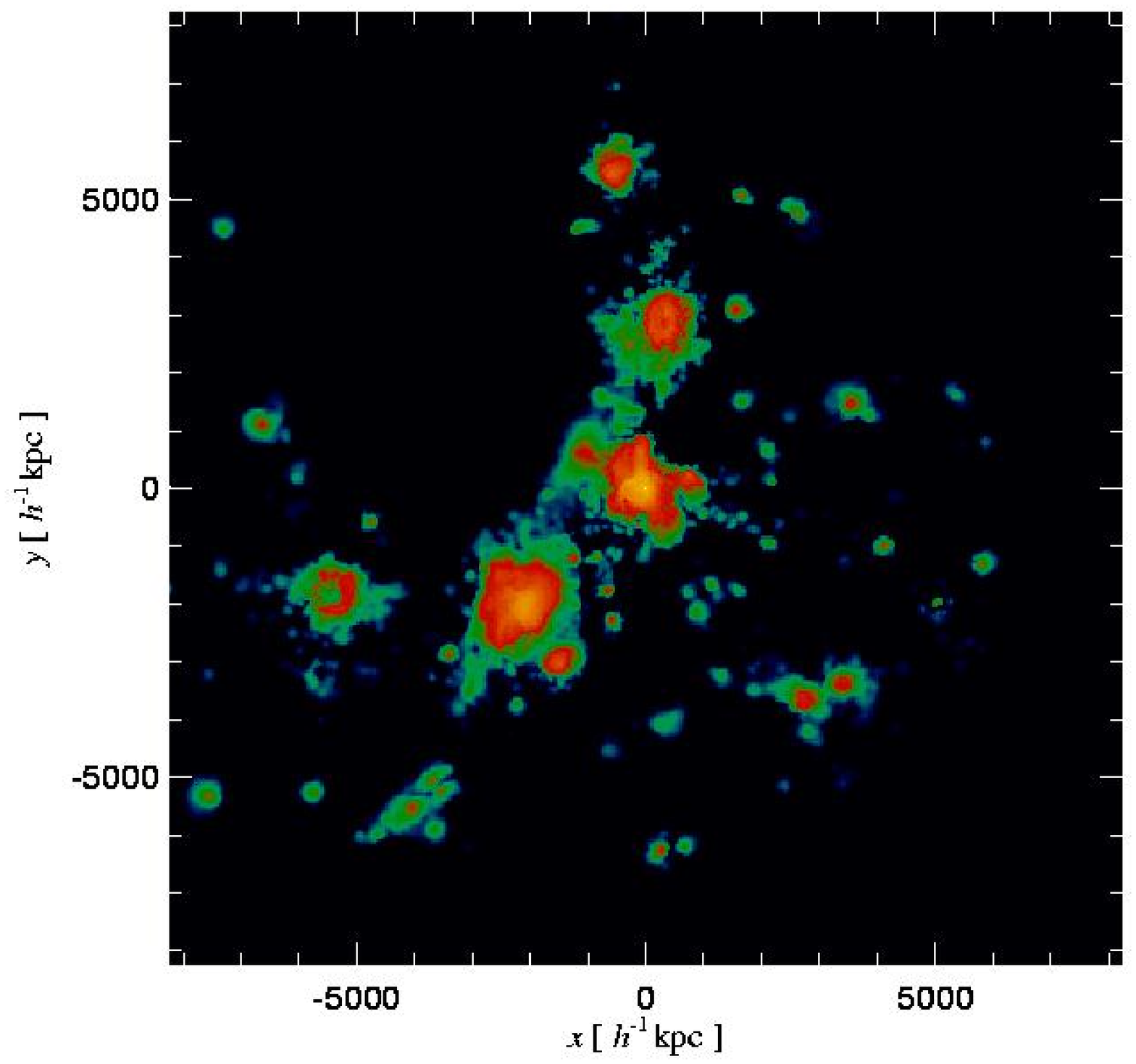}
  \end{minipage} \\%
%
%
  \begin{minipage}[c]{.3\textwidth}
   \centering \includegraphics[width=57mm]{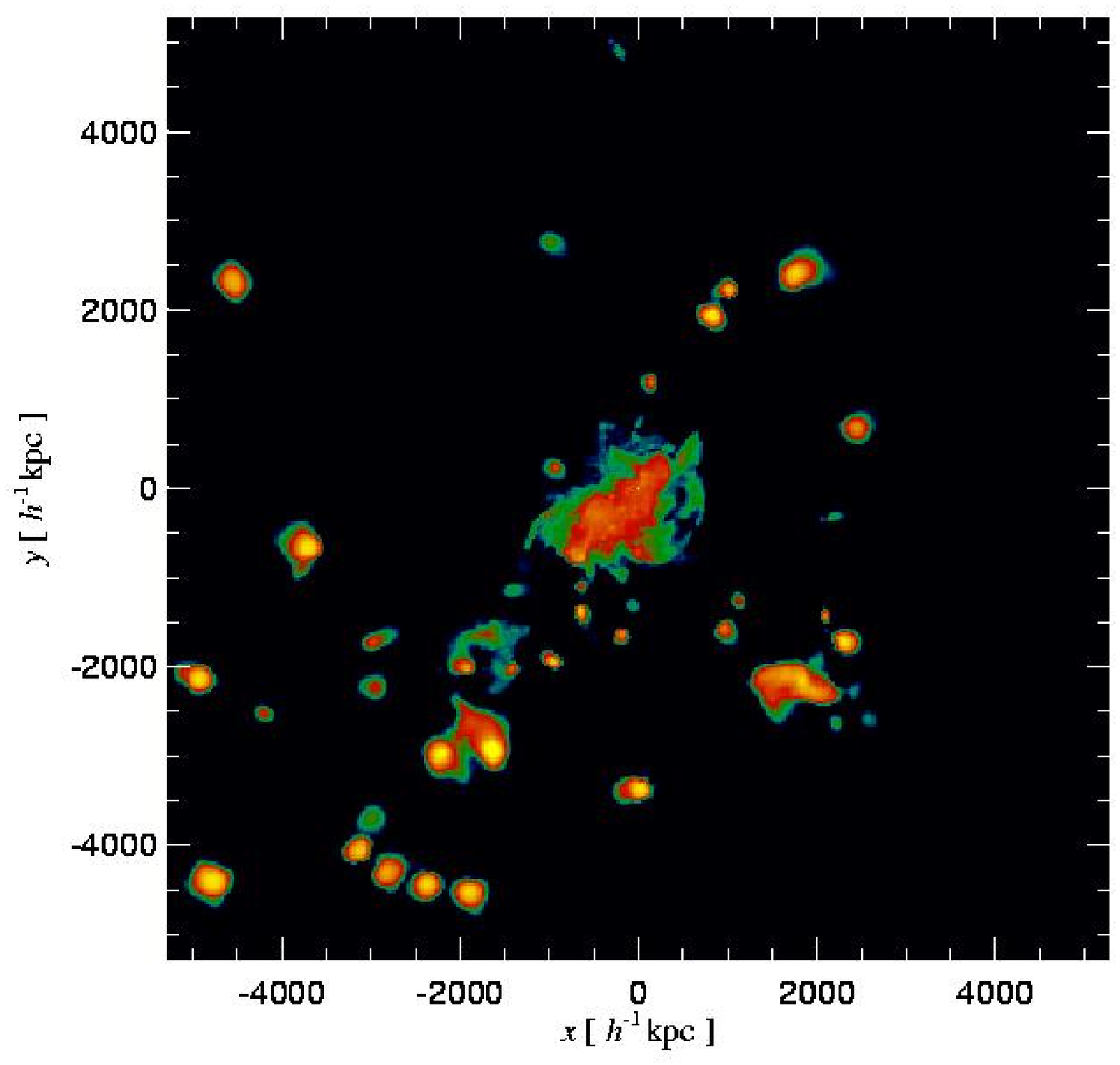}
  \end{minipage}%
  \hspace{4mm}
  \begin{minipage}[c]{.3\textwidth}
   \centering \includegraphics[width=57mm]{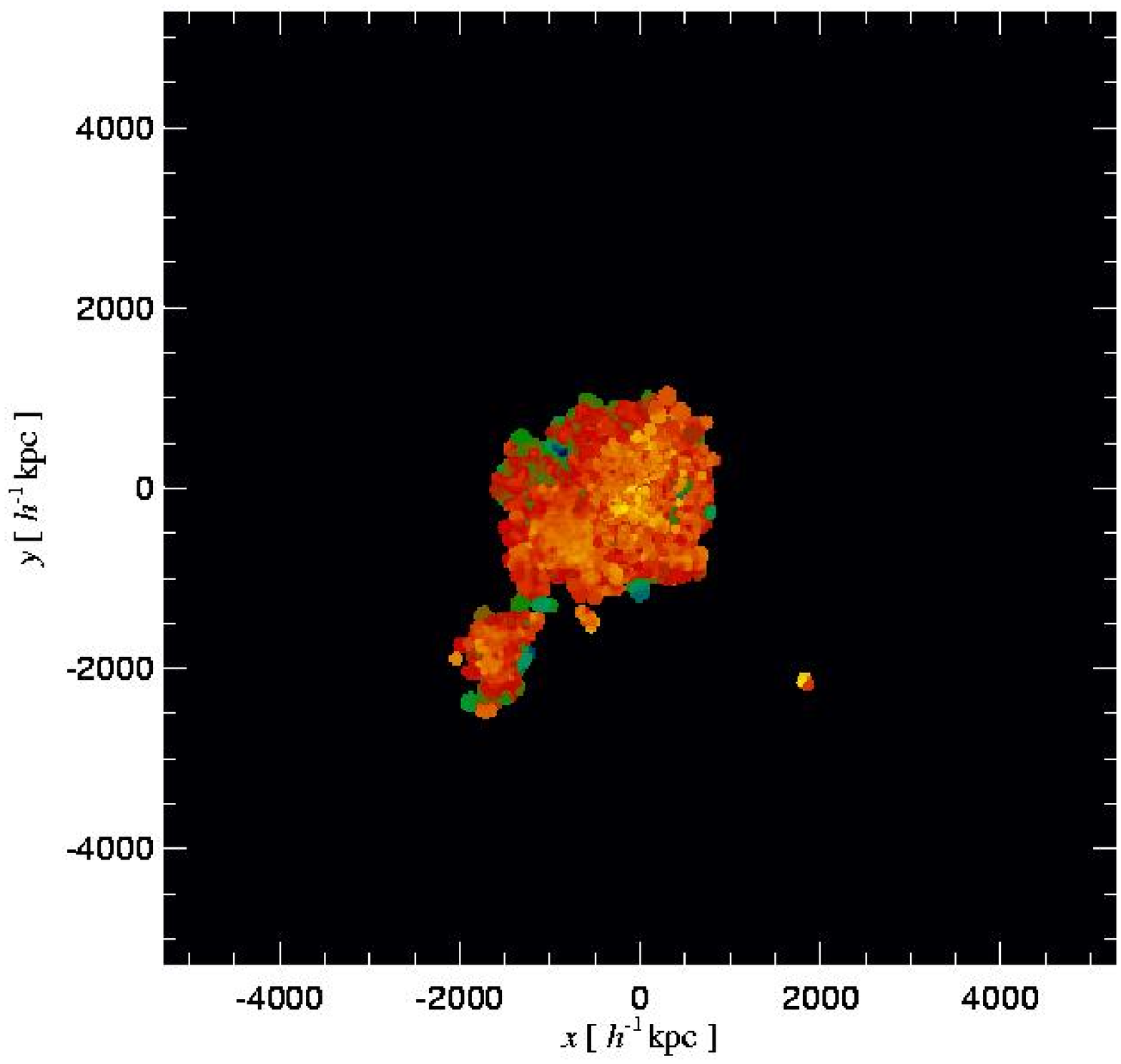}
  \end{minipage}%
  \hspace{4mm}
  \begin{minipage}[c]{.3\textwidth}
   \centering \includegraphics[width=57mm]{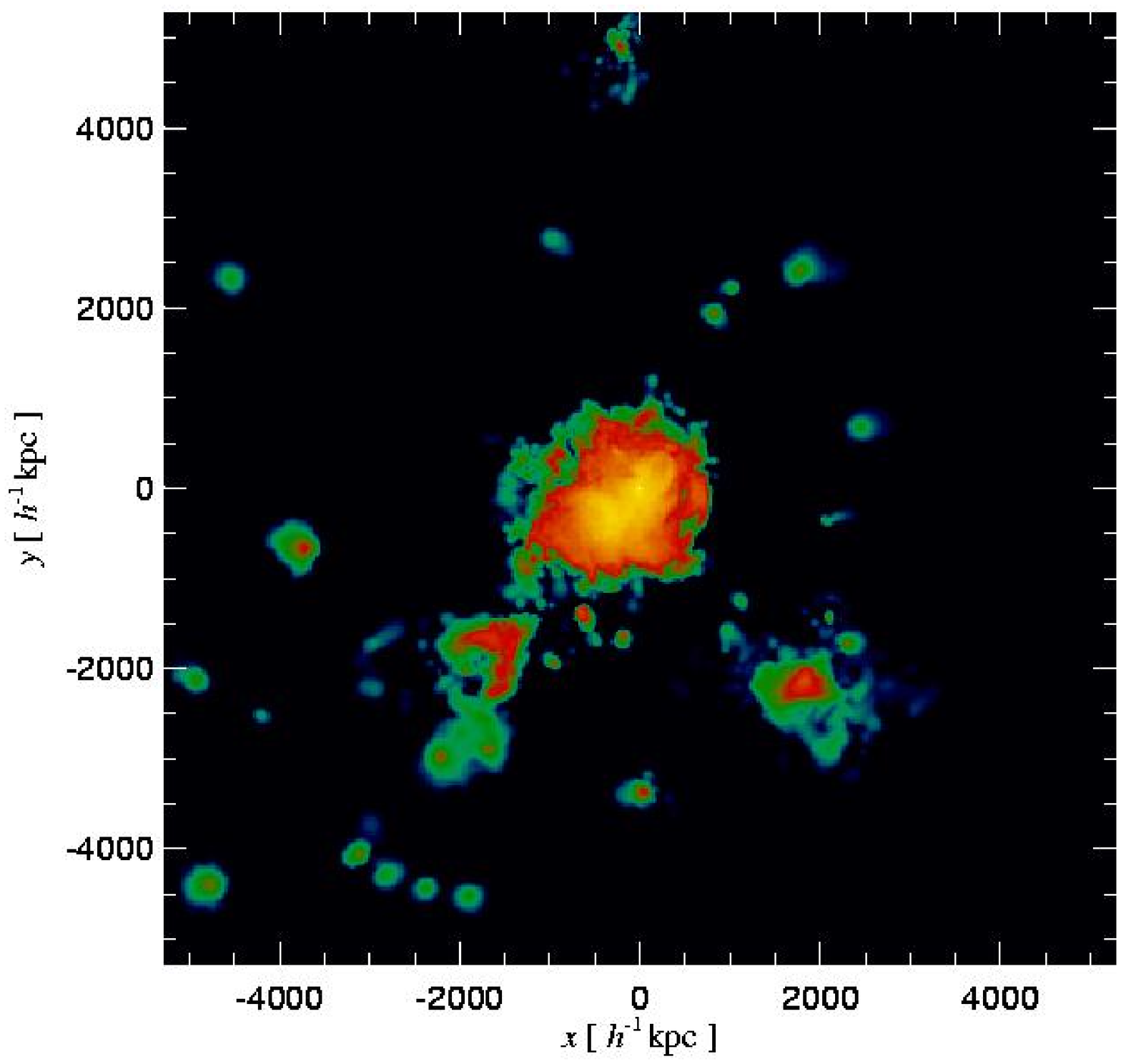}
  \end{minipage} \\%
%
%
  \begin{minipage}[c]{.3\textwidth}
   \centering \includegraphics[width=57mm]{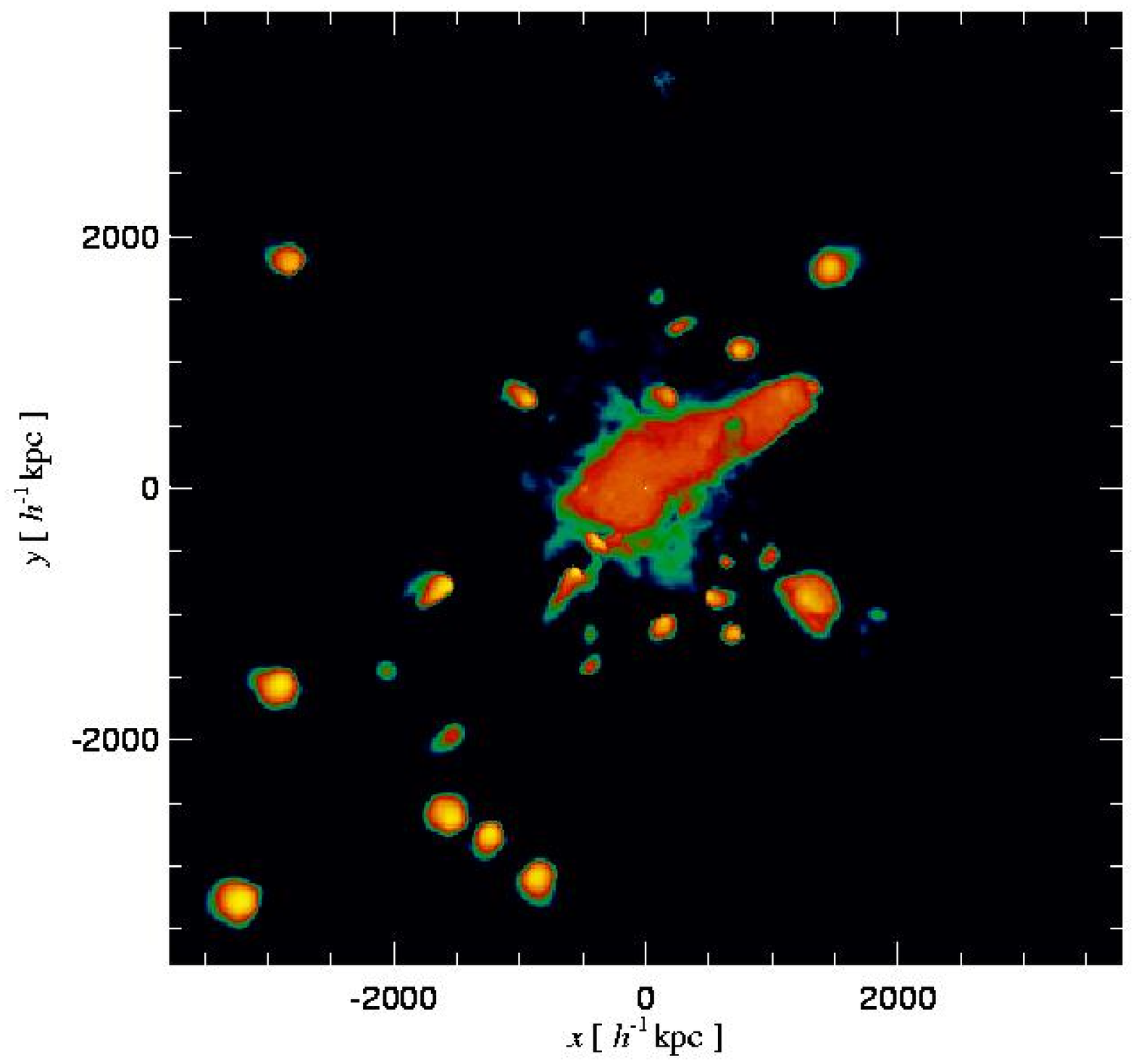}
  \end{minipage}%
  \hspace{4mm}
  \begin{minipage}[c]{.3\textwidth}
   \centering \includegraphics[width=57mm]{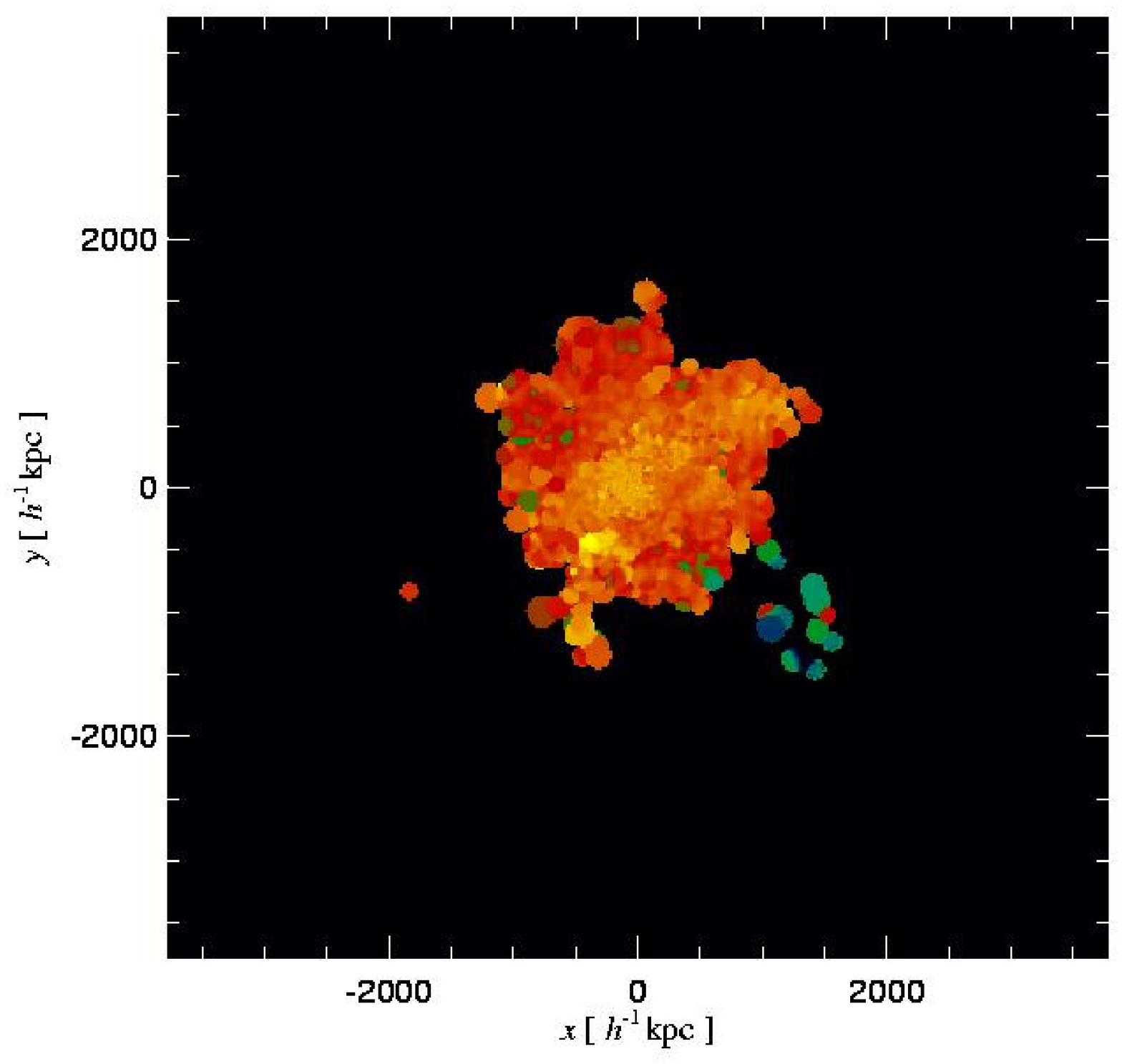}
  \end{minipage}%
  \hspace{4mm}
  \begin{minipage}[c]{.3\textwidth}
   \centering \includegraphics[width=57mm]{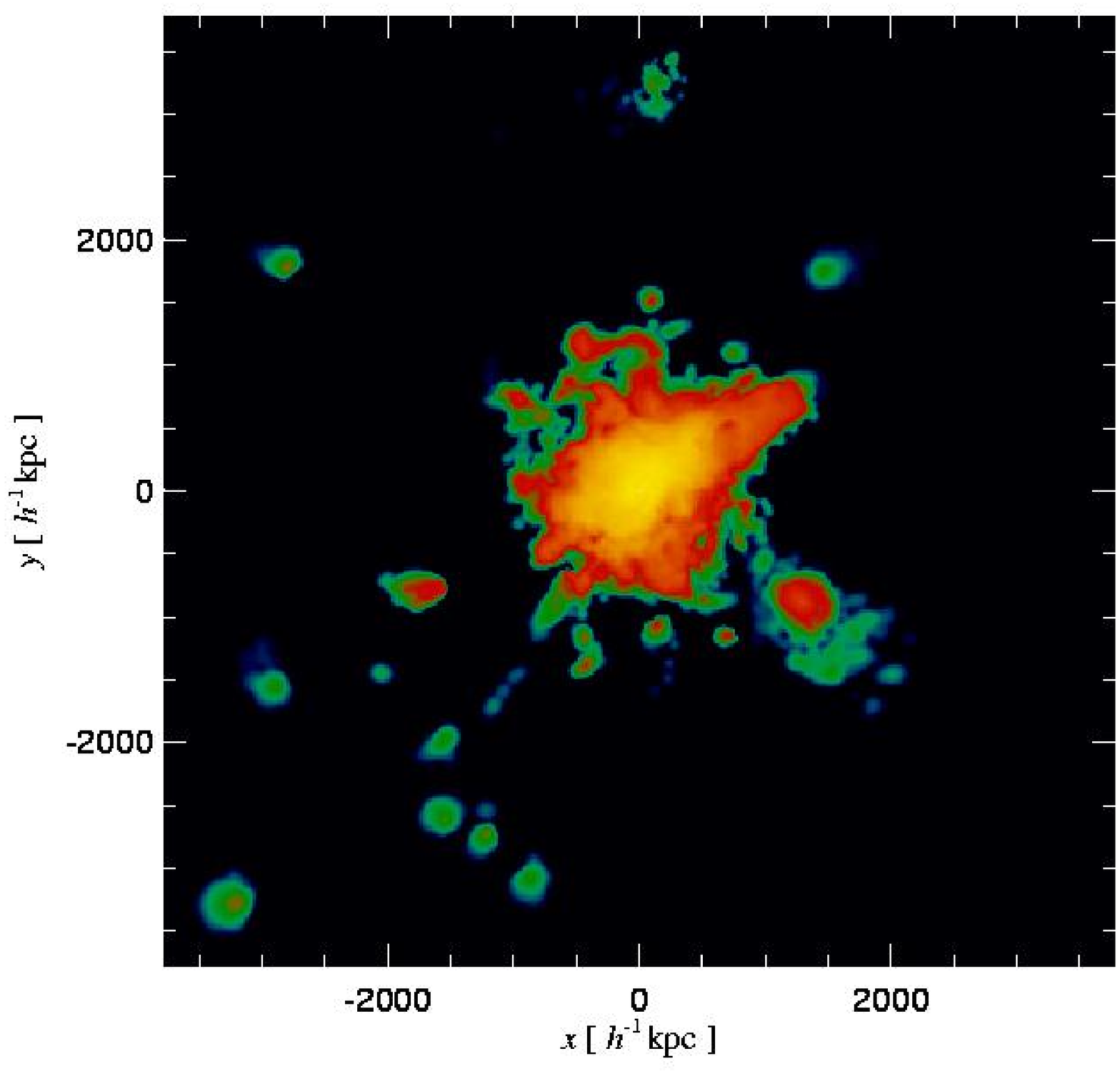}
  \end{minipage}\\%
 \begin{minipage}[c]{.71\textwidth}
   \centering \includegraphics[width=20mm,angle=270]{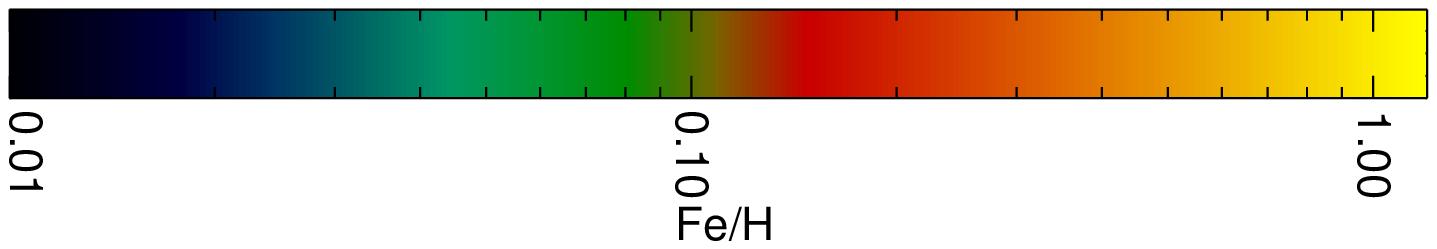}
  \end{minipage}%
 \begin{minipage}[c]{.29\textwidth}
   \centering \includegraphics[width=10mm,angle=270]{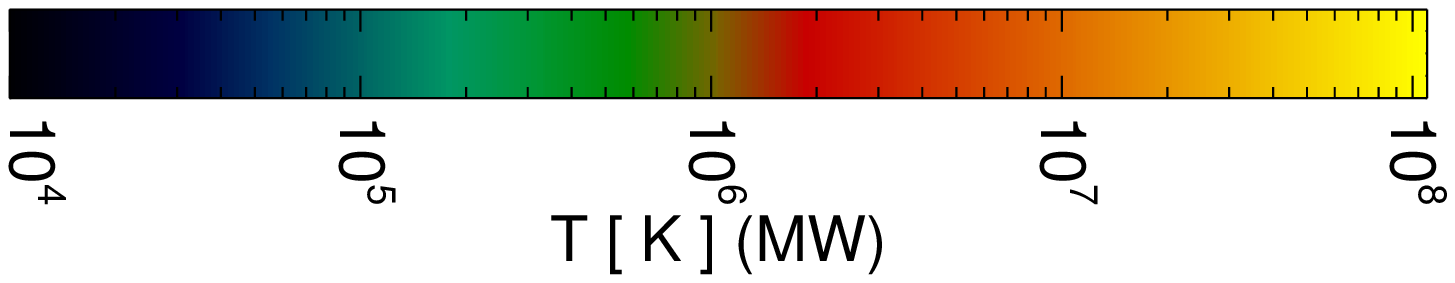}
  \end{minipage}
\caption{Evolution of iron abundance and temperature of gas particles
that lie within $500\, h^{-1}$ kpc from the cluster centre at $z=0$.
The plots show the projection of mass-weighted iron abundance by
number relative to hydrogen, Fe/H, respect to the solar value (left
column), the corresponding Fe ${\rm K}_{\alpha}$ 6.7 keV
emission-line-weighted abundance (middle column), and the
mass-weighted temperature of the gas (right column).  We give results
for three redshifts: $z \sim 1$ (lookback time of $7.8$ Gyr; first
row), $z \sim 0.5$ (lookback time of $4.9$ Gyr; second row), and $z
\sim 0.3$ (lookback time of $3.6$ Gyr; third row). 
At each refshift, the spatial coordinates are centred at the most massive 
progenitor of the cluster at $z=0$ and are expressed in comoving scales.} 
\label{figMaps3}
\end{figure*}

\begin{figure*}
  \centering
  \begin{minipage}[c]{.3\textwidth}
   \centering \includegraphics[width=57mm]{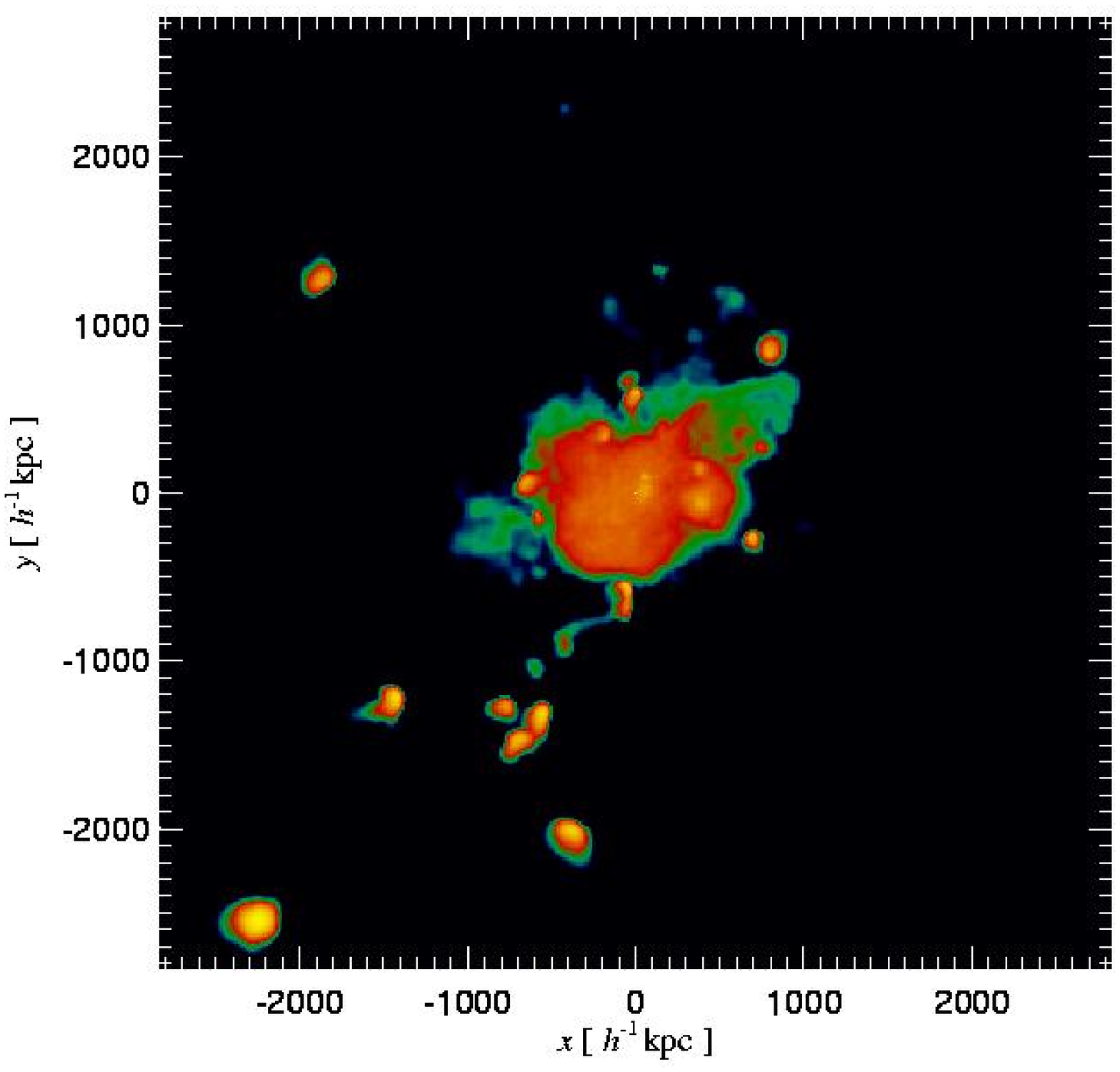}
  \end{minipage}%
  \hspace{4mm}
 \begin{minipage}[c]{.3\textwidth}
   \centering \includegraphics[width=57mm]{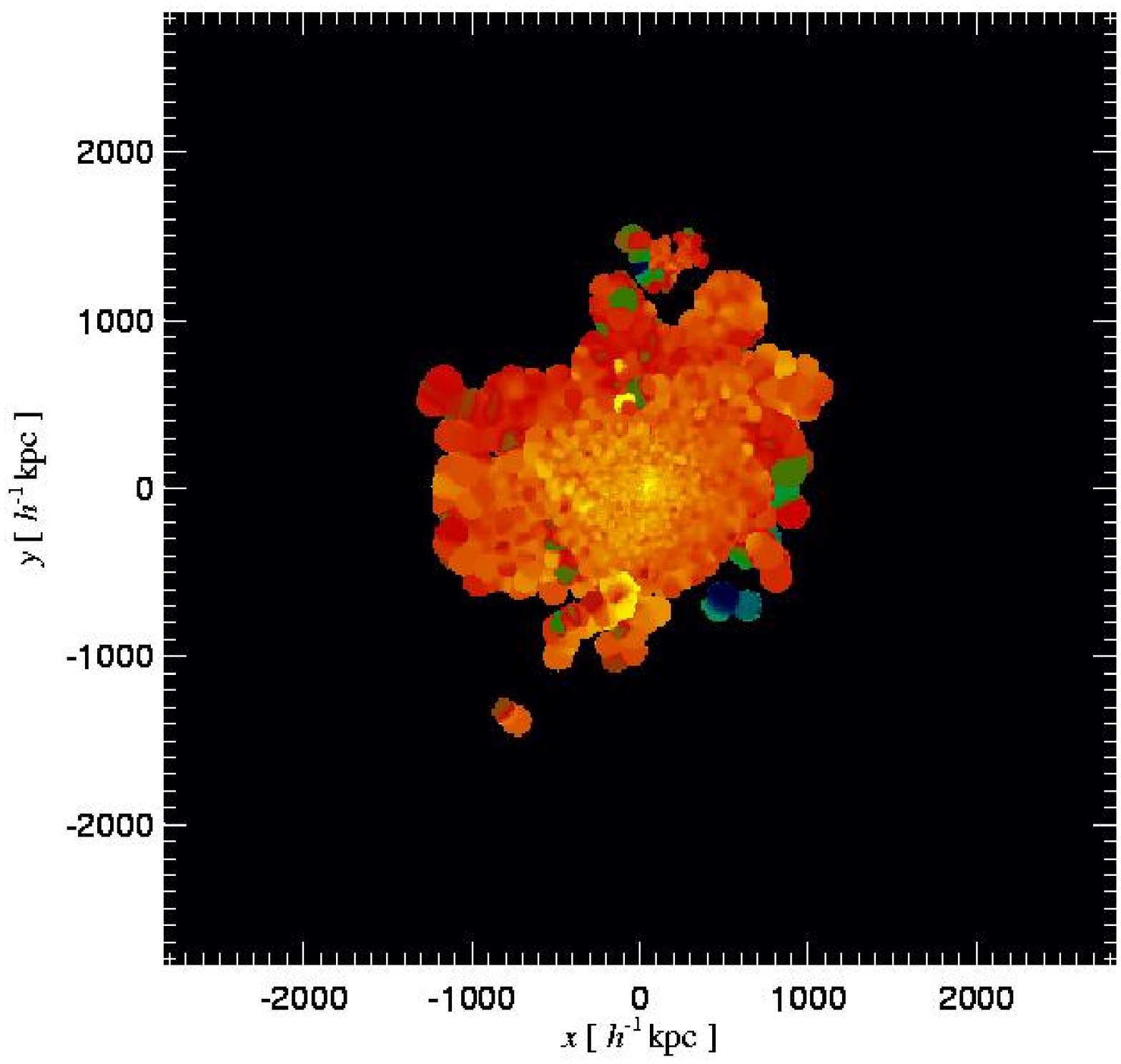}
  \end{minipage}%
  \hspace{4mm}
  \begin{minipage}[c]{.3\textwidth}
   \centering \includegraphics[width=57mm]{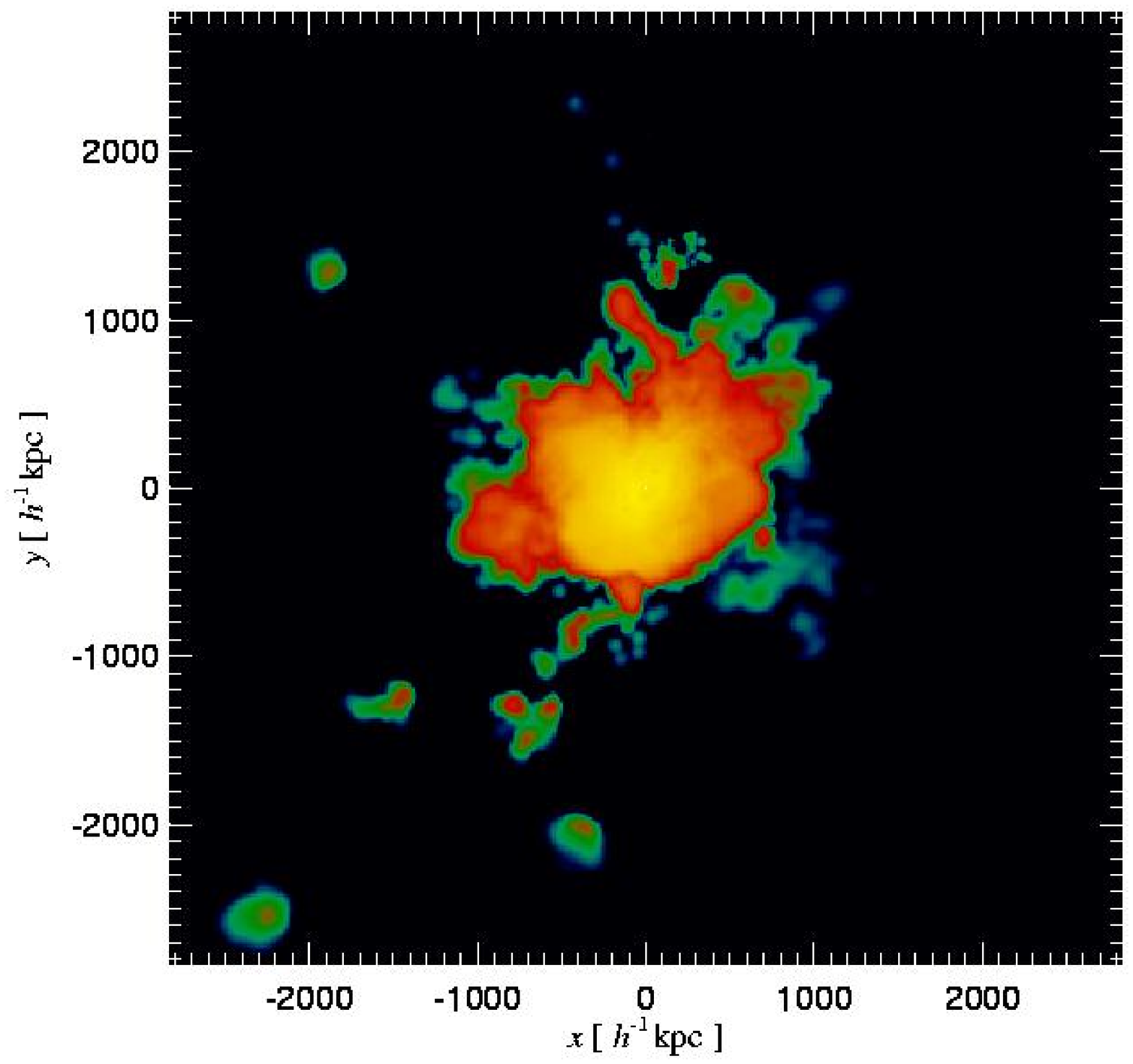}
  \end{minipage} \\%
%
%
  \begin{minipage}[c]{.3\textwidth}
   \centering \includegraphics[width=57mm]{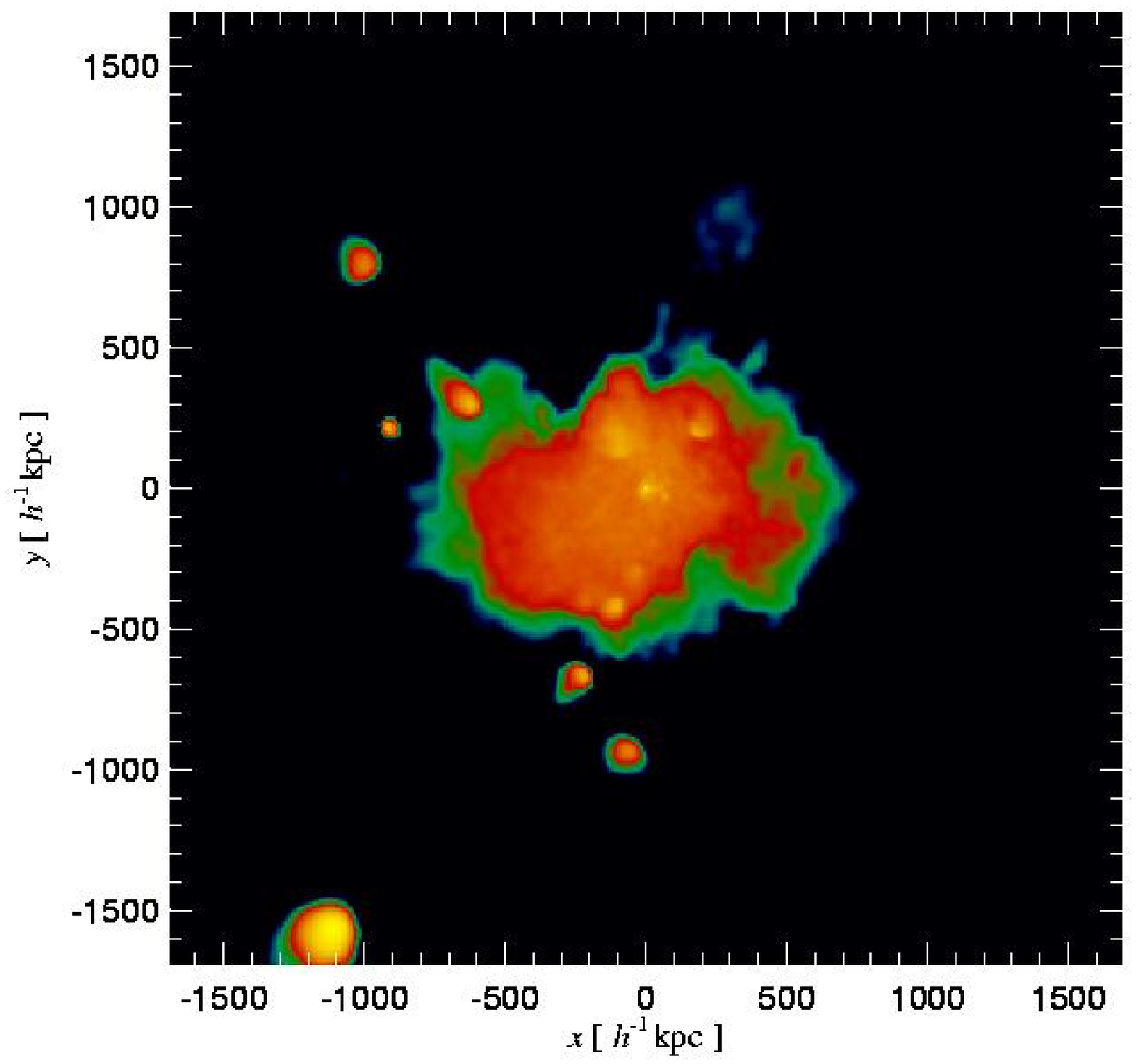}
  \end{minipage}%
  \hspace{4mm}
  \begin{minipage}[c]{.3\textwidth}
   \centering \includegraphics[width=57mm]{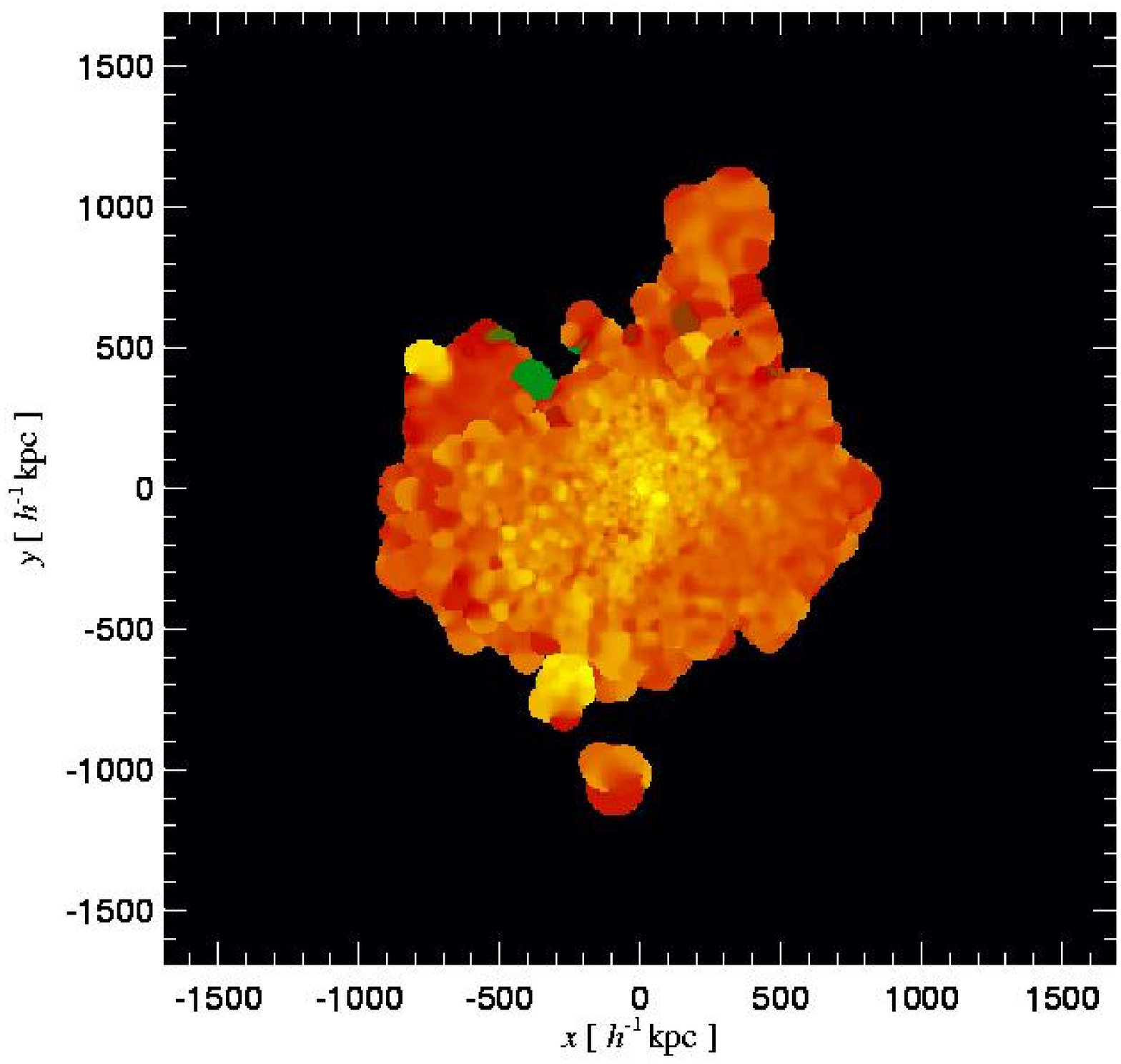}
  \end{minipage}%
  \hspace{4mm}
  \begin{minipage}[c]{.3\textwidth}
   \centering \includegraphics[width=57mm]{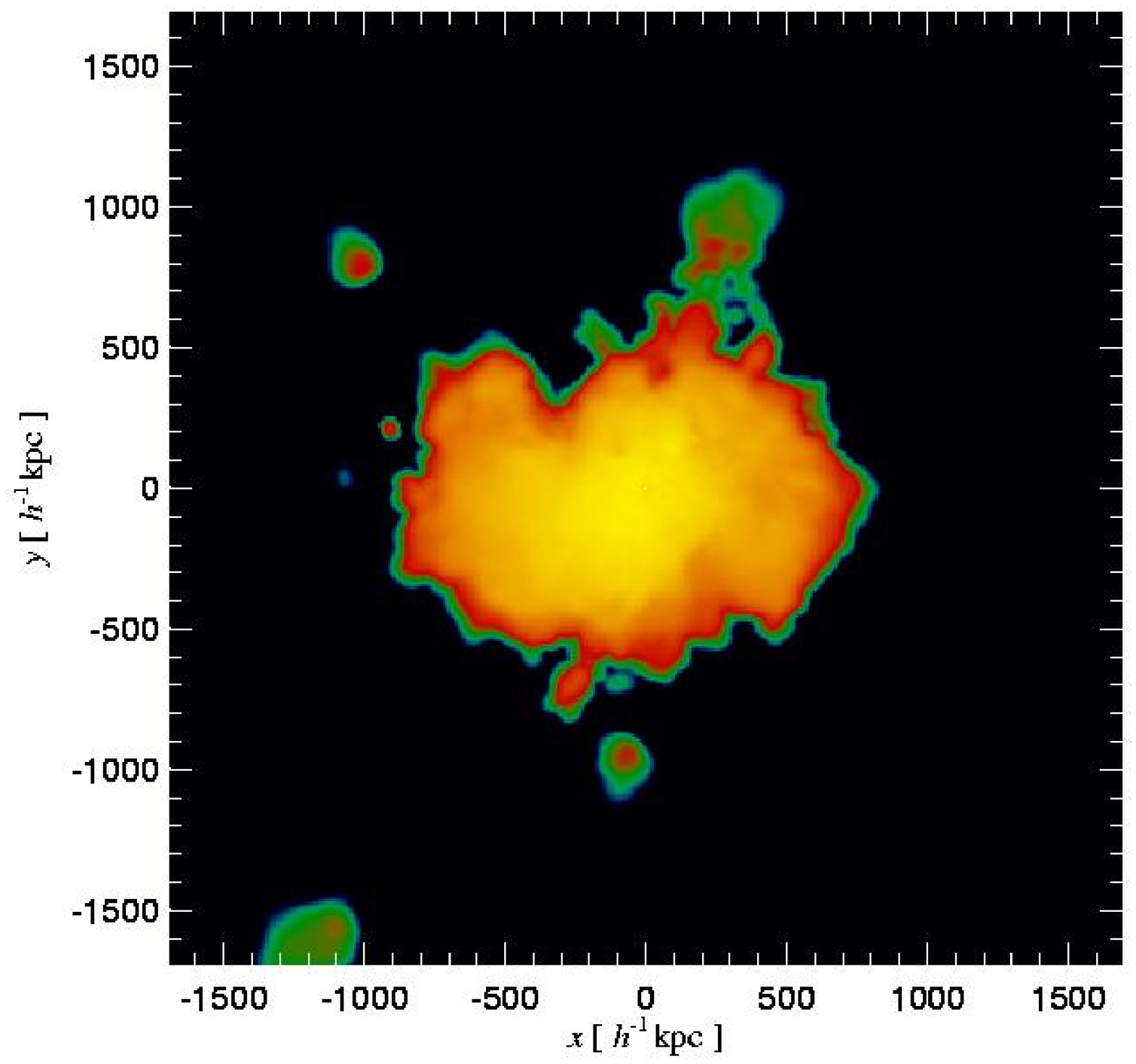}
  \end{minipage} \\%
 \begin{minipage}[c]{.71\textwidth}
   \centering \includegraphics[width=20mm,angle=270]{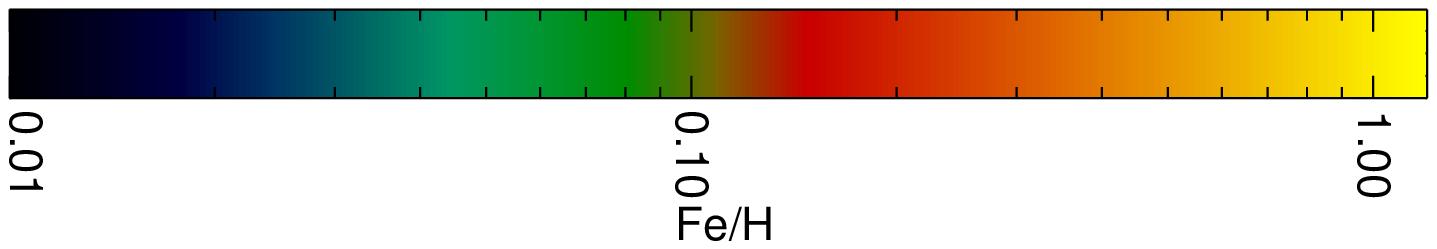}
  \end{minipage}%
 \begin{minipage}[c]{.29\textwidth}
   \centering \includegraphics[width=10mm,angle=270]{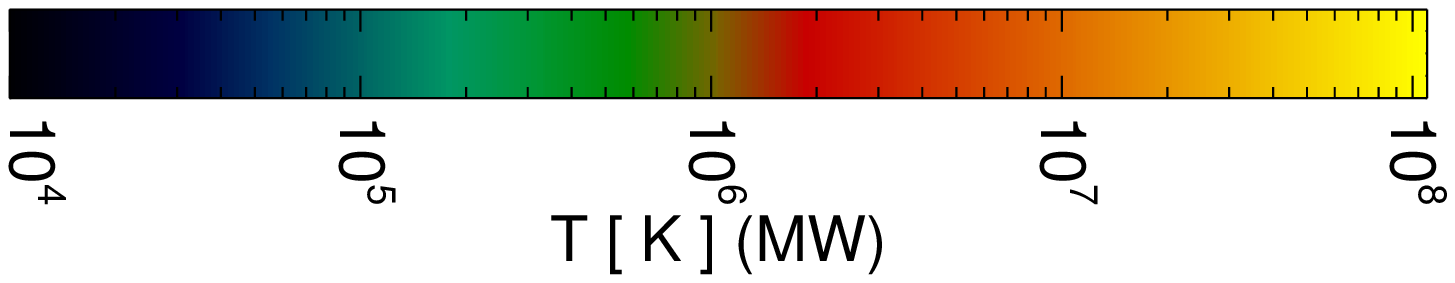}
  \end{minipage}
\caption{Same as Fig.~\ref{figMaps3}, but for redshifts: $z \sim
  0.2$ (lookback time of $2.4$ Gyr; first row), and $z \sim 0.1$
  (lookback time of $1.2$ Gyr; second row).}
\label{figMaps4}
\end{figure*}

For this
purpose, we first consider gas particles lying within $500 \, h^{-1}$~kpc
from the cluster centre at $z=0$ and trace their properties back in
time.  Thus, we follow the chemical enrichment of small substructures
forming at high redshift and study how they progressively become part
of the cluster and contribute to establishing its final distribution
of metals.
Figures~\ref{figMaps3} and \ref{figMaps4} show mass-weighted iron
abundances by number relative to the sun, Fe/H (left column), the
corresponding Fe ${\rm K}_{\alpha}$ 6.7 keV emission-line-weighted
abundances (middle column), and the mass-weighted temperature of the
gas (right column), for five different redshifts from $z \sim 1$
to $z \sim 0.1$.
At each redshift, the cluster centre is defined as the centre of the
most massive progenitor of the cluster at $z=0$, and the distances
from it are estimated using comoving coordinates.
We can see that the ICM at $z=0$ is formed by the accretion and merger
of small substructures that at early times ($z\sim 1$) are spread
within an extended region of radius $\sim 9\,h^{-1}$ Mpc.  These
gaseous building blocks are already considerably contaminated by that
time, with iron abundances that reach values of ${\rm Fe}/{\rm H}
\sim 0.3$.

\begin{figure}
    \centering \includegraphics[width=85mm]{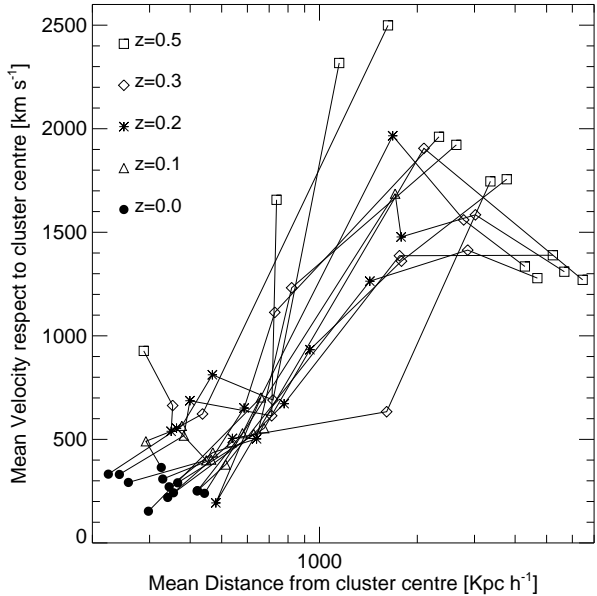}
\caption{
Mean velocities of gas particles contained within spherical 
shells of $0.5 \,h^{-1}$~Mpc of
thickness, centred at the most massive progenitor of the cluster
at $z \sim 0.5$, as a function of their mean clustercentric distances.
This dependence is estimated for different redshifts as denoted
by the symbols.
The lines connect the sets of gas
particles defined according to their clustercentric distance at $z \sim 0.5$.
}
\label{figVelDistz}
\end{figure}

From these figures, we can also infer interesting aspects about the
dynamics of the contaminated gaseous clumps. Those that
are at $\sim 6 \,h^{-1}$ Mpc from the cluster centre at $z \sim 0.5$
needs to move at more than $\sim 1000 \, {\rm km} \, {\rm s}^{-1}$ 
to get inside $0.5 \,h^{-1}$ Mpc by $z=0$. 
A quantitative analysis of the dynamics of
these gaseous clumps is presented in Figure~\ref{figVelDistz},
which shows 
the dependence of the velocity of gas particles with their 
clustercentric
distances and its evolution with redshift.
We identify the 
gas particles contained within spherical shells of $0.5 \,h^{-1}$ Mpc of 
thickness, centred at the most massive progenitor of the cluster
at $z \sim 0.5$, and estimate their mean velocities and clustercentric
distances at lower redshifts. In this way, we follow the evolution of these
properties as the cluster is being assembled.
The different symbols denote
the redshifts considered, and the lines connect the sets of gas
particles defined according to their clustercentric distance at $z \sim 0.5$.
There are many aspects to emphasize from this plot.
At $z \sim 0.5$, gas particles 
have very high velocities, which range from 
$\sim 1300$ to $\sim 2500 {\rm km} \, {\rm s}^{-1}$. 
The highest velocities correspond to gas particles lying 
between $\sim 1$ and $\sim 2 \,h^{-1}$ Mpc. This is the same for redshifts
$z \sim 0.3$ and $\sim 0.2$, although the velocities acquired are smaller.
If we follow the evolution with redshift of gas particles lying
more than $4 \,h^{-1}$ Mpc at $z \sim 0.5$, we see that they 
are accelerated till they achieve the highest velocities 
at $z \sim 0.3$ and $\sim 0.2$.
This occurs at a distance corresponding to the virial radius of the main 
progenitor, which is of the order $1 \,h^{-1}$ Mpc at $z\sim 0.5$ 
and increases to $\sim 1.5 \,h^{-1}$ Mpc at $z=0$.
Thus, despite shell crossing, it is evident that the
dynamics of gas particles
is characterised by a general
behaviour in which they are accelerated till they reach
the main cluster shock radius, 
at a distance of the order of its virial radius, and
once they have been incorporated to it they converge to the centre
with velocities of $\sim 300 \, {\rm km} \, {\rm s}^{-1}$.
This turbulent velocity reached inside the virial radius
and the infall that dominates the gas
outside it 
are consistent with the results
obtained by \citet{Sunyaev03} based on the analysis of cluster simulations
at $z=0$ \citep{Norman99}.
The complex dynamics of gas particles acquired during the
hierarchical formation of the structure helps to explain
the way in which the metallicity profiles develop, as we
discuss later in this section.

\begin{figure}
 \centering
    \centering \includegraphics[width=85mm]{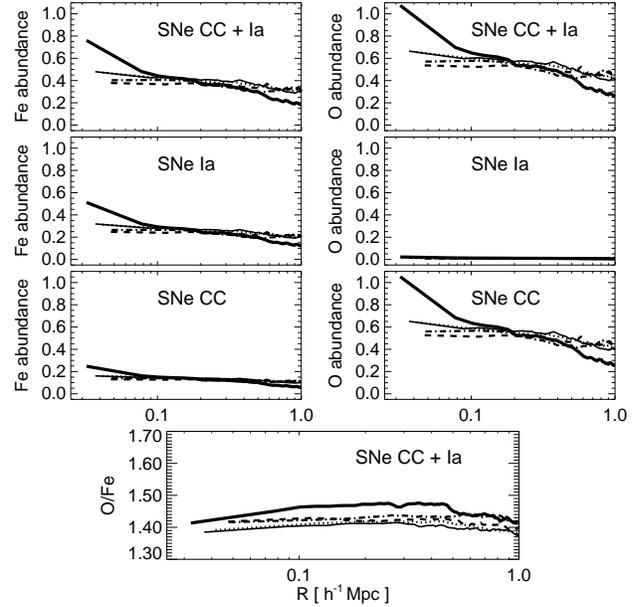}
\caption{Evolution of the radial profiles of iron and oxygen
abundances by number relative to hydrogen 
and of the profile of the O/Fe ratio, referred to the solar value.  
The profiles are
determined for the gas contained within the innermost comoving 
$1 \,h^{-1}$~Mpc at $z\sim 1$ (dash dot dot line), $z\sim 0.5$
(dash dot line), $z\sim 0.3$ (dashed line), $z\sim 0.2$ (dotted
line), $z\sim 0.1$ (thin solid line) and $z=0$ (thick solid line).
The first row shows the contributions of both types of SNe (CC and Ia)
to the Fe and O abundances, while the second and third rows show the
separate contributions of SNe Ia and CC, respectively.  The last row
presents the O/Fe abundance ratios when both types of SNe are
considered.}
\label{figProfEvol}
\end{figure}

Figure~\ref{figProfEvol} shows the evolution of iron and oxygen abundance 
profiles, which are determined for gas particles 
lying within the innermost comoving $1 \, h^{-1}$ Mpc
at $z=0$ and at each of the redshifts considered in 
Figures~\ref{figMaps3} and \ref{figMaps4}.
The first three rows show the shape of the profiles for the combined
effect of both types of SNe, and for the separate contributions of SNe
Ia and CC, respectively. The plot in the last row gives the evolution
of the O/Fe ratio profile resulting from the influence of the combined
effect of both types of SNe. The way in which the iron and oxygen
abundance profiles develop are quite similar, even though the former
is mainly determined by the large amount of iron ejected by SNe Ia,
while the latter is only determined by the SNe CC contribution.  This
fact leads to an almost flat O/Fe ratio profile at $z=0$, with a
slight increasing trend at large radii.
The central
abundances of both Fe and O, and the almost flat O/Fe ratio show that
both types of SNe contribute in a similar fashion to every part of the
cluster, that is, without a preference for SNe Ia to chemically enrich
the inner regions of the cluster, and for SNe CC to pollute its
outskirts, as inferred from some observations \citep{finoguenov01}.
The main change
in the profiles occurs in the central region ($\la 100 \, h^{-1}$ kpc) 
at late times by a progressive increase of the chemical abundances.  

\begin{figure}
    \centering \includegraphics[width=85mm]{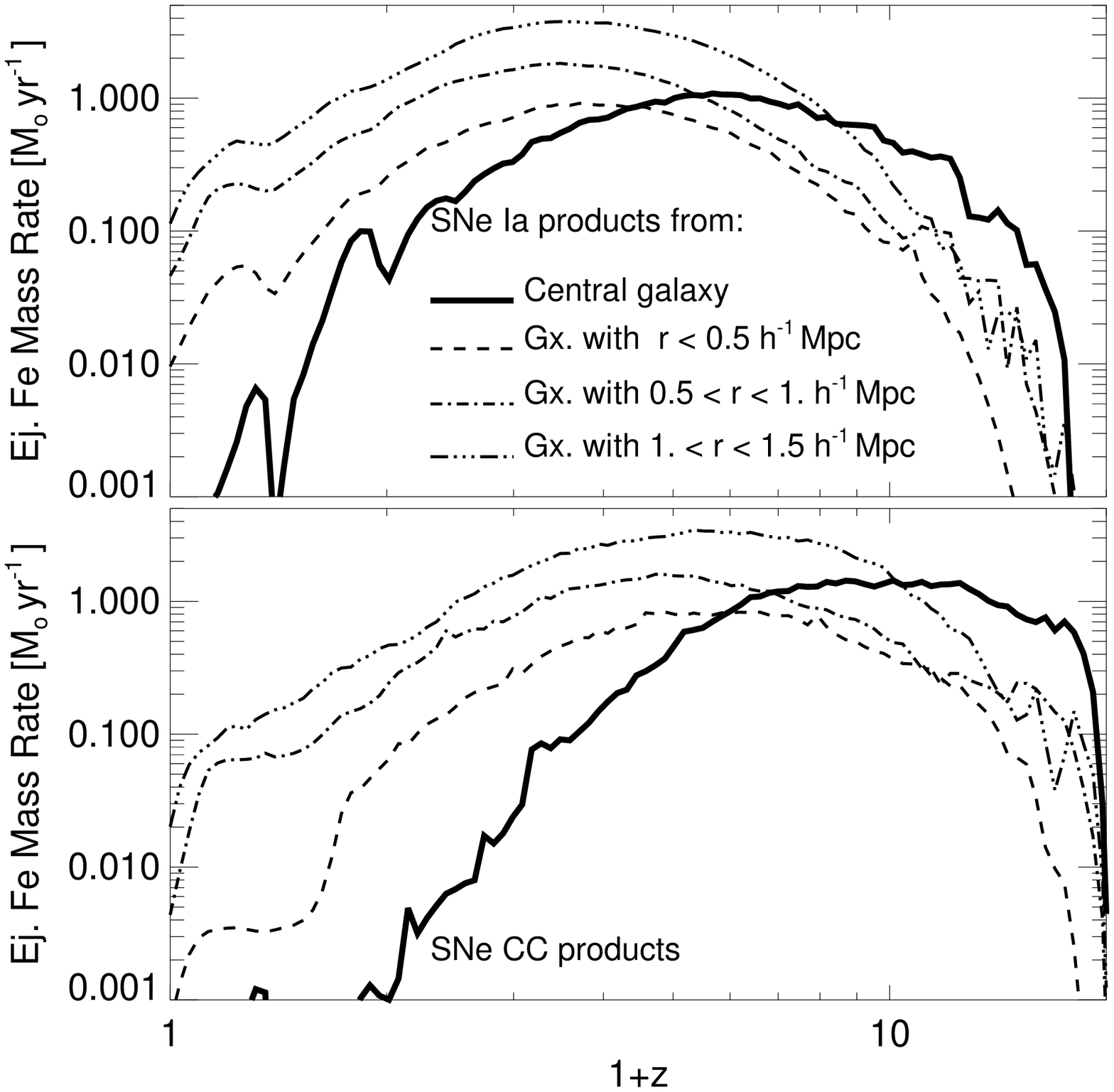}
\caption{Iron mass ejection rate as a function of redshift produced by
SNe Ia (top panel) and SNe CC (bottom panel), contained in galaxies
that lie within different distances from the cluster centre at $z=0$:
the dominant cluster galaxy (at the cluster centre, thick solid line),
galaxies within the spherical region with radius $0.5 \, h^{-1}$ Mpc
without the central galaxy (dashed line), galaxies within the
spherical shells between $0.5$ and $1. \, h^{-1}$ Mpc (dash dot line),
and between $1.0$ and $1.5 \, h^{-1}$ Mpc (dash dot dot dot line).  }
\label{figEjEvol2}
\end{figure}

The evolution of the radial abundance profiles of the main SNe Ia and CC
products (iron and oxygen, respectively) from $z \sim 1$ to $z=0$,
can be explained by taking into account the history of metal ejection
of the cluster galaxies as well as
the dynamics of the gas associated to them while galaxies are being accreted
onto the cluster.
                                                                                
Figure~\ref{figEjEvol2} shows the iron mass ejection rate
as a function of redshift produced by SNe Ia (top panel) and SNe CC
(bottom panel) that are contained in galaxies that lie within shells
of different distances from the cluster centre at $z=0$.
Since we are focusing on a set of galaxies in the $z=0$ cluster,
the contribution of SNe CC to the ICM chemical
enrichment peaks at higher redshifts ($z \sim 5 - 7$)
than the star formation rate
of the whole simulation (see figure 1 of \citet{ciardi03} for a comparison of
the redshift evolution of the star formation rate for field region simulations
and for cluster simulations).
As we have already mentioned, the SNe CC rate closely follows
the star formation
rate since the lifetimes of the stars involved are quite short.
Instead, the ejection rate of
elements produced by SNe Ia reaches a maximum at $z \sim 3 - 5$, due to
the return time distribution resulting from the adopted model for
SNe Ia explosions.
The peaks in the mass ejection rates
are followed by a strong decline
at lower redshifts for both types of SNe,
such that the ongoing chemical contamination is quite low at $z=0$.
This behaviour is more pronounced for those galaxies that are nearer
to the cluster centre at the present time.

Considering the accumulated mass of iron in the ICM at $z=0$,
we find that SNe Ia are
responsible for $\sim 75$ per cent of the iron content of the ICM.
Interpretation of observed profiles of abundance ratios extends this
percentage up to $\sim 80$ per cent \citep{gastaldello02}. However,
both results are quite dependent of the assumed SNe Ia model.
The point that we have to emphasize is that
the relative contribution of SNe CC and SNe Ia to the iron
content of the cluster is already one-third
at $z\sim 2$, when
the accumulated iron masses provided by both sources peack.
The lack of metal contribution by SNe explosions at late times
from galaxies that are near the cluster centre cannot explain
the increase of abundances in the inner $100 \, h^{-1}$~kpc
since $z\sim 0.1$, suggesting that dynamical processes
are playing an important role in the developement of
abundance profiles.

\begin{figure*}
  \centering
  \begin{minipage}[c]{.33\textwidth}
   \centering \includegraphics[width=58mm]{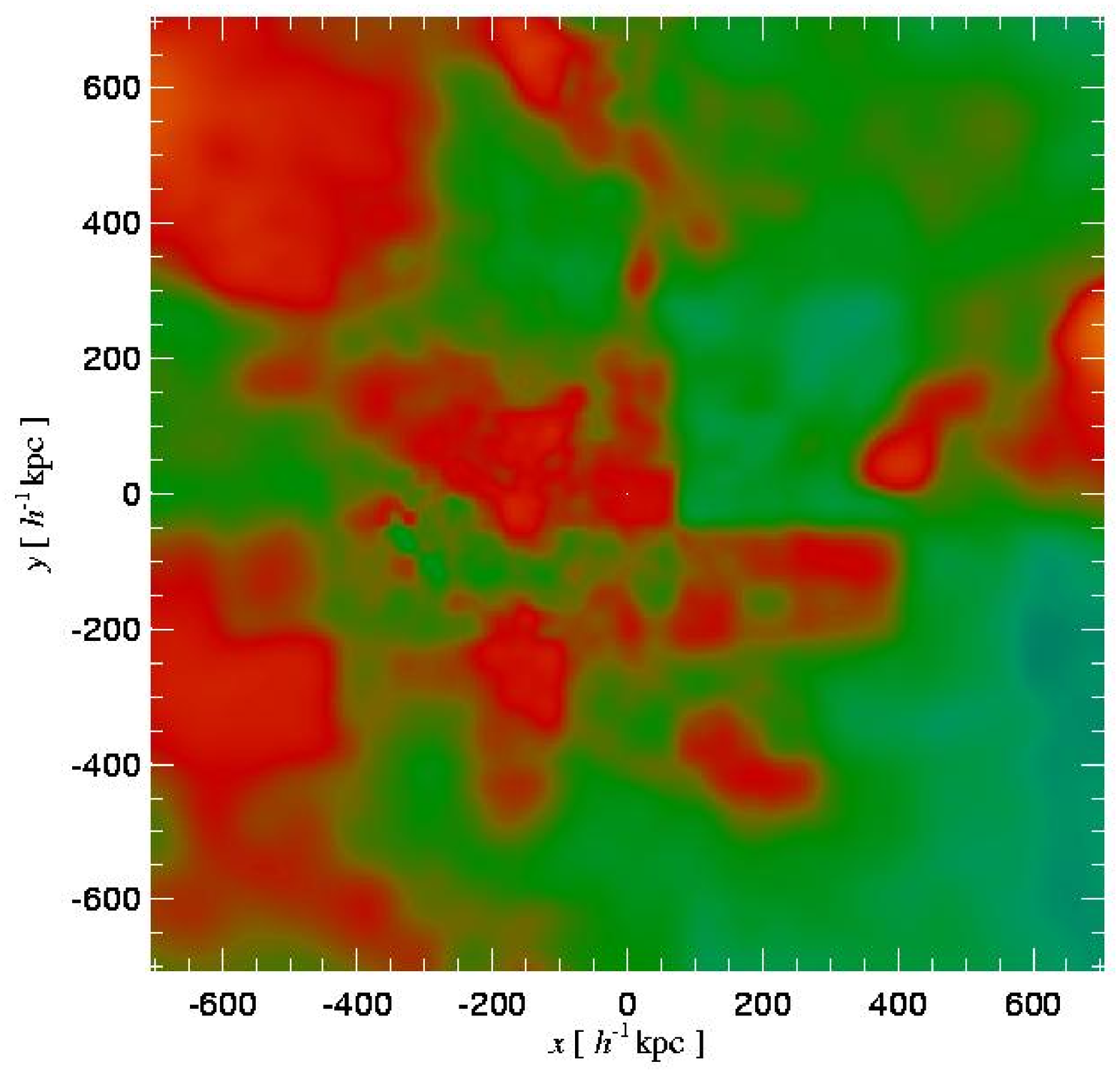}
  \end{minipage}%
  \begin{minipage}[c]{.33\textwidth}
   \centering \includegraphics[width=58mm]{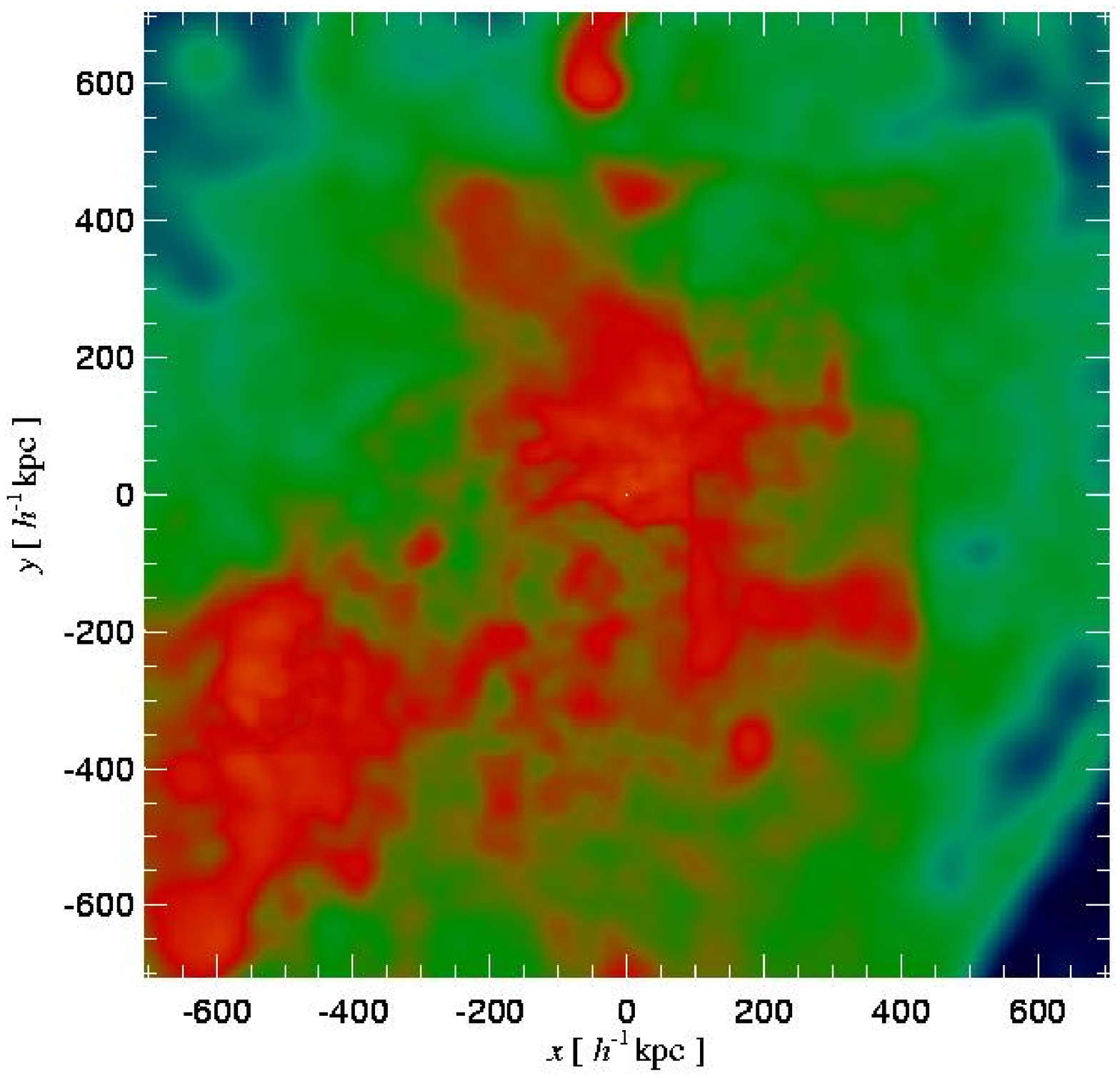}
  \end{minipage}%
  \begin{minipage}[c]{.33\textwidth}
   \centering \includegraphics[width=58mm]{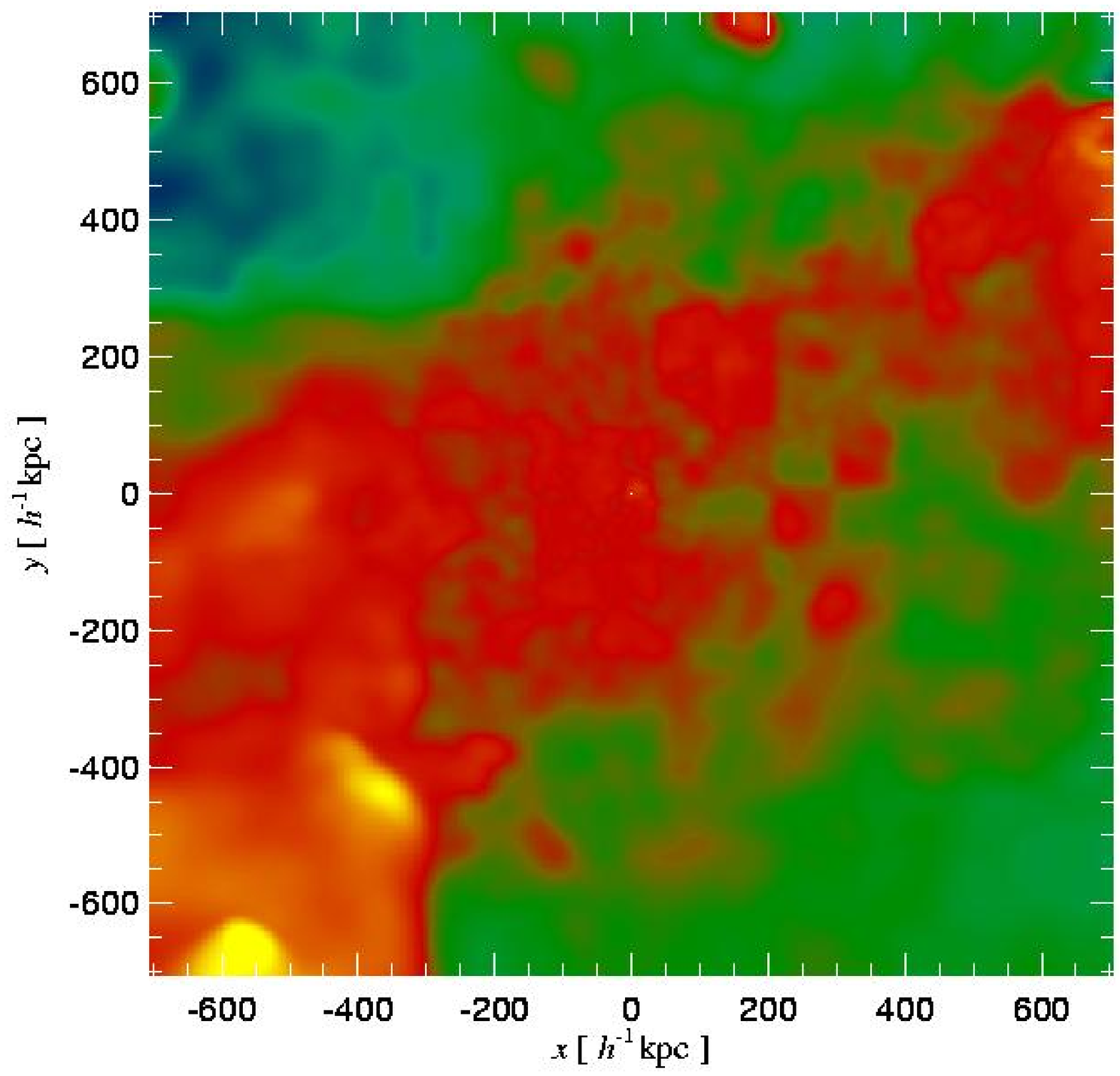}
  \end{minipage}\\
  \begin{minipage}[c]{.33\textwidth}
   \centering \includegraphics[width=58mm]{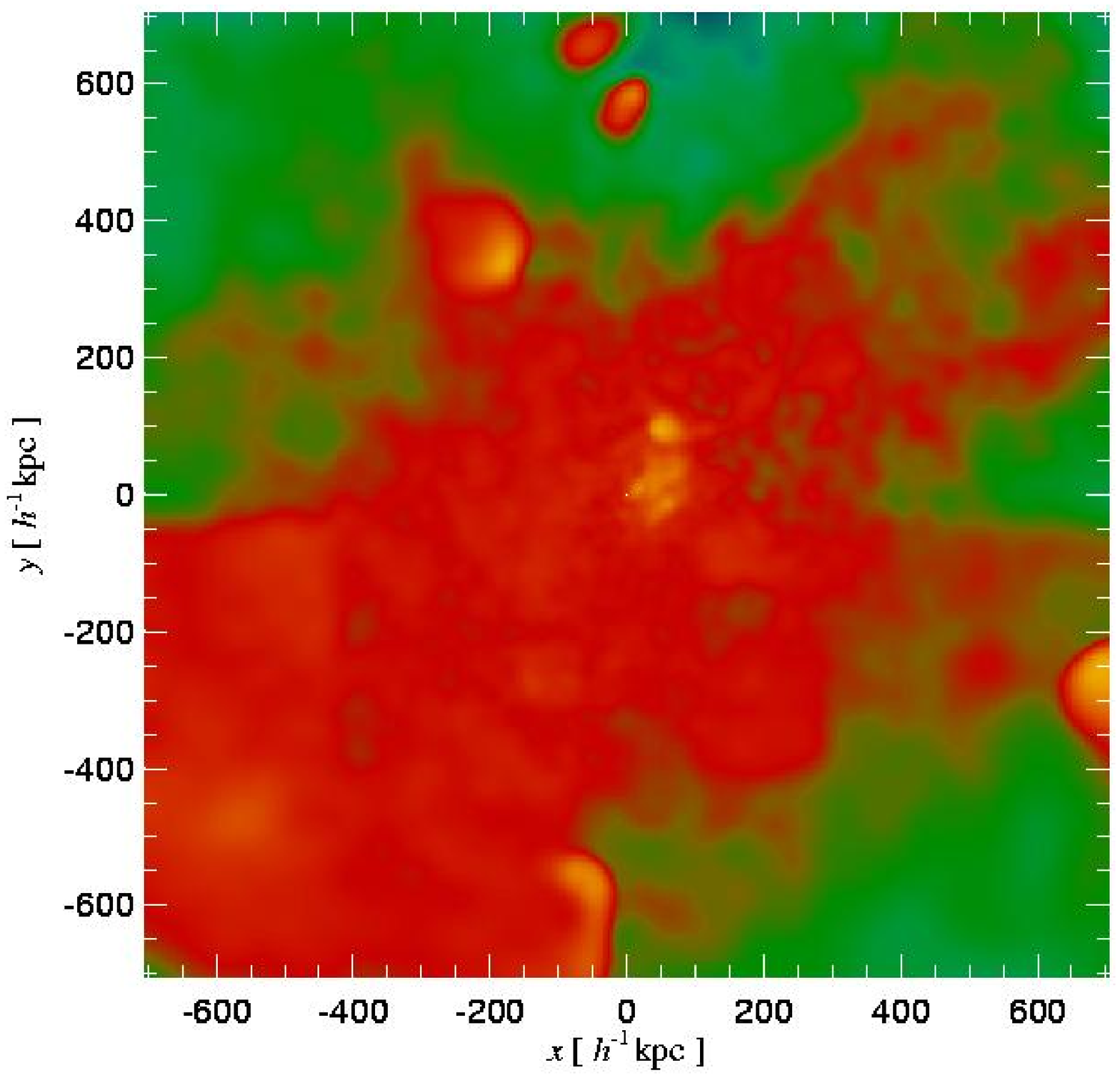}
  \end{minipage}%
  \begin{minipage}[c]{.33\textwidth}
   \centering \includegraphics[width=58mm]{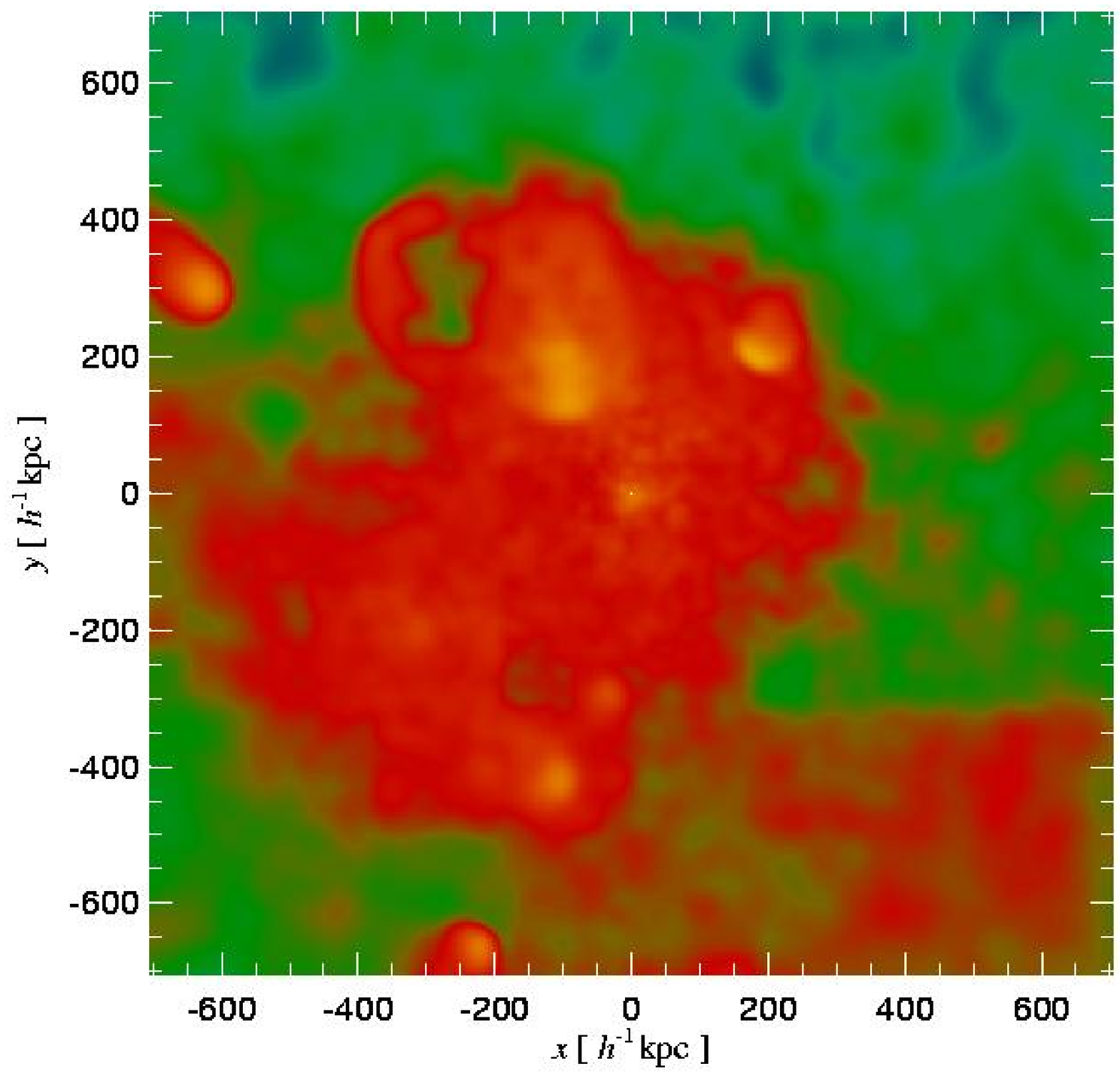}
  \end{minipage}%
  \begin{minipage}[c]{.33\textwidth}
   \centering \includegraphics[width=58mm]{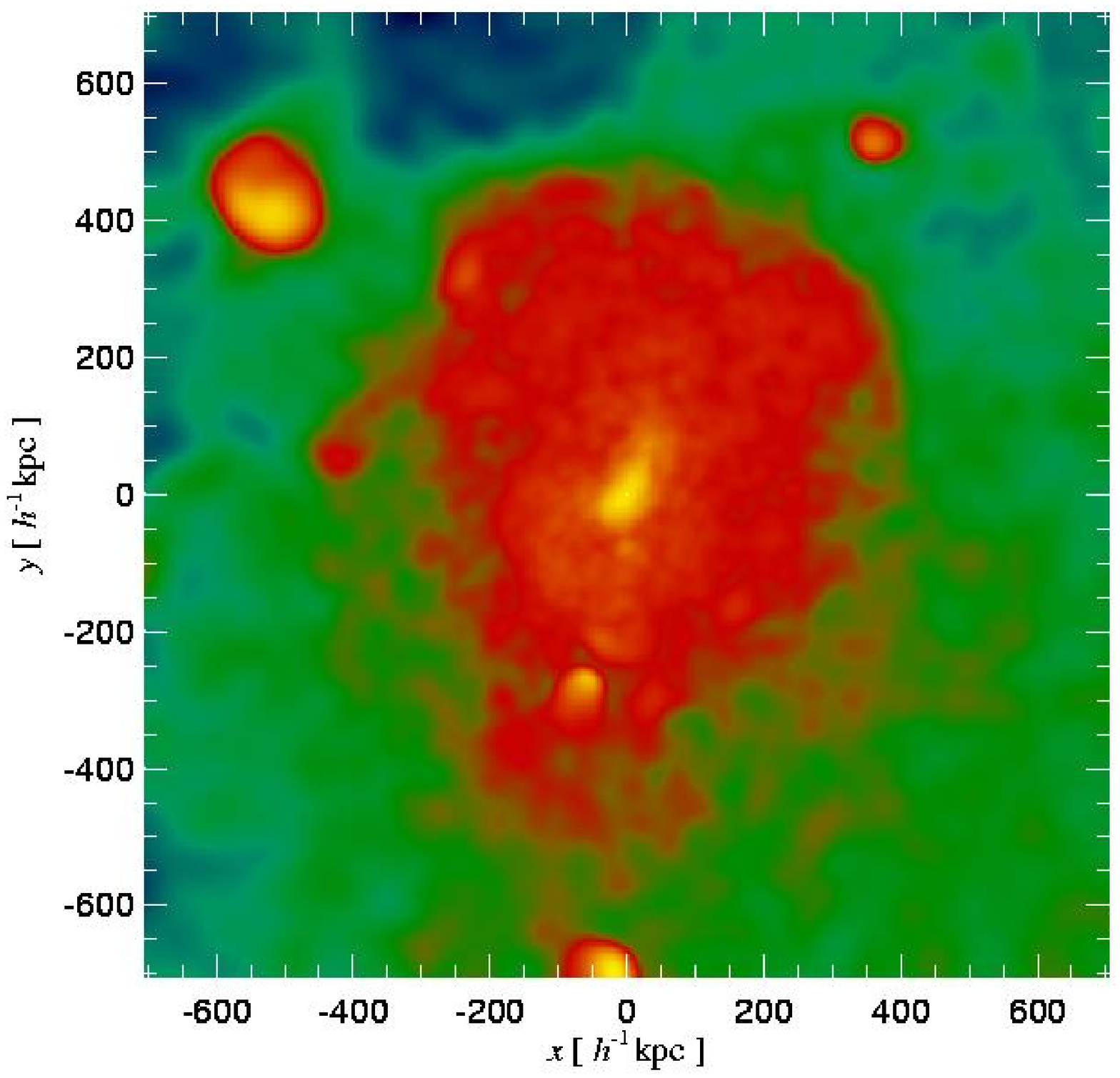}
  \end{minipage}\\
\hspace{6mm}
  \begin{minipage}[c]{.33\textwidth}
   \centering \includegraphics[width=10mm,angle=270]{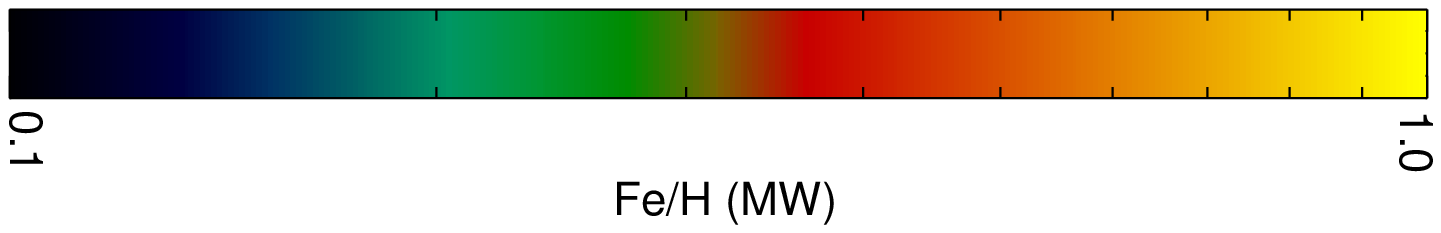}
  \end{minipage}%
\caption{Evolution of iron abundance of gas particles
that lie within $1\, h^{-1}$~Mpc from the cluster centre at 
redshifts: 
$z \sim 1$ (upper left panel),
$z \sim 0.5$ (upper middle panel), 
$z \sim 0.3$ (upper right panel),
$z \sim 0.2$ (lower left panel),
$z \sim 0.1$ (lower middle panel) and
$z = 0$ (lower right panel).
The plots show the projection of mass-weighted iron abundance by number relative
to hydrogen, Fe/H, respect to the solar value. 
At each refshift, the spatial coordinates are centred at the most massive
progenitor of the cluster at $z=0$ and are expressed in comoving scales.
}
\label{figMaps4b}
\end{figure*}

Figure~\ref{figMaps4b} shows the evolution of mass-weighted
iron abundances of gas particles contained within
$1\, h^{-1}$~Mpc from the cluster centre at redshifts
$z \sim 1$, $\sim 0.5$, $\sim 0.3$, $\sim 0.2$,
$\sim 0.1$ and $z=0$.
We clearly see from these plots that highly enriched gas clumps
are already present in the outskirts of the cluster at $z \sim 0.3$ and
that the iron abundance profile
starts to settle down at $z \sim 0.2$. The contaminated gas clumps
located near  
$y \sim 350 \, h^{-1}$~kpc 
at $\sim 0.2$ and
close to 
$y \sim 200 \, h^{-1}$~kpc 
at $\sim 0.1$ seem
to converge to the cluster centre at $z=0$ considerably increasing
the iron abundance in the inner $100 \, h^{-1}$~kpc.
Thus, gas particles bound to the dark matter substructures of `halo galaxies'
(see definition in section \ref{sec_SAM}) have been mainly enriched
at early epochs, when the metal ejection rates from
these galaxies were higher, and have then fallen to the inner regions
of the cluster driven by the haloes of infalling galaxies.
This dynamical effect is quite likely taking into account
the analysis of gas
dynamics based on figures~\ref{figMaps3}, \ref{figMaps4} and \ref{figVelDistz}.
A simple test of finding the position of the progenitor of
the central galaxy at different redshifts reveals that these 
galaxies
are at $\sim 100 \,h^{-1}$~kpc at $z \sim 0.5$, reinforcing our conclusion
based on gas dynamics. 

The scenario of chemical enrichment considered in our model
indicates that gas dynamical effects play an important role
in the developement of abundance patterns. However, 
there are other sources of chemical elements not
included in our model that may help to  
interprete observational results,
such as intracluster stellar populations and ram-pressure stripping.
There is increasing evidence of the 
presence of intracluster stars
based on observations of diffuse light 
(\citealt{feldmeier04,zibetti05}) and individual stars between cluster
galaxies (\citealt{feldmeier03,durrell02,galyam03}). 
High resolution {\em N}-Body/SPH simulations are starting to yield 
predictions of the spatial and kinematic 
distributions for these stars
(\citealt{willman04, murante04, sommerlarsen05}) 
which hold important information about the assembly 
history of a galaxy cluster.
These numerical works show that
a fraction of at least $\sim 10$~per cent of unbound stars 
accumulate within the cluster 
as a result of stripping events and infall of galaxy groups that already 
contain unbound stars.
The impact of intracluster stars in the
chemical enrichment of the ICM has been explored by
\citet{zaritsky04} by using a simple model. 
They demonstrate that this stellar component makes a significant
contribution to the iron content of the ICM. 
The metal pollution
due to ram-pressure stripping has been recently investigated by 
\citet{schindler05} and \citet{domainko05}. These authors argue
that the efficiency of this enrichment mechanism 
is higher than that of supernovae driven
galactic winds, producing a centrally concentrated
metal distribution in massive clusters.

\begin{figure*}
  \centering
  \begin{minipage}[c]{.33\textwidth}
   \centering \includegraphics[width=60mm]{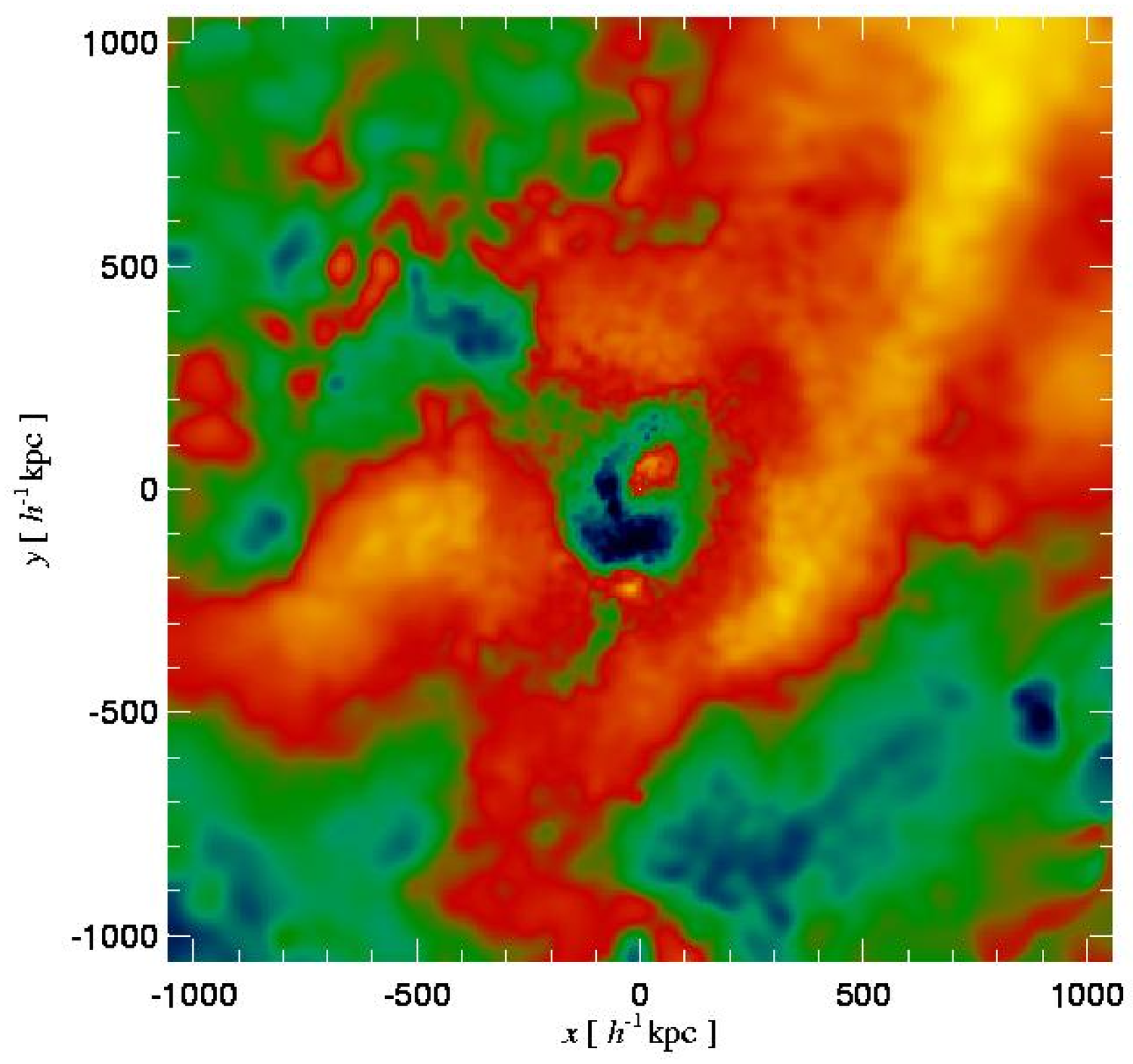}
  \end{minipage}%
  \begin{minipage}[c]{.33\textwidth}
   \centering \includegraphics[width=60mm]{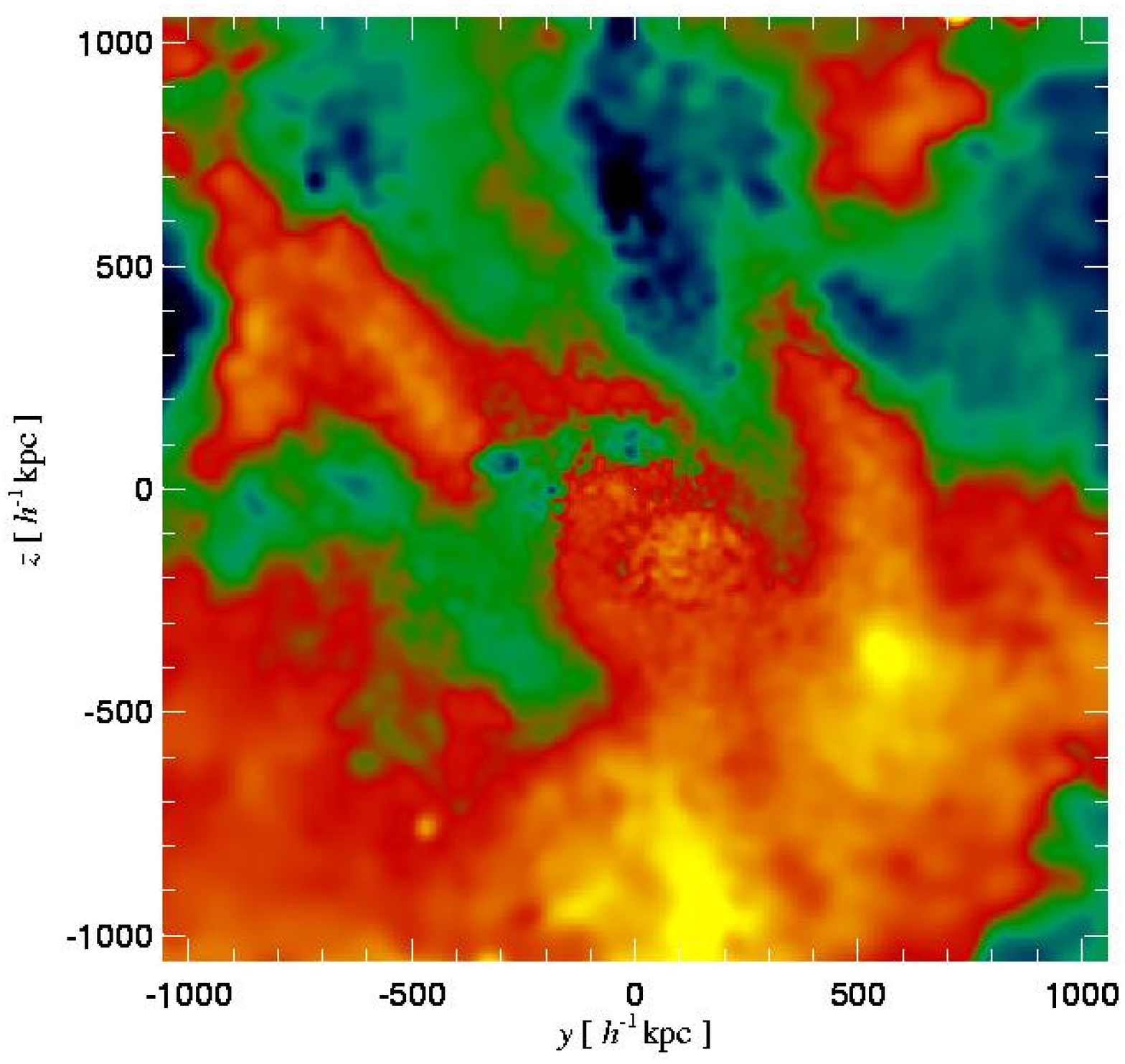}
  \end{minipage}%
  \begin{minipage}[c]{.33\textwidth}
   \centering \includegraphics[width=60mm]{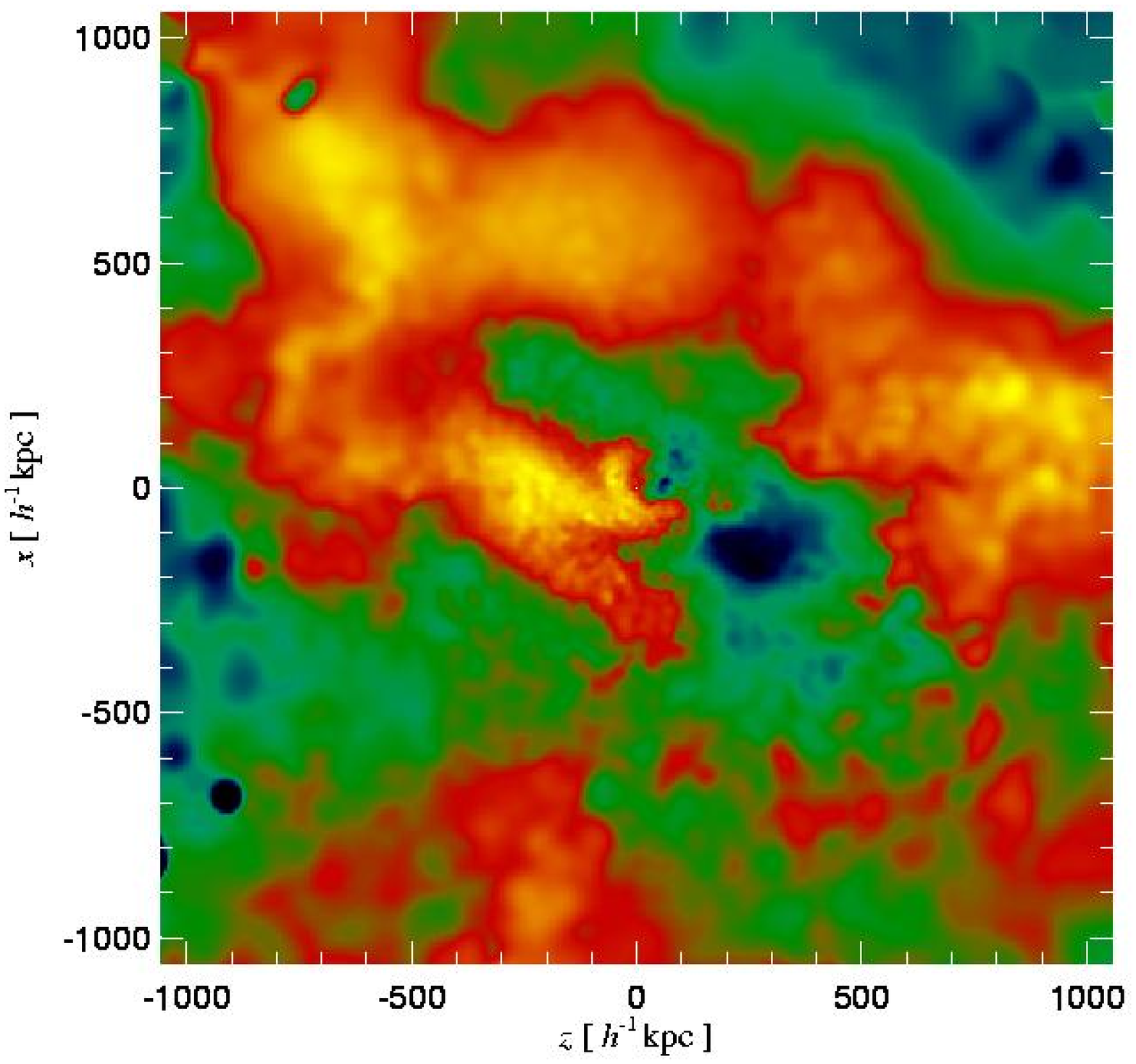}
  \end{minipage}\\
\hspace{6mm}
  \begin{minipage}[c]{.33\textwidth}
   \centering \includegraphics[width=10mm,angle=270]{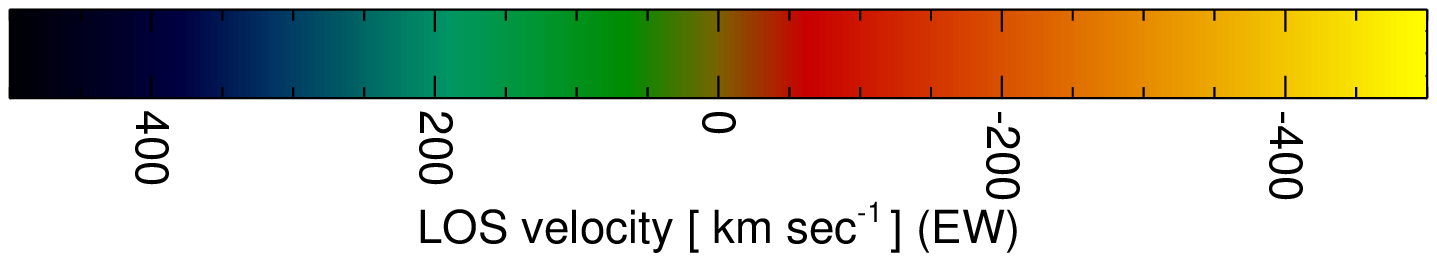}
  \end{minipage}%
\caption{Projection of Fe ${\rm K}_{\alpha}$ 6.7 keV emission-line-weighted 
line-of-sight (LOS) velocities of gas particles contained
within a sphere of radius $1\, h^{-1}\,{\rm Mpc}$ centred on the
dominant cluster galaxy: $x$-$y$ projection (left panel), $y$-$z$
projection (middle panel), and $z$-$x$ projection (right panel).}
\label{figMaps5}
\end{figure*}

\section{Probing the ICM dynamics with metals} \label{sec_Spectra}

In the previous section we have showed how the complex 
dynamical evolution of intracluster gas
influences the developement of the abundance patterns of the
ICM. We now discuss the potentiality of spectroscopic
observations for detecting such strong gas bulk motions,
sheding light on our understanding
of gas dynamics during the cluster formation.

The spatial and spectral resolution of X-ray telescope detectors
onboard of present satellite missions allow the measurement of
metallicity maps and of high resolution emission line spectra.  The
combination of these data contributes enormously to improving our
knowledge about the ICM properties. For example, Fe ${\rm K}_{\alpha}$
6.7 keV emission-line-weighted maps contain valuable information about
thermodynamical and chemical properties of the intracluster gas, as
discussed in Section~\ref{sec_Maps}.  We can also learn about the
kinematics of the intracluster gas and its spatial correlation with
other properties from maps of emission-line-weighted 
line-of-sight (LOS) velocities and from 
Fe ${\rm K}_{\alpha}$ 6.7 keV emission line spectra generated by the gas
along LOSs through the cluster.

Figure~\ref{figMaps5} shows projections of 
Fe ${\rm K}_{\alpha}$ 6.7
keV emission-line-weighted LOS velocities of the gas particles onto
the three orthogonal planes $x$-$y$, $y$-$z$ and $z$-$x$.  Velocities
are relative to the cluster centre, being colour-coded such that red
and yellow areas correspond to gas moving towards the observer, and green
and blue represent gas receding from it.  Only material contained
within a sphere of radius $1\,h^{-1}\,{\rm Mpc}$ centred on the
dominant galaxy of the cluster has been included.  These maps exhibit
steep azimuthal and radial gradients, revealing a much more complex
structure than those depicted by the density, temperature and iron
abundance distributions, indicating the presence of
large-scale turbulent motions generated by inhomogeneous infall.
This is consistent with our results on the evolution of
the dynamics of gas particles based on the analysis
of Figure~\ref{figVelDistz}.
The sharp edges defined by red and green
colours near the cluster centre are contact discontinuities between
gas that originated in different clumps.
Note that this is not a projection effect of gas
particles lying far from the cluster centre. These contact discontinuities
are also seen in the projection of particles well
inside the cluster ($\la 50 \, h^{-1}\,{\rm kpc}$).

The gradients visible in the velocity maps of Figure~\ref{figMaps5}
can be as large as $\sim 1000 \,{\rm km} \, {\rm s}^{-1}$ over
distances of a few hundred kpc.  They are of the same order of
magnitude as those obtained from accurate observations of the
Centaurus cluster (Abell 3526) \citep{dupke01}. We note that this
cluster is a very good candidate for this kind of studies, being one
of the closest X-ray bright clusters of galaxies with an optical
redshift of $0.0104$. Its core has been spatially resolved by the {\em
Chandra} Observatory, using the ACIS-S detector
\citep{sandersfabian02}, leading to temperature and abundance maps.
However, there is not enough information in the spectrum to isolate
spectral lines and perform accurate gas velocity measurements.  This
task is possible with the detector GIS on board of {\em ASCA}.  The
gas velocity distribution determined with this instrument
\citep{dupke01} shows a region associated with the subcluster Cen 45,
at $\sim 130 \, h^{-1} \, {\rm kpc}$ from the main group Cen 30
(centred on the cD galaxy), with a radial velocity higher than the
rest of the cluster by $\sim 1700 \, {\rm km} \,{\rm s}^{-1}$.
These radial velocity measurements obtained from X-ray observations
support the optically determined velocity difference between Cen 45
and Cen 30 of $\sim 1300\, {\rm km} \,{\rm s}^{-1}$
\citep{stein97}.

These gradients in radial velocities are observationally detected by
the Doppler shifting of the metal lines used to perform the spectral
fits, which are mainly driven by the Fe ${\rm K}_{\alpha}$ complex.
However, a higher spectral resolution would allow us to detect the
different components in which a certain line is split due to the bulk
motions of different magnitude along a line-of-sight through the
cluster.  Future X-ray missions promise a substantial increase in
energy resolution, paving the way for the construction of a full 3-D
picture of the ICM dynamics.  \citet{Sunyaev03} analyse this
possibility by calculating the synthetic spectra of the Fe ${\rm
K}_{\alpha}$ 6.7 keV emission line along sightlines through a
simulated cluster \citep{Norman99}.  They show that this emission line
is split into multiple components over a range of energy $\Delta E
\sim \pm 15\, {\rm eV}$, as a result of turbulent and bulk motions
of the gas. They estimate that a spectral resolution of 4 eV, similar
to the one planned for future missions, 
would be
enough for the detection of dynamical features in the ICM based on
emission line spectra.

Following the study made by \citet{Sunyaev03}, we analyse the way in
which the special features of the gas motion that become evident in
the velocity maps of Figure~\ref{figMaps5} are reflected in the Fe
${\rm K}_{\alpha}$ 6.7 keV emission line spectra generated by the gas
along lines-of-sight through the cluster.
The spectra are computed by dividing the sightlines in bins of width
$20 \, h^{-1} \, {\rm kpc}$ along directions perpendicular to the
projection planes.  The thermodynamical, kinematical and emission
properties assigned to each bin are obtained from the 64 closest gas
particles to the bin centres using the SPH smoothing technique.  The
energy of the iron line emitted from each bin along the sighline is
Doppler shifted proportionally to the smoothed LOS velocity inferred
from the gas around the bin.  Combining these shifted energy values
with the corresponding intensities of the line emission we obtain the
iron emission line spectra along the LOS.  On top of this, we consider
the broadening suffered by each line due to thermal motions of the
gas.

\begin{figure*}
\includegraphics[width=185mm]{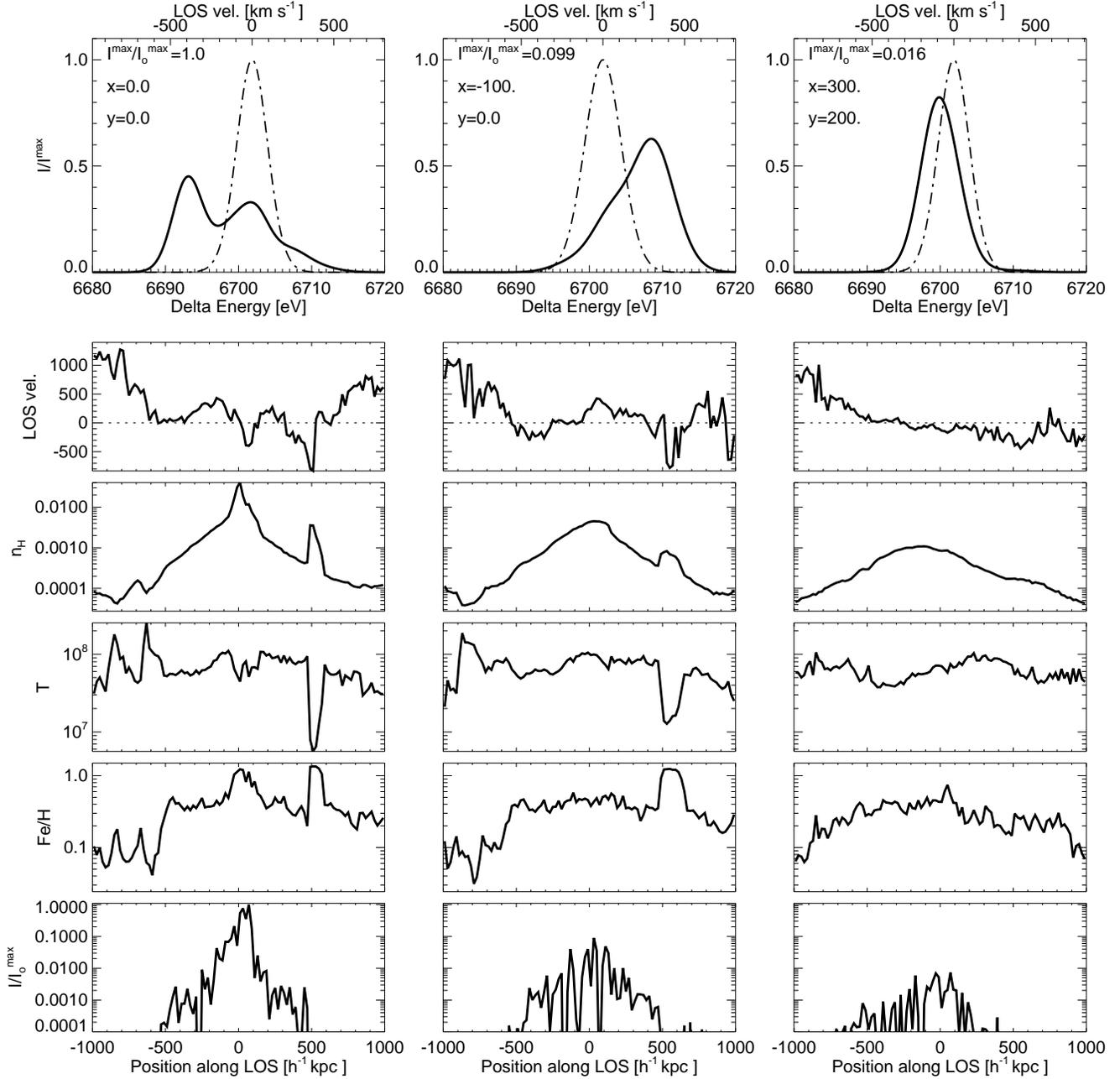}
 \caption{Fe ${\rm K}_{\alpha}$ 6.7 keV emission line spectra along
three lines of sight (LOSs) on the {\em z}-axis through the simulated
cluster, which is projected onto the {\em x}-{\em y} plane, and the
distribution along the corresponding LOS of kinematical,
thermodynamical, chemical and emissivity properties of the gas that
contributes to the spectra.  The upper panels show the effect of
thermal broadening on the Fe emission line (thin dash-dotted curve)
and the composition of thermal broadening and Doppler shift due to gas
bulk motion with respect to the cluster centre (thick solid line).
The maximum values of the broadened lines are used to normalise both
kinds of iron emission lines, with and without Doppler shift.  These
panels indicate the ratio of this maximum with respect to the one
corresponding to the cluster centre, $I^{\rm max}/I_{\rm o}^{\rm
max}$, and the {\em x}-{\em y} position where the LOS intercepts the
ICM. The five plots below each spectra show the distributions of gas
properties along the corresponding sightlines: emission-line weighted
LOS velocities used to generate the spectra, and mass-weighted
hydrogen density $n_{\rm H}$, temperature $T$, iron abundance {\rm
Fe/H}, and normalised emissivity of the ${\rm K}_{\alpha}$ 6.7 keV
emission line $I/I^{\rm max}$. 
Velocities are
expressed in ${\rm km} \, {\rm s}^{-1}$, $n_{\rm H}$ in ${\rm
cm}^{-3}$, $T$ in ${\rm K}$, and iron abundances are referred to the
solar value.  The positions along the LOSs are given
with respect to the cluster centre.}
\label{figSpectra1}
\end{figure*}

\begin{figure*}
\includegraphics[width=185mm]{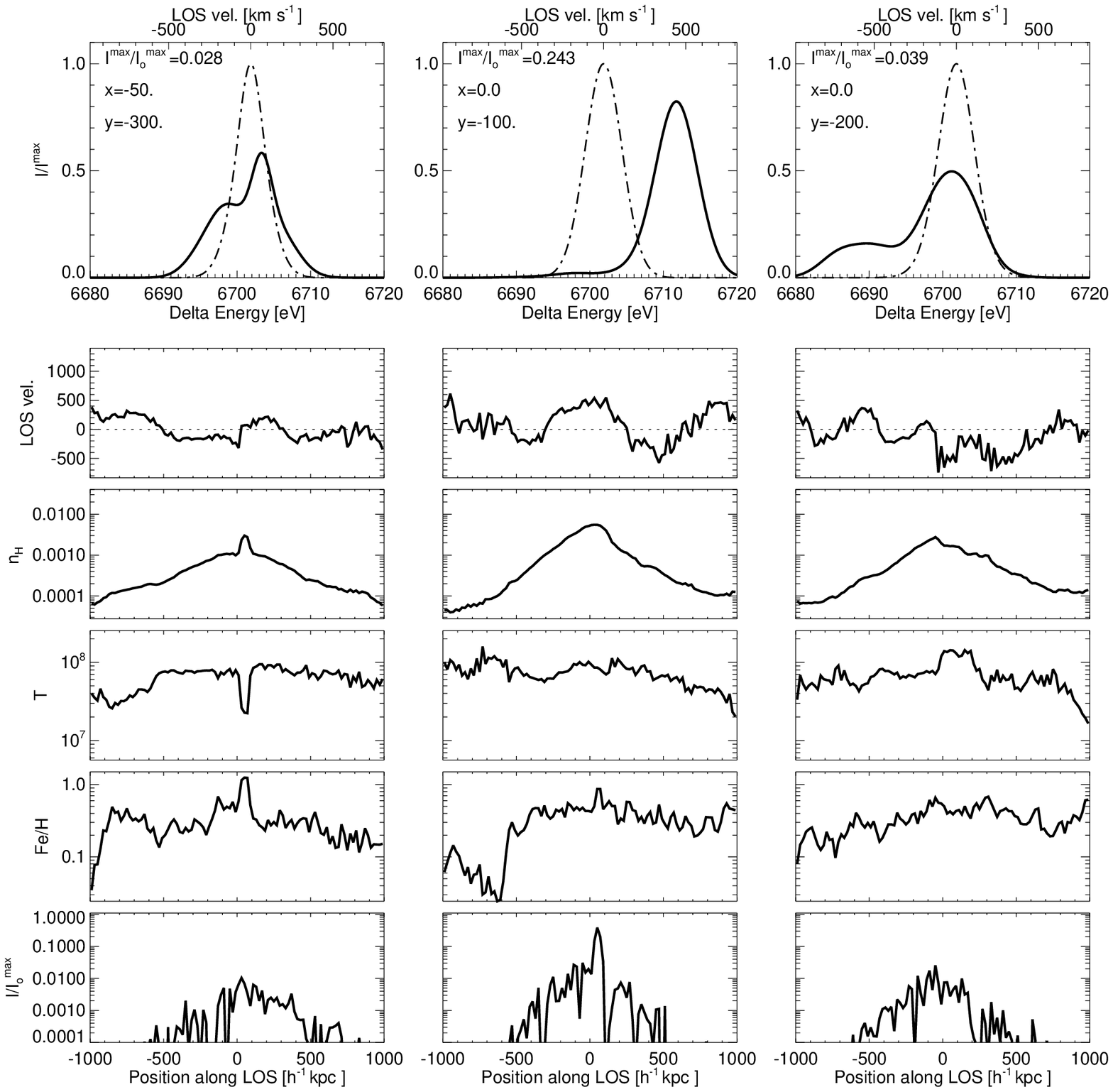}
 \caption{Same as figure~\ref{figSpectra1} for other set of LOSs
that intercept the ICM through areas with large offset velocities
present near the cluster centre as depicted by Figure~\ref{figMaps5}.}
\label{figSpectra2}
\end{figure*}

Figures~\ref{figSpectra1} and \ref{figSpectra2} show
the resulting synthetic spectra for
six LOSs along the {\em z}-axis through the {\em x-y} projection of
the simulated cluster.
Like in the construction of maps, the computation of the spectra only
comprises gas particles within the virial radius of the cluster.
Solid thick lines correspond to spectra that are affected by both Doppler
shifting and thermal broadening.  For comparison, we also include in
each panel the iron emission line affected only by thermal broadening
(dash-dotted lines).
The maximum values of these last ones are used to normalise both kinds
of iron emission lines, with and without Doppler shifting.  The ratios
of these maxima with respect to the one corresponding to the cluster
centre, referred to as $I^{\rm max}/I_{\rm o}^{\rm max}$, are also
given in each of the panels that show the spectra, together with the
location where the LOSs intercept the ICM.  Below these panels,
both Figures~\ref{figSpectra1} and \ref{figSpectra2} display 
the distribution of gas properties
along the corresponding sightlines: emission-line weighted LOS
velocities used to generate the spectra, and mass-weighted hydrogen
density, temperature, iron abundance and normalised emissivity of the
${\rm K}_{\alpha}$ 6.7 keV emission line.  These distributions allow
us to see how gas components characterised by different
thermodynamical and chemical properties move at different velocities,
as well as which parts of the line of sight contribute most to the
spectral lines.

An inspection of these distributions indicates that the gas reaches
high iron abundances at high density and low temperature regions, as
it is also shown by the maps of Figure~\ref{figMaps1}.  This situation is
especially clear in the cluster core and in a subclump moving at high
speed located at $\sim 500 \, h^{-1} \, {\rm kpc}$ from the cluster
centre, as it is evident from the first and second columns of
Figure~\ref{figSpectra1}.  The Doppler shifting and split of the
emission line at the cluster core are produced by the large-scale
turbulent motions of the gas,
which produces a LOS velocity gradient from
$\sim 200$ to $-400 \,{\rm km}\, {\rm s}^{-1}$ around the middle of
the sightline.  This motion is reflected in the spectra because the
intensity of the iron emission line is larger than $10$ per cent the central
value within a radius of $\sim 100 \, h^{-1} \, {\rm kpc}$ from the
cluster centre.  However, this intensity decays quite abruptly at
larger radii and becomes undetectable (see also the mass-weighted
emission map in Figure~\ref{figMaps1}).  For this reason, the subclump
moving at $-800 \,{\rm km}\, {\rm s}^{-1}$ from the cluster centre
does not produce any signature in the first two Doppler shifted
spectra shown. If the emissivity were higher at the location of this
substructure, a peak would appear at $\sim 6,685 {\rm eV}$.
The second and third spectra lines in figure~\ref{figSpectra1}
are produced by sightlines that
intercept the cluster at a distance of $100$ and $360 \, h^{-1} \,
{\rm kpc}$ from its centre, respectively.  Note how the ratio $I^{\rm
max}/I_{\rm o}^{\rm max}$ decreases as we get further from the cluster
core.  The second spectrum shows a clear Doppler shifting and
broadening of the line, while the third one is only slightly altered
with respect to the reference broadened line, in agreement with the
LOS velocity distributions within the inner $\sim 100 \, h^{-1} \,
{\rm kpc}$.

Figure~\ref{figSpectra2} presents three more synthetic spectra generated from
lines of sight that intercept the cluster through areas with large
offset velocities near its centre (see figure~\ref{figMaps5}).
The first one shows a split of the emission line in two components
arising from the central velocity gradient from
$\sim -200$ to $100 \,{\rm km}\, {\rm s}^{-1}$.
The line in the second spectrum is widely Doppler shifted as
a result of high bulk motions of $\sim 500 \,{\rm km}\, {\rm s}^{-1}$
in the central parts of the sightline. 
In the third emission line spectrum, a second component becomes apparent 
as a result of gas moving
at $\sim -500 \,{\rm km}\, {\rm s}^{-1}$.
As we noted in the previous
set of spectra in Figure~\ref{figSpectra1}, 
the strong gas motions of $\sim -500 \,{\rm km}\, {\rm s}^{-1}$
and $\sim -700 \,{\rm km}\, {\rm s}^{-1}$
occurring at $\sim 500 \, h^{-1}$~Kpc from the centre of the lines
of sight considered in the second and third columns 
are not imprinted in the corresponding spectra because of the low
emissivity that characterises this region. 
The shifts and splits observed in all these spectra are of the order
of $\sim 10 - 15 \, {\rm eV}$.  This effect is too small to be
resolved by present instruments, since, for example, the spectral
resolution of the EPIC pn detector for the Fe ${\rm K}_{\alpha}$ 6.7
keV emission line is 150 eV. 

From this analysis we can conclude that only gas motions that produce
absolute values of LOS velocities higher than $\sim 400 \,{\rm
km}\, {\rm s}^{-1}$ around the cluster core, where the iron line
emissivity is larger than $10$ per cent the central value, could become
observable by future X-ray missions through the shifting and split of
the line in two components due to Doppler effects.
Lower energy lines like the Fe L-blend centred around $1$~keV
could be also used for the present purpose. In principle, they should show
much more structure since they are strongly emitted in lower temperature
gas. However, gas with this characteristic lies far from the centre
of our hot cluster and the low line emissivity in these areas
prevent us from getting more information from these lines than that
provided by the Fe ${\rm K}_{\alpha}$ line at 6.7 keV. These lines
are probably much more usefull in a lower temperature cluster.
                                                                                
We would like to note that the hydrodynamical simulation considered by
\citet{Sunyaev03} ignores radiative cooling, which affects the thermal
structure in cluster cores, and they assume constant iron abundances
to calculate the iron emission line spectra. Our simulation suffers
from the first drawback as well; however, the critical
advantage of our hybrid model 
over approaches like the one used
by \citet{Sunyaev03} is that the chemical abundances of gas
particles are dynamically
and consistently generated from the stars in the galaxies.

Taking into account the information supplied by LOS velocity maps and
by spectra along selected sightlines, we can connect the global
features of bulk motions in the gas with the multiple components in
the spectra that originate as a result of these motions.  Note that
the latter feature can be more easily observationally detected and
quantified.  Thus, our results can constitute valuable help for the
interpretation of observational data.

\section{Summary and conclusions}\label{sec_Conclu}

We have presented an hybrid model for the chemical enrichment of the
intracluster gas that allows a detailed analysis of the spatial
distribution of chemical, thermodynamical and kinematical properties
of the ICM. Our model combines a cosmological non-radiative
hydrodynamical {\em N}-Body/SPH simulation of a cluster of galaxies,
and a semi-analytic model of galaxy formation.  
The spatial distribution of metals in the ICM can be accounted for by the
chemical enrichment of the gas particles in the underlying 
{\em N}-Body/SPH simulation.
This enrichment results from the metals ejected by
the galaxy population, which are generated by applying the
semi-analytic model to the the dark matter haloes detected in the
outputs of the simulation. 
This link between semi-analytic model results and the chemical
enrichment of gas particles 
leads to a consistent chemical enrichment scheme for the intracluster
gas, being the key ingredient to carry out the present project.

The parameters that characterise our model are chosen according to
several observational constraints.
The proper circulation of metals among the different baryonic components
leads to a spatial distribution of chemical elements in the intracluster gas
that closely resemble observed abundance patterns, as can be
seen from the comparison of the radial abundance profiles
with data from \citet{tamura04} (Figure~\ref{figRP1}). The corresponding  
mean iron abundance of the intracluster medium at $z=0$
is $[{\rm Fe/H}] \sim 0.28 $. It has been 
reached with the contribution of SNe Ia that provide $\sim 75$ per cent of the
total ICM iron content at $z=0$. 
While this percentage is quite dependent on the parameters that 
characterise the chemical model implemented, what reflects the
circulation of metals is the fraction of elements from different
sources that end up in the ICM as opposed to in stars.
The fraction of iron originated in SNe Ia that is
contained in the ICM is $\sim 0.78$ while that locked in the stars
of cluster galaxies is $\sim 0.21$.
These fractions are slightly different for the iron ejected by SNe II,
being  
$\sim 0.72$ and $\sim 0.28$ for hot gas and stars, respectively.
The increment of the proportion of SNe II products in stars reflects
the relative time-scales for star formation and metal pollution by 
different types of supernovae.

After showing that 
our model reproduces reasonably well
the distribution of metals in the ICM at $z=0$,  
we analyse the evolution of chemical
and dynamical properties of the intracluster gas
in order to understand 
the physical processes involved in the
determination of its final abundance patterns. The aim of this study
is also 
to provide hints that helps in gaining information about
the characteristics of gas bulk motions from spectroscopic observations. 
We analyse the rich information provided by our hybrid model in three
ways,
\begin{enumerate}
\item by constructing radial abundance profiles of the ICM,
\item by building projected mass-weighted and Fe ${\rm K}_{\alpha}$
  6.7 keV emission-line-weighted maps of chemical, thermodynamical and
  kinematical properties of the diffuse gas,
\item and by calculating synthetic spectra of the Fe ${\rm
K}_{\alpha}$ 6.7 keV emission line along sightlines through the
simulated cluster.
\end{enumerate}

The spatial distribution of chemical elements leads to radial
abundance profiles with negative gradients, irrespective of the type
of source from where they originated. The central enhancement of the
Fe and O abundance profiles and the flat profile of the O/Fe ratio
support the notion that the intracluster gas is polluted in the same
way by both types of SNe.
From the analysis of the evolution of these radial abundance profiles and  
the projected mass-weighted iron abundance of gas particles at different
redshifts we find that a high level of enrichment of the ICM is already 
reached at $z\sim 1$. Despite the delay time for the ejection of 
metals inherent to SNe Ia, the contribution of both types of SNe    
to the ICM chemical enrichment peaks at $z\sim 2$.

The lack of evolution of mean iron abundances up to $z \sim 1$ is
in agreement with observations of distant clusters \citep{tozzi03}.
However, the final abundance profile of the cluster gets steeper
within the inner $\sim 100 h^{-1}$~Kpc during the last Gyr
(see the increments
between $z\sim 0.1$ and $z=0$ in Figure~\ref{figProfEvol}).
This cannot be accounted for the metal production of the
cluster galaxies.
The peaks in the rates of the ejected iron mass from 
these galaxies are followed by
a strong decline towards lower redshifts for both types of SNe, such that
the ongoing chemical contamination at $z=0$ is quite low; this
behaviour is more pronounced for galaxies that are nearer to the cluster
centre at the present time.
The combination of all these results indicates that the enhancement of
central abundances produced at low redshift cannot be
explained by metals ejected during this time interval from galaxies
that lie close to the cluster centre.  Indeed, galaxies within a
sphere of radius $500 \, h^{-1} \, {\rm kpc}$ centred on the dominant
cluster galaxy do not contribute to the metal enrichment at
recent epochs.  Instead, a scenario appears favoured in which gas
particles are primarily enriched at high redshifts, and are
subsequently driven to the cluster centre by bulk motions in the
intracluster gas.
Hence, the main cause that contribute to this central abundances
enhancement is the
dynamical evolution of gas particles during the cluster formation.

Gas particles develop very high velocities ranging from 
$\sim 1300$ to $\sim 2500 \,{\rm km} \,{\rm s}^{-1}$
when they are farther than $\sim 1 \,h^{-1}$~Mpc from the cluster centre
at $z > 0.2$. Once these high velocity gas clumps cross this typical cluster
shock radius, 
which correspond approximately to the
virial radius of the most massive progenitor of the cluster
at these redshifts, they slow down ending up within
the inner $\sim 500 h^{-1}$~Kpc of the cluster at $z=0$.
The inhomogeneous gas infall produces contact discontinuities 
that are manifested as sharp edges in the 
emission-line-weighted line-of-sight velocity maps. 
These maps provide useful
information about the gas dynamics in the ICM and its dependence on
spatial position within the cluster.  They reveal a
quite complex structure of the gas dynamics, including the presence of
strong radial and azimuthal gradients with values as large as $\sim
1000 \,{\rm km}\, {\rm s}^{-1}$. Gradients of this magnitude are also
seen in velocity distributions along sightlines through the cluster
centre.  Doppler shifting and broadening suffered by the Fe ${\rm
K}_{\alpha}$ 6.7 keV emission line along sightlines could be used to
probe these gas bulk motions when they are produced within an area
characterised by high iron line emissivity.  Velocities of $\sim
400 \,{\rm km}\, {\rm s}^{-1}$, typically found within a radius of
$100 \, h^{-1} \, {\rm kpc}$ from the cluster core, produce Doppler
shifts of the line of the order of $\sim 10 - 15 \, {\rm eV}$.  The
line can be split in multiple components depending on the structure of
the velocity field.  We find that the higher velocities reached by the
intracluster gas beyond $500 \, h^{-1} \, {\rm kpc}$, or those aquired
by infalling subclumps located at such large distances from the
cluster centre do not produce a detectable signature in the spectral
lines because of the rather low iron line emissivity in the outskirts
of the cluster.

The combination of all these results 
strongly supports the fact that abundance patterns
of the ICM are the result of gas mixing
because of hydrodynamical processes during cluster formation,
being a considerable amount of the gas mainly enriched at high redshifts, 
before the cluster have
time to virialise.
The central enhancement of Fe and O abundance profiles might be considered as
produced by a particular situation taking place in our 
simulation,
in which two big highly enriched gas subclumps converge at the cluster
centre at the present epoch. 
However, this kind of behaviuor is consistent 
with the hierarchical formation scenario, in which accretion and merging
of substructures constitute the main processes driving the formation
and evolution of clusters of galaxies. Thus, the specific shape of
metallicity profiles might well be a consequence of the combination of two
time-scales, one related to the metal production of galaxy clusters
and the other to the
metal mixing. In this last case, both the effect of gas dynamics,
galaxy mergers and the
efficiency in the diffusion of metals due to energetic and chemical
feedback associated with the star formation process play an important
role.
Other enrichment processes 
not taken into account in our model, such as
metal pollution by intracluster stars and ram-pressure stripping, 
may contribute significantly to recover the observed abundance patterns 
reinforcing the
effect of gas dynamics discussed here. 

Our analysis techniques in this study yield a 3-D picture of the
chemical and dynamical properties of the ICM. The information about
the latter can be extracted from the metal content of the intracluster
gas.  Our results show that gas bulk motions are imprinted in the
shape of X-ray lines of heavy ions.  Since thermal broadening is
small, these features could be detected with the high-resolution
spectrographs of future X-ray missions
({\em CONSTELLATION-X} and {\em XEUS}).
Such observations should be
very important in determining the physical processes involved in the
formation and evolution of galaxy clusters, contributing with clues
that allow us to verify 
the enrichment scenario inferred from our model in which gas dynamics
has a crucial effect.

Finally, even though all the processes considered in our hybrid model
are tuned to satisfy numerous observational constraints, the implied
global level of iron enrichment we find is on the low side of the
generally accepted observed range, which is around one-third the solar
value.  This may in part be due to the uncertainties in the stellar
yields and to our poor knowledge of the strength and nature of the
feedback processes involved, but could also point to a systematic
problem with the observational estimates. Our model so far only
accounts for the effect of SNe driven outflows, but it should be
possible to include other sources of feedback, like those from AGN, as
well. This is left for future work, and may perhaps provide a solution
to some of the puzzles still present in understanding the observed
cluster metallicities.

\section*{Acknowledgments}

Simon White and Volker Springel are warmly thanked for making possible
this project and for very useful comments and discussions. Volker
is specially acknowledged for providing the simulation and
postprocessing codes.
We are grateful to the anonymous referee for useful comments and suggestions
that improved the presentation of this work.
We thank Gabriella De Lucia, Patricia Tissera, Felix Stoehr, 
Amina Helmi, Mariano
M\'endez, Marcus Br\"uggen, Diego Garc\'{\i}a Lambas, Hern\'an Muriel,
Stefano Borgani, Fabio Gastaldello, Luca Tornatore and
Naoki Yoshida for helpful and stimulating discussions.  
Gabriella is
also acknowledged for making available the version of the
semi-analytic code used in DL04, on which our model is based.  We
thank Laura Portinari for providing the tables of stellar
yields, and for useful comments. Part of the calculations included in
this work were done with the publicly available programs {\small
CLOUDY} and {\small MICE}. Fundaci\'on
Antorchas is gratefully acknowledged for the external postdoctoral 
fellowship that allowed S.A.C. to start this project and for the Reentry
Grant awarded when returning to her home institution. Part of this
project was financially supported by a grant from Consejo Nacional de
Investigaciones Cient\'{\i}ficas y T\'ecnicas. S.A.C. thanks the
hospitality of MPA during her postdoctoral stay at the institute.

\bsp

\label{lastpage}

\end{document}